\documentclass[11pt,a4paper]{article}
 \pdfoutput=1
\usepackage{jcappub}

\usepackage{amsmath,amssymb}
\usepackage{graphicx}
\usepackage{dcolumn}
\usepackage{bm,color}
\usepackage{hyperref}
\usepackage{accents}
\usepackage{amssymb,float}
\usepackage{amsmath}
\usepackage{multirow}

\hypersetup{
    colorlinks=true,
    linkcolor=red,
    filecolor=magenta,      
    citecolor=blue
}

\newcommand{\udt}[3]{#1^{#2}_{\phantom{#2}#3}}

\newcommand{\dut}[3]{#1_{#2}^{\phantom{#2}#3}}

\newcommand{\lc}[1]{\accentset{\circ}{#1}}

\title{\boldmath Reconstructing teleparallel gravity with cosmic structure growth and expansion rate data}

\author[a,b]{Jackson Levi Said,}
\author[a]{Jurgen Mifsud\footnote{Author to whom any correspondence should be addressed.},}
\author[c]{Joseph Sultana,}
\author[a,b,d]{and Kristian Zarb Adami}

\affiliation[a]{Institute of Space Sciences and Astronomy, University of Malta, Msida, MSD 2080, Malta}
\affiliation[b]{Department of Physics, University of Malta, Msida, MSD 2080, Malta}
\affiliation[c]{Department of Mathematics, University of Malta, Msida, MSD 2080, Malta}
\affiliation[d]{Department of Physics, University of Oxford, Denys Wilkinson Building, Keble Road, Oxford OX1 3RH, UK}

\emailAdd{jackson.said@um.edu.mt}
\emailAdd{jurgen.mifsud@um.edu.mt}
\emailAdd{joseph.sultana@um.edu.mt}
\emailAdd{kristian.zarb-adami@um.edu.mt}

\abstract{
In this work, we use a combined approach of Hubble parameter data together with redshift--space--distortion $(f\sigma_8)$ data, which together are used to reconstruct the teleparallel gravity (TG) Lagrangian via Gaussian processes (GP). The adopted Hubble data mainly comes from cosmic chronometers, while for the Type Ia supernovae data we use the latest jointly calibrated Pantheon compilation. Moreover, we consider two main GP covariance functions, namely the squared--exponential and Cauchy kernels in order to show consistency (to within 1$\sigma$ uncertainties). The core results of this work are the numerical reconstructions of the TG Lagrangian from GP reconstructed Hubble and growth data. We take different possible combinations of the datasets and kernels to illustrate any potential differences in this regard. We show that nontrivial cosmology beyond $\Lambda$CDM falls within the uncertainties of the reconstructions from growth data, which therefore indicates no significant departure from the concordance cosmological model.
}

\begin{document}

\maketitle
\flushbottom

\section{\label{sec:intro}Introduction}
The $\Lambda$CDM cosmological model is overwhelmingly supported by a wide range of cosmological evidence at all cosmological scales \cite{dodelson2003modern,Clifton:2011jh}, which incorporates a significant portion of particle physics beyond the standard model \cite{weinberg_1995}. In this setting, galactic structures are stabilised by cold dark matter \cite{Baudis:2016qwx,Bertone:2004pz}, and dark energy produces the observed late-time accelerated expansion \cite{Riess:1998cb,Perlmutter:1998np} through a cosmological constant $\Lambda$ \cite{Peebles:2002gy,Copeland:2006wr}. Despite decades of work, internal inconsistencies remain present in $\Lambda$CDM cosmology \cite{Adler:1995vd,RevModPhys.61.1}, while the possibility of direct measurements of dark matter continues to become more elusive \cite{Gaitskell:2004gd}. \medskip

Recent observations have called into question the predictive power of $\Lambda$CDM cosmology, as its effectiveness in confronting cosmological tensions becomes an increasingly contentious issue \cite{DiValentino:2020zio}. Primarily, this has taken the form of differences in the values of the Hubble parameter with local, and cosmological model-independent, measurements from type Ia supernovae \cite{Riess:2019cxk} and strong lensing by distant quasars \cite{Wong:2019kwg} giving much higher values when compared with early Universe, and cosmology model-dependent, observations \cite{Aghanim:2018eyx,Ade:2015xua}. While measurements from the tip of the red giant branch (TRGB, Carnegie-Chicago Hubble Program) and gravitational wave astronomy \cite{Baker:2019nia,2017arXiv170200786A,Barack:2018yly} point to a lower $H_0$ tension, a number of other works point to an even larger tension in the Hubble parameter \cite{Riess:2020sih,Pesce:2020xfe,deJaeger:2020zpb}. A similarly important quantity that appears to be expressing a growing tension between local and early Universe measurements is that of the growth rate of large scale structures \cite{Aghanim:2018eyx,DiValentino:2020vvd}. In this context, the growth rate from a particular cosmological model $f(a)=\mathrm{d}\ln\delta (a)/\mathrm{d}\ln a$ (where $\delta (a)=\delta\rho/\rho$ is the linear matter overdensity), and the matter power spectrum normalised on scales of $8 h^{-1} {\rm Mpc}$ $\sigma_8^{} (a)$ (see section \ref{sec:f_T}) are combined in the parameter $f\sigma_8^{} (a) = f(a) \sigma_8^{}(a)$ \cite{Kazantzidis:2018rnb,Douspis:2018xlj} with $a$ being the cosmic scale factor. This parameter naturally emerges in cosmological theories of gravity through linear scalar perturbations and has been used in a number of analyses in competing theories of gravity beyond GR \cite{Clifton:2011jh,Lambiase:2018ows}. \medskip

Efforts continue to resolve these puzzling cosmological tensions within $\Lambda$CDM through further analyses of the observations and considerations of more exotic particle physics beyond the standard model such as extended neutrino species. However, a full resolution may require the consideration of cosmological models beyond $\Lambda$CDM \cite{DeFelice:2010aj,Capozziello:2011et}. These are by and large described within the framework of curvature-based theories of gravity through the Levi-Civita connection \cite{Misner:1974qy}. More broadly, these theories of gravity can be classed as extensions to GR because they limit to an Einstein-Hilbert scenario in some parameter setting \cite{Capozziello:2011et}. On the other hand, there is a growing body of work which considered gravity as an expression of torsion rather than curvature \cite{Aldrovandi:2013wha}. Torsion can be used to replace curvature in the geometric description of gravity by considering teleparallel gravity (TG) \cite{Cai:2015emx,Krssak:2018ywd} in which the teleparallel connection replaces the Levi-Civita connection. \medskip

TG represents a large class of theories in which the teleparallel connection is adopted \cite{Weitzenbock1923} to describe gravitational interactions, which is torsion-full while being curvature-less and satisfying metricity. Thus, all curvature based measures identically vanish once this connection replaces the Levi-Civita connection. For instance, the Ricci scalar $\lc{R}$ (over-circles denote all quantities that are calculated using the Levi-Civita connection) will vanish for the teleparallel connection, namely $R \equiv 0$. Another important matter to point out is that a torsion scalar, $T$, can be defined (see section \ref{sec:f_T}) that is dynamically identical to the Einstein-Hilbert action in that it produces the same equations of motion. For this reason, a linear torsion scalar action generates the so-called \textit{teleparallel equivalent of general relativity} (TEGR), which differs from GR by a boundary term in its Lagrangian density. The boundary, or total divergence, term in TEGR has a meaningful impact on possible extensions to TEGR and their relationship to extended gravity in GR \cite{Gonzalez:2015sha,Bahamonde:2019shr}. \medskip

Using the same reasoning as in $\mathcal{F} (\lc{R})$ gravity \cite{DeFelice:2010aj,Capozziello:2011et}, TEGR can be arbitrarily generalised to $\mathcal{F} (T)$ gravity \cite{Cai:2015emx,Ferraro:2006jd,Ferraro:2008ey,Bengochea:2008gz,Linder:2010py,Chen:2010va,Bahamonde:2019zea,Ualikhanova:2019ygl}. This is a second-order theory that has shown in its confrontation with observations \cite{Cai:2015emx,Nesseris:2013jea,Farrugia:2016qqe,Finch:2018gkh,Farrugia:2016xcw,Iorio:2012cm,Ruggiero:2015oka,Deng:2018ncg,Yan:2019gbw,LeviSaid:2020mbb,Paliathanasis:2017htk}. Also, $\mathcal{F} (T)$ gravity is fundamentally distinct from $\mathcal{F} (\lc{R})$ gravity in that to recover models in the latter case one must also include the boundary term (more details on this in Section \ref{sec:f_T_cosmo}) as in $\mathcal{F} (T,B)$ gravity \cite{Bahamonde:2020bbc,Bahamonde:2020lsm,Bahamonde:2015zma,Capozziello:2018qcp,Bahamonde:2016grb,Paliathanasis:2017flf,Farrugia:2018gyz,Bahamonde:2016cul,Bahamonde:2016cul,Wright:2016ayu}. To fully recover an $\mathcal{F} (\lc{R})$ model, a specific subset of $\mathcal{F} (T,B)$ models must be selected, namely $\mathcal{F} (T,B) = \mathcal{F} (-T+B) = \mathcal{F} (\lc{R})$. $\mathcal{F} (T,B)$ gravity also has been tested against a number of observational phenomena with significant progress being made \cite{Farrugia:2020fcu,Capozziello:2019msc,Farrugia:2018gyz,Bahamonde:2015zma,Paliathanasis:2017flf,Bahamonde:2016grb,Bahamonde:2016cul,Bahamonde:2015zma,Escamilla-Rivera:2019ulu,Franco:2020lxx,Rave-Franco:2021yvu}. Another interesting extension to TEGR is $\mathcal{F} (T,T_G)$ gravity where $T_G$ is the teleparallel equivalent of the Gauss-Bonnet term \cite{Kofinas:2014daa,Capozziello:2016eaz,delaCruz-Dombriz:2018nvt,delaCruz-Dombriz:2017lvj}. TG has also recently been used to construct a teleparallel analogue of Horndeski gravity \cite{Bahamonde:2019shr,Bahamonde:2019ipm,Bahamonde:2020cfv} which is promising because it can bypass the problems related to the speed of gravitational waves in standard gravity. Observationally, $f(T)$ gravity has shown promise in confronting cosmological data. In Refs.~\cite{Nesseris:2013jea,Anagnostopoulos:2019miu,LeviSaid:2020mbb} it is shown that a number of viable models of $f(T)$ gravity model cosmological data well for small deviations from $\Lambda$CDM parameter values. \medskip

The problem with such larges classes of cosmological theories is that it is not intuitively clear which models have the potential to meet the observational challenges such as the recently reported cosmological tensions. Gaussian processes (GP) have an advantage in this regards in that they do not require one to have a specific parametrised cosmological model \cite{10.5555/1162254}, and have been used in a number of diverse settings \cite{Elizalde:2018dvw,Benisty:2020kdt,Benisty:2020otr}. In terms of cosmological models in $\mathcal{F} (T)$ gravity, GP provides a model-independent approach that can lead to constraints on the form of the arbitrary Lagrangian that would normally require a specific model to be chosen. \medskip

In Ref. \cite{Cai:2019bdh}, this was first used to reconstruct the $\mathcal{F} (T)$ functional form using Hubble observational data. While in Ref. \cite{Briffa:2020qli}, the concept was expanded to include more data sets and prior values for the Hubble data. Also, this work provides a number of other GP reconstructed cosmological parameters such as the dark energy equation of state and the cosmological deceleration parameter. In the present work, this approach is extended into the perturbative regime through the $f\sigma_{8}^{}(a)$ parameter. In section \ref{sec:f_T} we give a brief review of $\mathcal{F}(T)$ gravity and its cosmology. While in section \ref{sec:methodology}, we introduce the concepts behind GP reconstructions and apply them to both Hubble and redshift--space--distortion (RSD) $f\sigma_8^{}(a)$ data. Section \ref{sec:results} then contains the core results of this work in which we present constraints on the arbitrary function in $\mathcal{F}(T)$ cosmology. An interesting point of distinction between the present work and Ref. \cite{Briffa:2020qli} is that in this other work an approximation had to be taken to produce initial conditions on the propagation equation for the $\mathcal{F}(T)$ Friedmann equation, whereas in the present study the growth equation does not require any restrictions to produce initial conditions. Finally, we summarise our results in section \ref{sec:conc}. Throughout the manuscript, Latin indices are used to refer to tangent space coordinates, while Greek indices refer to general manifold coordinates.

\section{\label{sec:f_T}\texorpdfstring{$\mathcal{F}(T)$}{ft} gravity}
In this section, we show how $f(T)$ gravity can be tested with observables related to its background and linear (scalar) perturbation equations. We connect the foundations of the theory with expansion and growth observables.

\subsection{\texorpdfstring{$\mathcal{F}(T)$}{} Cosmology}\label{sec:f_T_cosmo}
The curvature associated with GR is expressed through the Levi-Civita connection $\lc{\Gamma}^{\sigma}_{\mu\nu}$ (over-circles are used throughout to denote quantities determined using the Levi-Civita connection) rather than the metric tensor which acts as the fundamental variable of the theory. The core distinction that TG represents is the exchange of the Levi-Civita connection with the teleparallel connection $\Gamma^{\sigma}_{\mu\nu}$, which characterises gravitation through torsion while being curvature-less and satisfying metricity \cite{Hayashi:1979qx,Aldrovandi:2013wha}. The transition from the Levi-Civita to the teleparallel connection renders all curvature quantities such as the Riemann tensor organically zero (when calculated with the teleparallel connection). Thus, to construct theories of gravity in TG requires an entire new formulation of gravitational tensors (see reviews in Refs. \cite{Krssak:2018ywd,Cai:2015emx,Aldrovandi:2013wha}). \medskip

In TG, another important point to highlight is that the metric $g_{\mu\nu}$ is no longer the fundamental dynamical object and is now derived from the tetrad $\udt{e}{a}{\mu}$ (and its inverses $\dut{e}{a}{\mu}$) through
\begin{align}\label{metric_tetrad_rel}
    g_{\mu\nu}=\udt{e}{a}{\mu}\udt{e}{b}{\nu}\eta_{ab}\,,& &\eta_{ab} = \dut{e}{a}{\mu}\dut{e}{b}{\nu}g_{\mu\nu}\,,
\end{align}
where Latin indices represent coordinates on the tangent space while Greek indices represent coordinates on the general manifold. Tetrads also appear in GR but are reserved for specific usages \cite{Chandrasekhar:1984siy}. These tetrads must also satisfy orthogonality conditions
\begin{align}
    \udt{e}{a}{\mu}\dut{e}{b}{\mu}=\delta^a_b\,,&  &\udt{e}{a}{\mu}\dut{e}{a}{\nu}=\delta^{\nu}_{\mu}\,,
\end{align}
for internal consistency. The teleparallel connection can then be defined as \cite{Weitzenbock1923}
\begin{equation}
    \Gamma^{\sigma}_{\nu\mu}:= \dut{e}{a}{\sigma}\partial_{\mu}\udt{e}{a}{\nu} + \dut{e}{a}{\sigma}\udt{\omega}{a}{b\mu}\udt{e}{b}{\nu}\,,
\end{equation}
where $\udt{\omega}{a}{b\mu}$ is a flat spin connection that appears to incorporate the local Lorentz transformation (LLT) (namely Lorentz boosts and rotations) invariance of the theory \cite{Krssak:2015oua}. In GR, spin connections are not flat and are largely hidden within the internal structure of the theory (and so do not contribute to the GR equations of motion) \cite{misner1973gravitation}, whereas in TG they can have an important impact on the ensuing equations of motion (meaning that the spin connection explicitly appears in the generic field equations of the theory). For a particular choice of the Weitzenb\"{o}ck gauge, all of these components vanish \cite{Krssak:2018ywd}. In this context, a torsion tensor can be written as
\cite{Cai:2015emx}
\begin{equation}
    \udt{T}{\sigma}{\mu\nu} := -2\Gamma^{\sigma}_{[\mu\nu]}\,,
\end{equation}
where square brackets denote an antisymmetric operator, and where this tensor represents the field strength of gravity in the TG framework \cite{Aldrovandi:2013wha}. The torsion tensor transforms covariantly under both diffeomorphisms and LLTs. By suitable contractions of the torsion tensor, the torsion scalar can be written through the representation \cite{Krssak:2018ywd,Cai:2015emx,Aldrovandi:2013wha}
\begin{equation}
    T:=\frac{1}{4}\udt{T}{\alpha}{\mu\nu}\dut{T}{\alpha}{\mu\nu} + \frac{1}{2}\udt{T}{\alpha}{\mu\nu}\udt{T}{\nu\mu}{\alpha} - \udt{T}{\alpha}{\mu\alpha}\udt{T}{\beta\mu}{\beta}\,,
\end{equation}
which is entirely dependent on the teleparallel connection in the same way that the Ricci scalar depends only on the Levi-Civita connection. \medskip

For curvature-based gravity, the Levi-Civita connection produces the Ricci scalar $\lc{R}$ which naturally vanishes when calculated using the curvature-less connection, meaning $R\equiv0$ (where $R = R(\Gamma^{\sigma}_{\mu\nu})$ and $\lc{R}=\lc{R}(\lc{\Gamma}^{\sigma}_{\mu\nu})$). By construction, the torsion scalar is built to be equivalent to the Ricci scalar up to a total divergence term $B$, which can be represented as \cite{Bahamonde:2015zma,Farrugia:2016qqe}
\begin{equation}\label{LC_TG_conn}
    R=\lc{R} + T - B = 0\,.
\end{equation}
This relation guarantees that the GR equations of motion are reproduced by an action based on a linear torsion scalar alone. Thus, making this the TEGR Lagrangian. Now, taking the same reasoning as in many extensions to GR, such as $\mathcal{F}(\lc{R})$ gravity \cite{DeFelice:2010aj,Capozziello:2011et}, the TEGR Lagrangian can be arbitrarily generalised to $\mathcal{F}(T)$ gravity through \cite{Ferraro:2006jd,Ferraro:2008ey,Bengochea:2008gz,Linder:2010py,Chen:2010va}
\begin{equation}\label{f_T_ext_Lagran}
    \mathcal{S}_{\mathcal{F}(T)}^{} =  \frac{1}{2\kappa^2}\int \mathrm{d}^4 x\; e\left(-T + \mathcal{F}(T)\right) + \int \mathrm{d}^4 x\; e\mathcal{L}_{\text{m}}\,,
\end{equation}
where $\kappa^2=8\pi G$, $\mathcal{L}_{\text{m}}$ is the matter Lagrangian, and $e=\det\left(\udt{e}{a}{\mu}\right)=\sqrt{-g}$ is the tetrad determinant. The core difference between $\mathcal{F}(\lc{R})$ and $\mathcal{F}(T)$ is that the total divergence term in Eq. (\ref{LC_TG_conn}) leads to fourth-order equations of motion in $\mathcal{F}(\lc{R})$ gravity, while $\mathcal{F}(T)$ gravity is generically second-order which is advantageous for numerous reasons, such as being organically Gauss-Ostrogadsky ghost free and being more amenable to numerical approaches. In Eq. (\ref{f_T_ext_Lagran}), the functional form of $\mathcal{F}(T)$ appears as an extension to TEGR, meaning that $\Lambda$CDM would correspond to a constant for the function in this setting. Finally, by performing variation of the $\mathcal{F}(T)$ action with respect to the tetrads, we arrive at the following field equations
\begin{align}\label{ft_FEs}
    e^{-1} &\partial_{\nu}\left(e\dut{e}{a}{\rho}\dut{S}{\rho}{\mu\nu}\right)\left(-1 + \mathcal{F}_T\right) - \dut{e}{a}{\lambda} \udt{T}{\rho}{\nu\lambda}\dut{S}{\rho}{\nu\mu} \left(-1 + \mathcal{F}_T\right) + \frac{1}{4}\dut{e}{a}{\mu}\left(-T + \mathcal{F}(T)\right) \nonumber\\
    & + \dut{e}{a}{\rho}\dut{S}{\rho}{\mu\nu}\partial_{\nu}\left(T\right)\mathcal{F}_{TT}  + \dut{e}{b}{\lambda}\udt{\omega}{b}{a\nu}\dut{S}{\lambda}{\nu\mu}\left(-1 + \mathcal{F}_T\right) = \kappa^2 \dut{e}{a}{\rho} \dut{\Theta}{\rho}{\mu}\,,
\end{align}
where subscripts denote derivatives ($\mathcal{F}_T=\partial \mathcal{F}/\partial T$ and  $\mathcal{F}_{TT}=\partial^2 \mathcal{F}/\partial T^2$), and $\dut{\Theta}{\rho}{\nu}$ is the regular energy-momentum tensor. \medskip

In $\mathcal{F}(T)$ gravity, a flat homogeneous and isotropic Universe can be considered through the tetrad choice \cite{Krssak:2015oua,Tamanini:2012hg}
\begin{equation}
    \udt{e}{a}{\mu} = \text{diag}\left(1,\,a(t),\,a(t),\,a(t)\right)\,,
\end{equation}
where $a(t)$ is the scale factor in cosmic time $t$, and which reproduces the regular flat Friedmann--Lema\^{i}tre--Robertson--Walker (FLRW) metric \cite{misner1973gravitation}
\begin{equation}\label{FLRW_metric}
     \mathrm{d}s^2=-\mathrm{d}t^2+a^2(t) \left(\mathrm{d}x^2+\mathrm{d}y^2+\mathrm{d}z^2\right)\,.
\end{equation}

In this setting, and taking the Hubble parameter as $H=\dot{a}/a$ where over-dots refer to derivatives with respect to cosmic time, we show all relevant cosmological scalar values in curvature and torsion based theories in Table.~\ref{tab:TG_scalars}.

\begin{table}[h]
\centering
\begin{tabular}{|c|c|}
\hline
Scalars & FLRW Value\\
\hline
$T$ & $6H^2$ \\
\hline
$B$ & $6\left(3H^2 + \dot{H}\right)$ \\
\hline
$\lc{R}$ & $6\left(\dot{H} + 2H^2\right)$ \\
\hline
\end{tabular}
\caption{\label{tab:TG_scalars} Each gravitational scalar is presented with their (flat) FLRW value where Eq.~\eqref{LC_TG_conn} was used.}
\end{table}


The tetrad corresponding to the FLRW metric can then be used to find the $\mathcal{F}(T)$ Friedmann equations as
\begin{align}
    H^2 - \frac{T}{3}\mathcal{F}_T + \frac{\mathcal{F}}{6} &= \frac{\kappa^2}{3}\rho\,,\label{eq:Friedmann_1}\\
    \dot{H}\left(1 - \mathcal{F}_T - 2T\mathcal{F}_{TT}\right) &= -\frac{\kappa^2}{2} \left(\rho + p \right)\label{eq:Friedmann_2}\,,
\end{align}
where we denote the energy density and pressure of the matter content by $\rho$ and $p$, respectively.

\subsection{\label{sec:fT_cosmology_perturbations}Linear perturbations}

Cosmological probes provide a number of measurements related to the evolution of cosmological perturbations. We will be using the growth rate measurements of $f\sigma_8^{}(z)$ from RSD, as described in section \ref{sec:data}. By definition, the growth rate of cosmic structure $f(z)$, is given by the derivative of the logarithm of the matter perturbation $\delta(z)$ with respect to the logarithm of the cosmic scale factor, namely
\begin{equation}\label{eq:fz_growth}
    f(z)=\frac{\mathrm{d}\ln\delta(z)}{\mathrm{d}\ln a}=-(1+z)\frac{\mathrm{d}\ln\delta(z)}{\mathrm{d}z}=-(1+z)\frac{\delta^\prime(z)}{\delta(z)}\,,
\end{equation}
where a prime denotes a derivative with respect to redshift $z=a^{-1}-1$. Moreover, the linear theory root--mean--square mass fluctuation within a sphere of radius $8h^{-1}$ Mpc, where $h=100\,\mathrm{km}\,\mathrm{s}^{-1}\mathrm{Mpc}^{-1}/H_0^{}$ is the dimensionless Hubble constant, is given by
\begin{equation}\label{eq:s8}
    \sigma_8^{}(z)=\sigma_{8,0}^{}\frac{\delta(z)}{\delta_0^{}}\,,
\end{equation}
where a 0--subscript denotes the respective value at $z=0$. Hence, the growth rate of structure can be written as follows,
\begin{equation}\label{eq:fs8}
    f\sigma_8^{}(z)=-(1+z)\frac{\sigma_{8,0}^{}}{\delta_0^{}}\delta^\prime(z)\,.
\end{equation}
Consequently, from the $f\sigma_8(z)$ measurement, one can easily find the normalised evolution of $\delta^\prime(z)$ via
\begin{equation}\label{eq:delta_prime}
    \frac{\delta^\prime(z)}{\delta_0^{}}=-\frac{1}{\sigma_{8,0}^{}}\frac{f\sigma_8^{}(z)}{(1+z)}\,.
\end{equation}
By integrating Eq. (\ref{eq:delta_prime}), we can then determine the redshift evolution of the normalised matter density contrast
\begin{equation}\label{eq:delta}
    \frac{\delta(z)}{\delta_0^{}}=1-\frac{1}{\sigma_{8,0}}\int_0^z\frac{f\sigma_8^{}(\tilde{z})}{(1+\tilde{z})}\,\mathrm{d}\tilde{z}\,,
\end{equation}
while from Eq. (\ref{eq:delta_prime}) we also get
\begin{equation}\label{eq:delta_prime_prime}
    \frac{\delta^{\prime\prime}(z)}{\delta_0^{}}=-\frac{1}{\sigma_{8,0}^{}}\left[\frac{(1+z)\,f\sigma_8^\prime(z)-f\sigma_8^{}(z)}{(1+z)^2}\right]\,.
\end{equation}
From the above formalism, we can now determine the evolution of the derivative of $f$ with respect to $z$, which is given by
\begin{equation}\label{eq:f_prime}
    f^\prime(z)=-\left(\frac{\delta^\prime(z)}{\delta_0^{}}\right)\left(\frac{\delta_0^{}}{\delta(z)}\right)-(1+z)\left[\left(\frac{\delta^{\prime\prime}(z)}{\delta_0^{}}\right)\left(\frac{\delta_0^{}}{\delta(z)}\right)-\left(\frac{\delta^\prime(z)}{\delta_0^{}}\right)^2\left(\frac{\delta_0^{}}{\delta(z)}\right)^2\right]\,.
\end{equation}
In the subhorizon limit, the equation that governs the evolution of linear matter perturbations in the context of General Relativity and the majority of modified gravity models, is specified by
\begin{equation}\label{eq:delta_dot}
    \ddot{\delta}+2H\dot{\delta}=4\pi G_\mathrm{eff}\,\rho\,\delta\,,
\end{equation}
where $G_\mathrm{eff}$ is the effective Newton's constant which is in general a function of $z$ and cosmic wavenumber $k$. However, in this analysis $G_\mathrm{eff}$ will only be a function of redshift and $k$--independent, since the adopted data is not sensitive to the $k$--dependence \cite{amendola2010,Hernandez:2016xci}. We now write Eq. (\ref{eq:delta_dot}) in terms of redshift, where we get
\begin{equation}\label{eq:delta_prime2}
    \delta^{\prime\prime}(z)+\left(\frac{H^\prime(z)}{H(z)}-\frac{1}{1+z}\right)\delta^\prime(z)=\frac{3}{2}\frac{G_\mathrm{eff}(z)}{G_N^{}}\left(\frac{H_0^{}}{H(z)}\right)^2\Omega_{m,0}^{}\,(1+z)\,\delta(z)\,,
\end{equation}
such that $\Omega_{m,0}^{}$ denotes the current matter fractional density, $H_0^{}$ is Hubble's constant, and $G_N^{}$ is Newton's gravitational constant. We can also express Eq. (\ref{eq:delta_prime2}) in terms of the growth rate, which simplifies to the following
\begin{equation}
    f^2(z)+\left[2-(1+z)\frac{H^\prime(z)}{H(z)}\right]f(z)-(1+z)f^\prime(z)=\frac{3}{2}\mathcal{Q}(z)\left(\frac{H_0}{H(z)}\right)^2\Omega_{m,0}^{}(1+z)^3\,.
\end{equation}
where we define the fractional change in the effective Gravitational coupling constant by 
\begin{equation}
    \mathcal{Q}(z):=\frac{G_\mathrm{eff}^{}(z)}{G_N^{}}\,,
\end{equation}
such that in General Relativity we have $\mathcal{Q}(z)=1$ for any arbitrary redshift. In the case of $\mathcal{F}(T)$ gravity, the evolution equation (\ref{eq:delta}) is satisfied with \cite{Nunes:2018xbm,Golovnev:2018wbh,LeviSaid:2020mbb,Sahlu:2019bug}
\begin{equation}\label{eq:Qz}
    \mathcal{Q}(z)=\frac{1}{1+\mathcal{F}_T^{}(z)}\,.
\end{equation}
Hence, from Eq. (\ref{eq:Qz}) one could infer the evolution of $\mathcal{F}_T^{}(T)$, which when combined with Eq. (\ref{eq:Friedmann_1}) we could then determine the functional form of $\mathcal{F}(T)$ via
\begin{equation}\label{eq:f(T)_rec}
    \mathcal{F}(T)=6H_0^2\Omega_{m,0}^{}(1+z)^3+6H^2(z)\left[2\mathcal{F}_T^{}(T)-1\right]\,.
\end{equation}

\section{\label{sec:methodology}Methodology}

We devote this section to present and describe the adopted technique of GP, along with the data sets which will be used in the below analyses. We briefly mention the model dependencies and assumptions behind these data sets, which will therefore ensure the correct understanding to the range of validity of our results.

\subsection{\label{sec:GP}Gaussian processes}

Similar to the definition of a multivariate normal distribution, which is specified by a vector of mean values and a covariance matrix, a Gaussian process \cite{10.5555/971143,10.5555/1162254} is defined via a mean function $\mu(z)$, together with its associated two--point covariance function $\mathcal{C}(z,\tilde{z})$, such that we get a continuous realisation
\begin{equation}
	\xi(z)\sim\mathcal{GP}\left(\mu(z),\mathcal{C}(z,\tilde{z})\right)\,,
\end{equation}
and its corresponding uncertainty $\Delta\xi(z)$, leading to the realisation region of $\xi(z) \pm \Delta \xi(z)$. The realisation region thus forms the posterior GP which is formed through a Bayesian iterative process in which a suitable covariance function is determined to model the datasets. Thus, for Gaussian distributed data the posterior distribution of the reconstructed function can be expressed via the joint Gaussian distribution of different data. A key element for this GP reconstruction is the kernel function $\mathcal{K}(z,\tilde{z})$ which incorporates the uncertainties and correlations from observational data, along with the strength of the correlations between the reconstructed data points at distinct redshifts $z$ and $\tilde{z}$ (see Refs. \cite{Seikel:2013fda,Gomez-Valent:2018hwc} for further details on its construction). There exists a number of kernel function templates \cite{10.5555/1162254,Seikel:2013fda}, and in this work we will be considering the customary infinitely differentiable squared--exponential kernel function
\begin{equation}\label{eq:square_exp}
	\mathcal{K}\left(z,\tilde{z}\right) = \sigma_f^2 \exp\left[-\frac{\left(z-\tilde{z}\right)^2}{2l_f^2}\right]\,,
\end{equation}
along with the Cauchy kernel function,
\begin{equation}
    \mathcal{K}\left(z,\tilde{z}\right) = \sigma_f^2 \left[\frac{l_f}{\left(z-\tilde{z}\right)^2 + l_f^2}\right]\,.\label{eq:cauchy}
\end{equation}
We consider two kernels to understand better any dependence on the choice of kernel function (we also define a Mat\'{e}rn kernel in appendix \ref{sec:H0_HW_Matern} which is used for comparison purposes) \cite{Colgain:2021ngq}. The so--called latent parameters or hyperparameters $\sigma_f^{}$ and $l_f^{}$, characterise the smoothness and overall profile of the GP reconstructed function \cite{2012JCAP...06..036S}, such that $\sigma_f^{}$ controls the uncertainties in the vertical direction as it compares the off--diagonal with the diagonal contributions, whereas $l_f^{}$ adjusts the characteristic correlations' length--scale in $z$. Consequently, a large value of $l_f^{}$ leads to a smoother GP function, whereas a higher value of $\sigma_f^{}$ is characterised by a lower signal--to--noise ratio. Hence, although the hyperparameters appear as constants, their values point to the behaviour of the underlying function, rather than a model that mimics this behaviour. 

In order to find the properly suited values of the hyperparameters, we must make use of the observational data, which is itself a subset of realisations of the GP. Thus, the optimal values of the hyperparameters are derived from the maximisation of the probability of the GP to generate our considered set of data, that is implemented \cite{Busti:2014dua,Verde:2014qea,Seikel:2013fda,Li:2015nta} via the minimisation of the GP marginal likelihood, which is similar to the hierarchical Bayesian approach. GP have now been exhaustively used for the reconstruction of cosmological functions, particularly related to the late--time cosmic accelerated expansion observables \cite{Shafieloo:2012ht,Seikel:2013fda,Cai:2015zoa,Cai:2015pia,Wang:2017jdm,Zhou:2019gda,Cai:2019bdh,Mukherjee:2020vkx,Gomez-Valent:2018hwc,Zhang:2018gjb,Aljaf:2020eqh,Li:2019nux,Liao:2019qoc,Yu:2017iju,Briffa:2020qli,Yennapureddy:2017vvb}. We should remark that although GP are independent from any cosmological model, GP rely on the choice of the kernel function which governs the correlations between distinct points in the GP reconstructed function, and hence its profile. This is particularly noticeable at those locations which lack observational data points. As discussed above, we therefore consider different kernel functions in order to address this kernel dependent characteristic.

\subsection{\label{sec:data}Observational data sets}

We here discuss the adopted cosmological data which incorporates probes of the expansion rate of the Universe, as well as the growth rate of cosmic structure. We also illustrate the respective direct GP reconstructions of $H(z)$ and $f\sigma_8^{}(z)$, which will be further exploited in section \ref{sec:results}.  

\subsubsection{Hubble parameter data}
We first focus on the Hubble parameter data points which we get from cosmic chronometers (CC), along with a Type Ia supernovae (SN) compilation data set. CC data allows us to obtain direct information about the Hubble function at several redshifts, up to around $z\lesssim2$, contrary to other cosmological probes which have to derive the value of $H(z)$ from other observables. Since this technique is primarily based on measurements of the age difference between two passively--evolving galaxies that formed at the same time but are separated by a small redshift interval (from which one can compute $\Delta z/\Delta t$), CC were found to be more reliable than any other method based on an absolute age determination for galaxies \cite{Jimenez:2001gg}. Our CC data points coincide with the considered data sets of Refs. \cite{Gomez-Valent:2018hwc,Yu:2017iju,Briffa:2020qli}, which were compiled from Refs. \cite{2014RAA....14.1221Z,Jimenez:2003iv,Moresco:2016mzx,Simon:2004tf,2012JCAP...08..006M,2010JCAP...02..008S,Moresco:2015cya}. We should remark that these CC measurements are independent of the Cepheid distance scale and from any cosmological model, although they rely on the modelling of stellar ages, which depend on robust stellar population synthesis techniques (see, for instance, Refs. \cite{Gomez-Valent:2018hwc,Lopez-Corredoira:2017zfl,Lopez-Corredoira:2018tmn,Verde:2014qea,2012JCAP...08..006M,Moresco:2016mzx} for analyses related to CC systematics).

\begin{figure*}[t!]
\begin{center}
    \includegraphics[width=0.455\columnwidth]{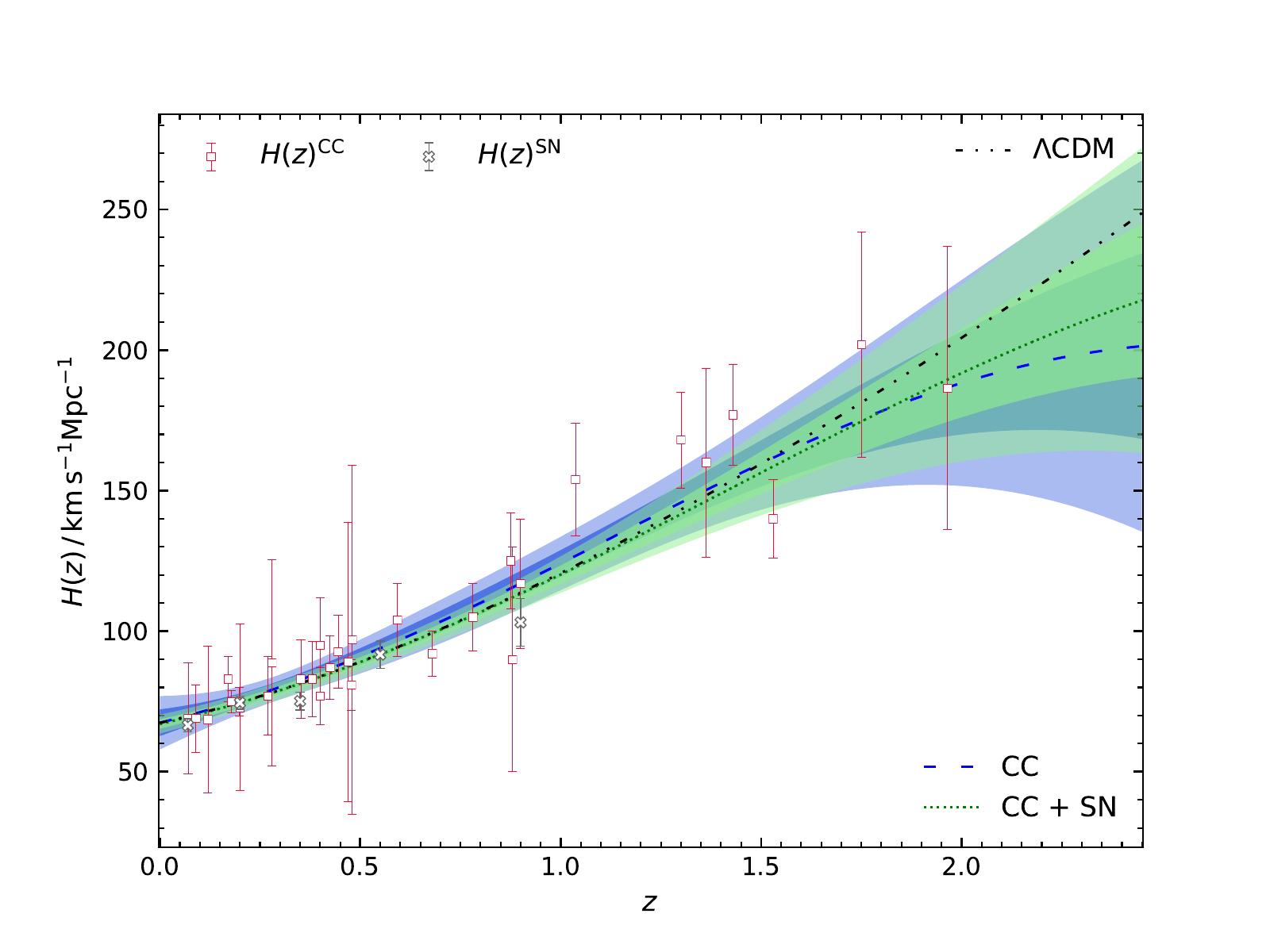}
    \includegraphics[width=0.45\columnwidth]{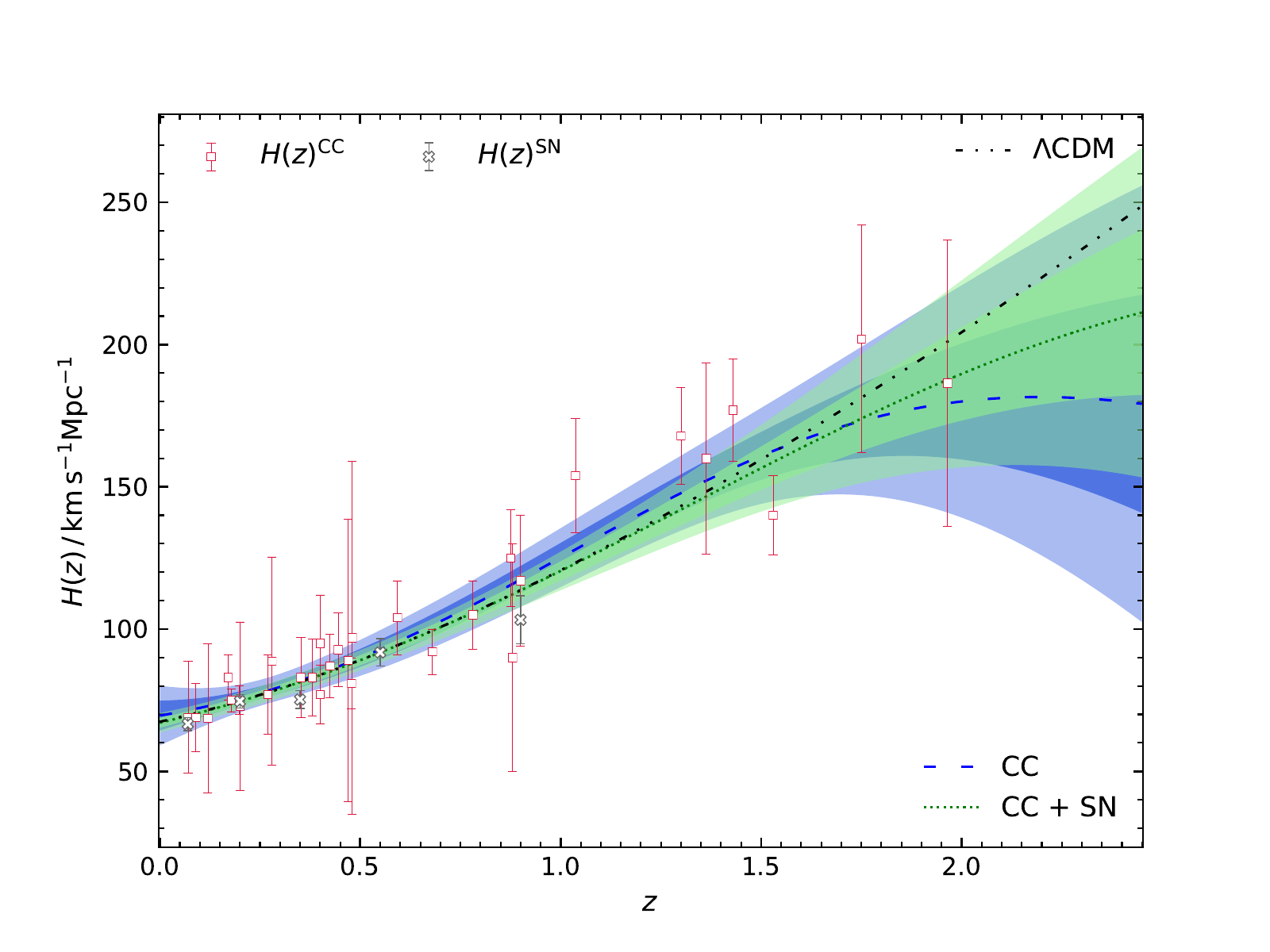}
    \includegraphics[width=0.45\columnwidth]{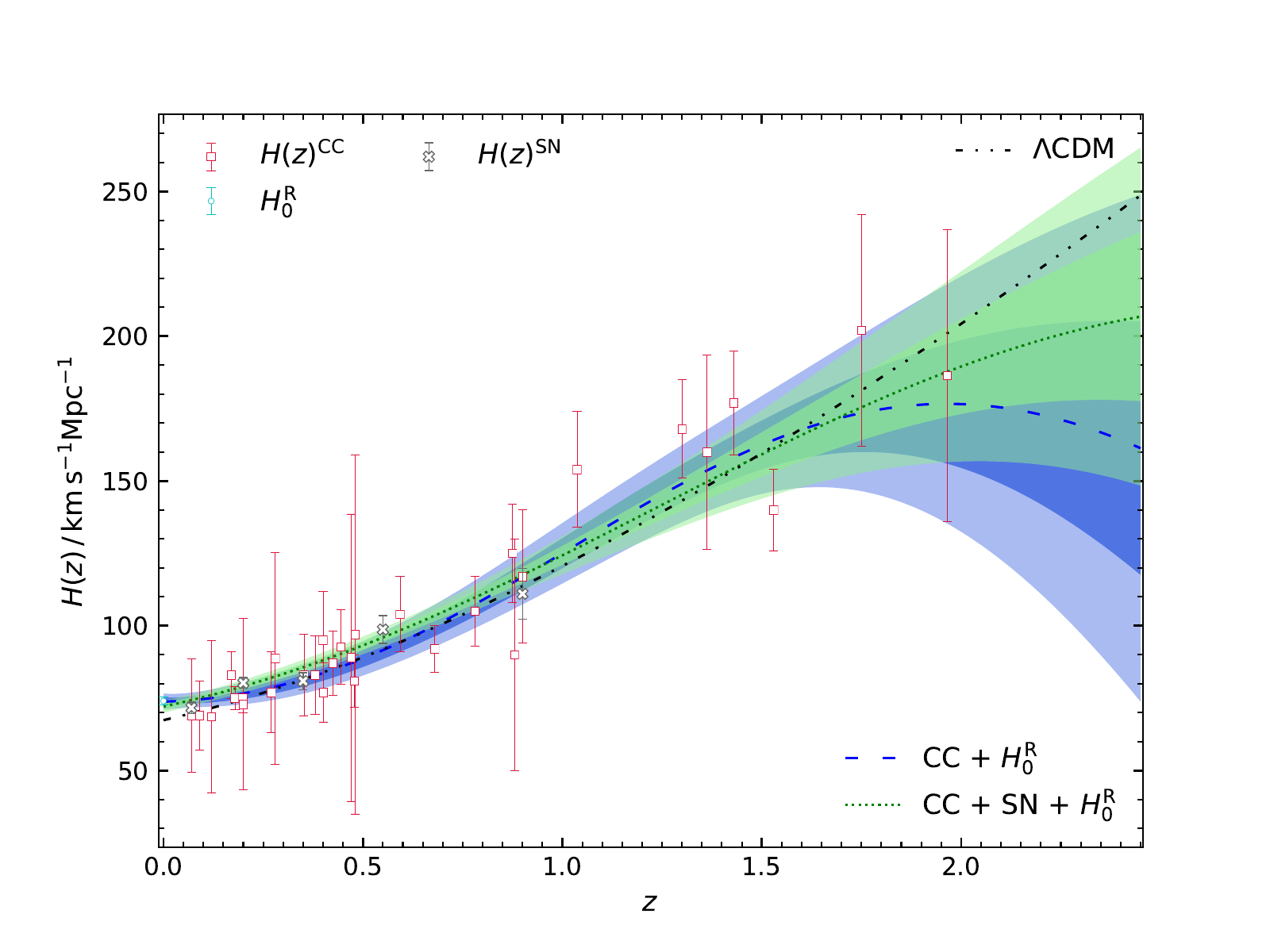}
    \includegraphics[width=0.45\columnwidth]{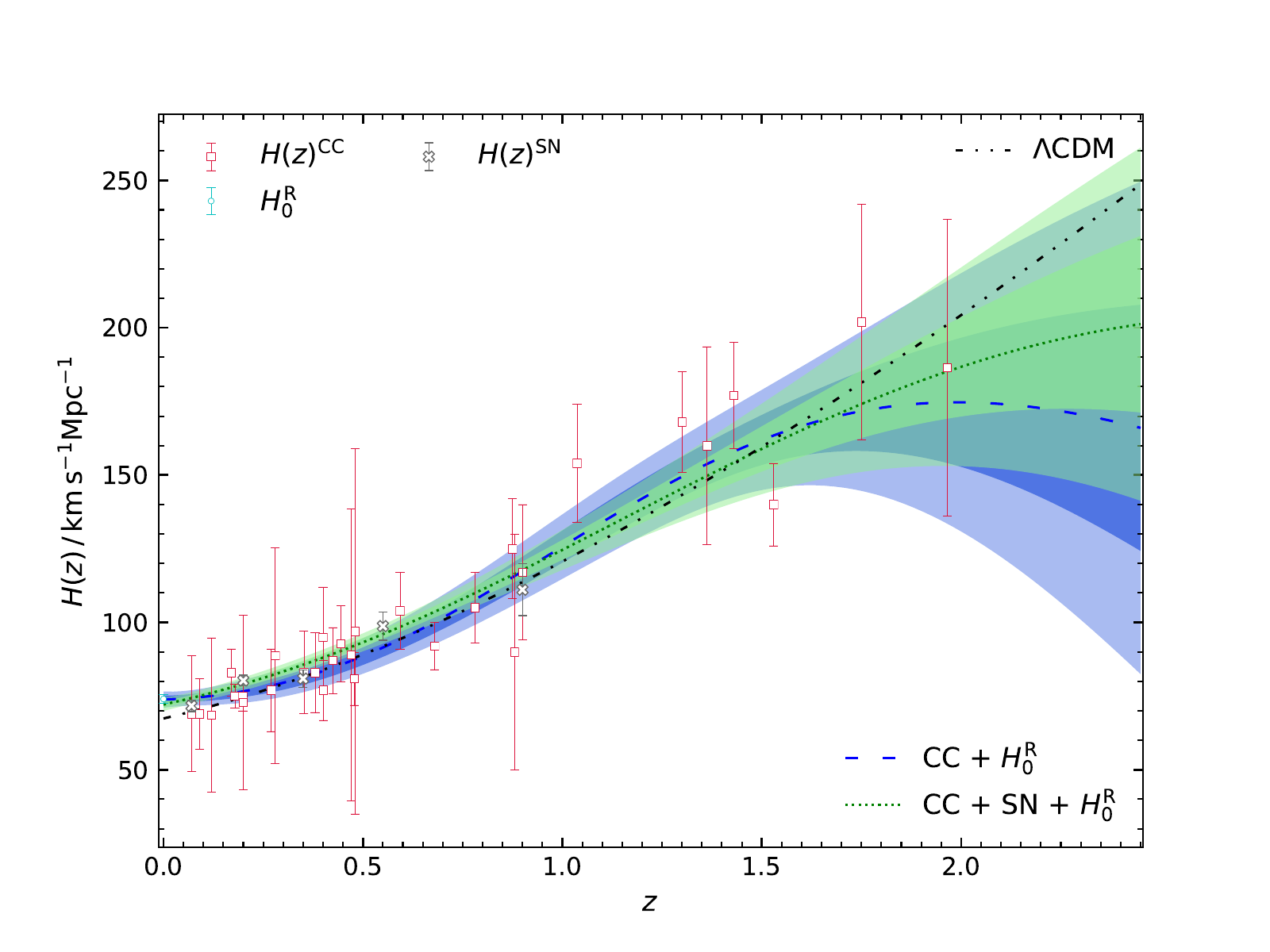}
    \includegraphics[width=0.45\columnwidth]{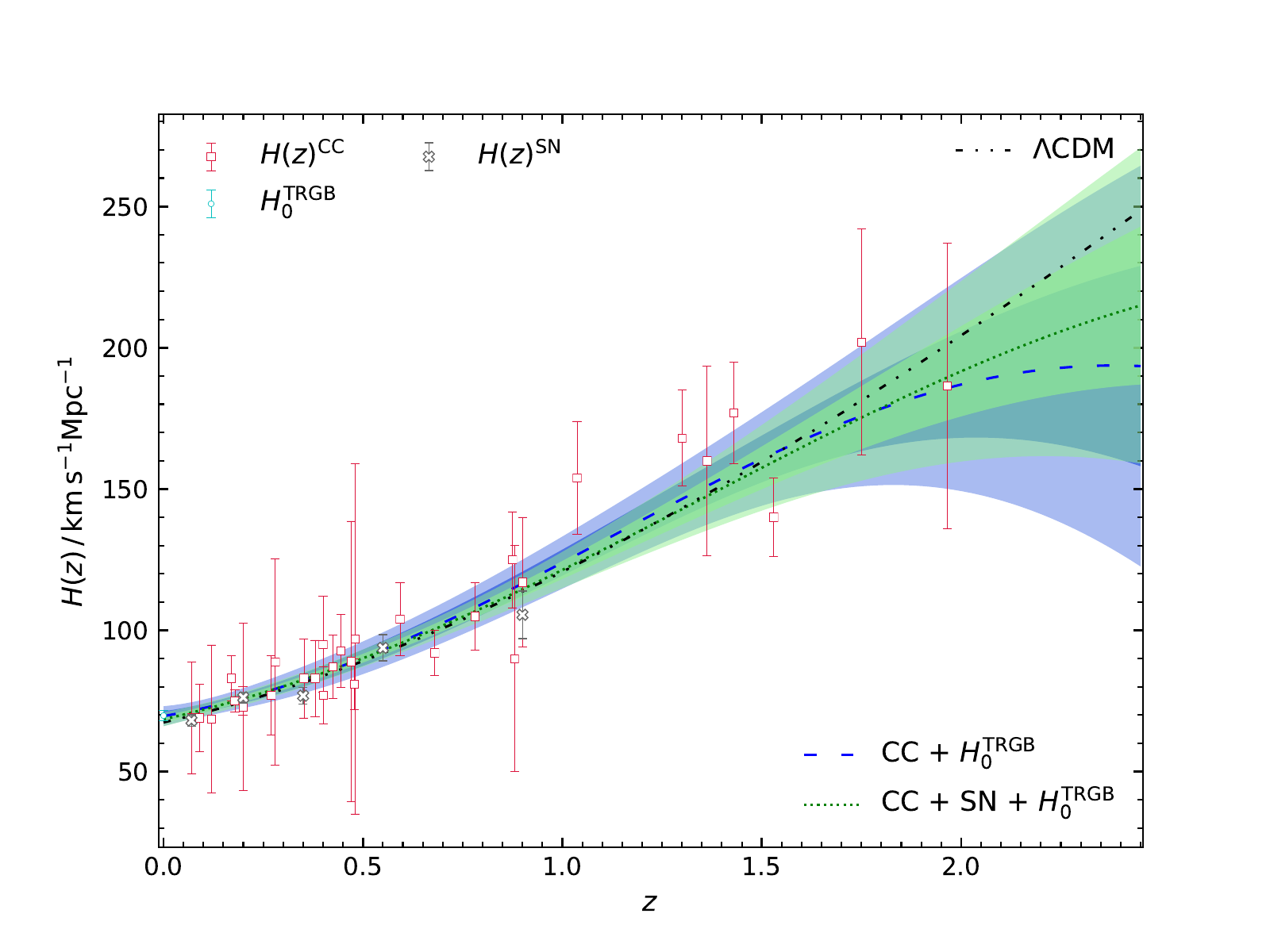}
    \includegraphics[width=0.45\columnwidth]{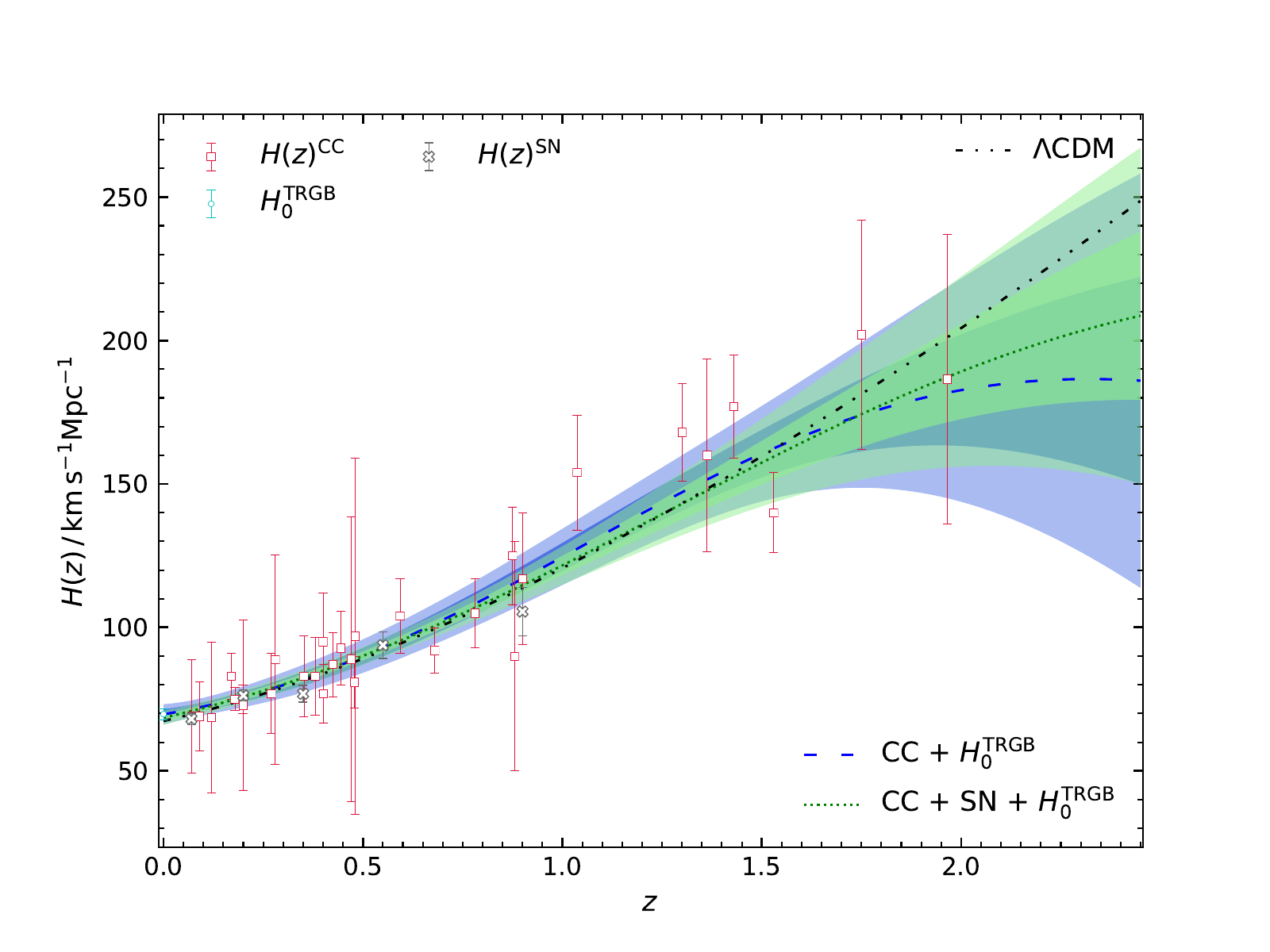}
    \caption{\label{fig:H0_rec_sq_ca}{GP reconstructions of $H(z)$ with CC (blue) and CC + SN (green) data sets. We illustrate the reconstructions without an $H_0^{}$ prior (top), and with the $H_0^{\rm R}$ (middle) and $H_0^{\rm TRGB}$ (bottom) priors. We also compare the GP reconstructions utilising the squared--exponential (left) and the Cauchy (right) kernel functions.}}
\end{center}
\end{figure*}

For the SN data, we use a jointly calibrated and compressed data set \cite{Riess:2017lxs} consisting of the Pantheon compilation \cite{Scolnic:2017caz}, which includes $\sim1050$ SN at $z<1.5$, along with another 15 SN at $z>1$ from the CANDELS and CLASH Multi--Cycle Treasury programs \cite{Riess:2017lxs}. We use the reported Hubble rate parameter measurements of $E(z)\equiv H(z)/H_0$ in the redshift range of $z\in[0.07,\,1.5]$, along with the corresponding correlation matrix, where only five of the reported six data points are adopted since the $z=1.5$ data point is not Gaussian--distributed (similar to Refs. \cite{Gomez-Valent:2018hwc,Briffa:2020qli}). We should mention that the considered SN inferred $E(z)$ measurements could not be considered as a model independent data set, since a spatially flat Universe was assumed in Ref. \cite{Riess:2017lxs}. However, this reasonable assumption has a negligible impact on our analyses, as we are assuming a spatially flat Universe, and the best constraints on spatial curvature \cite{Ade:2015xua,Aghanim:2018eyx} are in very good agreement with this assumption. Moreover, at these low redshifts, the deviation in $H(z)$ due to a minute spatial curvature contribution is much less than the current uncertainties of the adopted SN data points. CC and SN data provides an interesting contrast since they originate from distinct physical processes. In the literature, CC data alone has not been enough to adequately constrain various $f(T)$ models and so, we also include the SN data for comparative purposes.

In order to incorporate the SN data set in our GP analyses, we make use of an iterative numerical procedure \cite{Gomez-Valent:2018hwc,Briffa:2020qli} to determine an $H_0$ value from GP. Basically, we first infer an $H_0$ value by applying GP to the CC data set only, and then we promote the SN $E(z)$ data points to the corresponding $H(z)=H_0E(z)$ values via a Monte Carlo routine. A number of successive GP reconstructions are applied on the combined CC + SN data set, until the resulting value of $H_0$ and its uncertainty converge to $\lesssim10^{-4}$.

We further consider a number of recently reported local measurements of $H_0^{}$. We will be adopting the currently highest value of $H_0^{\rm R} = 74.22 \pm 1.82 \,{\rm km\, s}^{-1} {\rm Mpc}^{-1}$ \cite{Riess:2019cxk} determined via long period observations of Cepheids in the Large Magellanic Cloud, along with $H_0^{\rm TRGB} = 69.8 \pm 1.9 \,{\rm km\, s}^{-1} {\rm Mpc}^{-1}$ \cite{Freedman:2019jwv} which was based on  the Tip of the Red Giant Branch (TRGB) as a standard candle. Another recent determination of $H_0^{}$ was reported by the strong lensing H0LiCOW Collaboration \cite{Wong:2019kwg}, with $H_0^{\rm HW} = 73.3^{+1.7}_{-1.8} \,{\rm km\, s}^{-1} {\rm Mpc}^{-1}$. We illustrate the GP reconstructions of $H(z)$ with the mentioned $H_0^{}$ priors and different kernel functions in Fig. \ref{fig:H0_rec_sq_ca}, while we depict similar results with the $H_0^{\rm HW}$ prior in appendix \ref{sec:H0_HW_Matern}. 

While other measurements exist (for instance, as reported in Refs. \cite{PhysRevD.98.043526,Aghanim:2018eyx,Abbott:2017xzu}), the aforementioned measurements are the most representative model independent values. We should further remark that we decided to use the above mentioned values in light of the current $H_0^{}$--tension conundrum \cite{DiValentino:2020zio,Jedamzik:2020zmd,Verde:2019ivm,Kenworthy:2019qwq,Vagnozzi:2019ezj,Ivanov:2020mfr}, such that we can analyse the impact of a chosen $H_0^{}$ prior on our GP reconstructions of section \ref{sec:results}.

\begin{figure*}[t!]
\begin{center}
    \includegraphics[width=0.49\columnwidth]{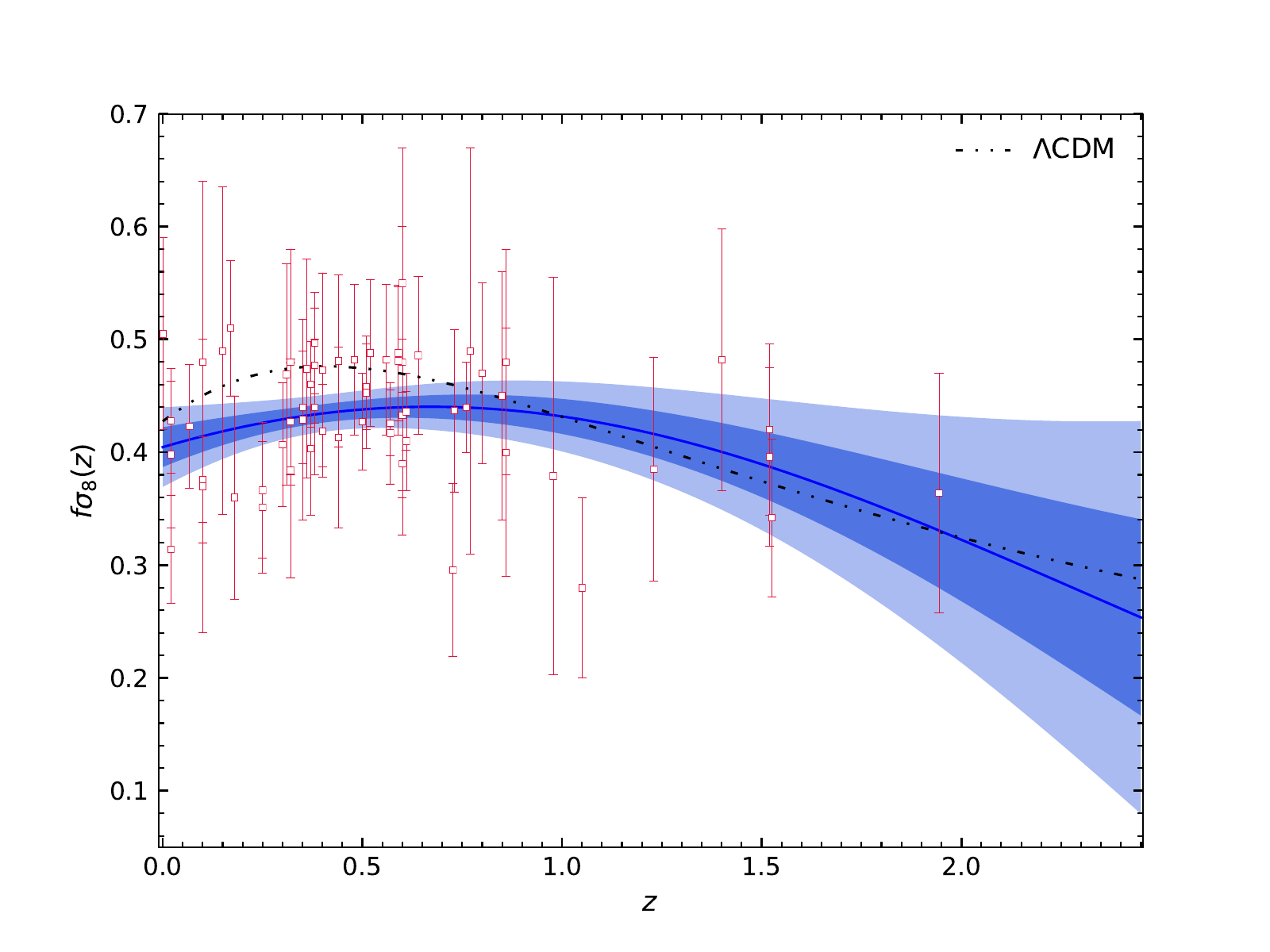}
    \includegraphics[width=0.49\columnwidth]{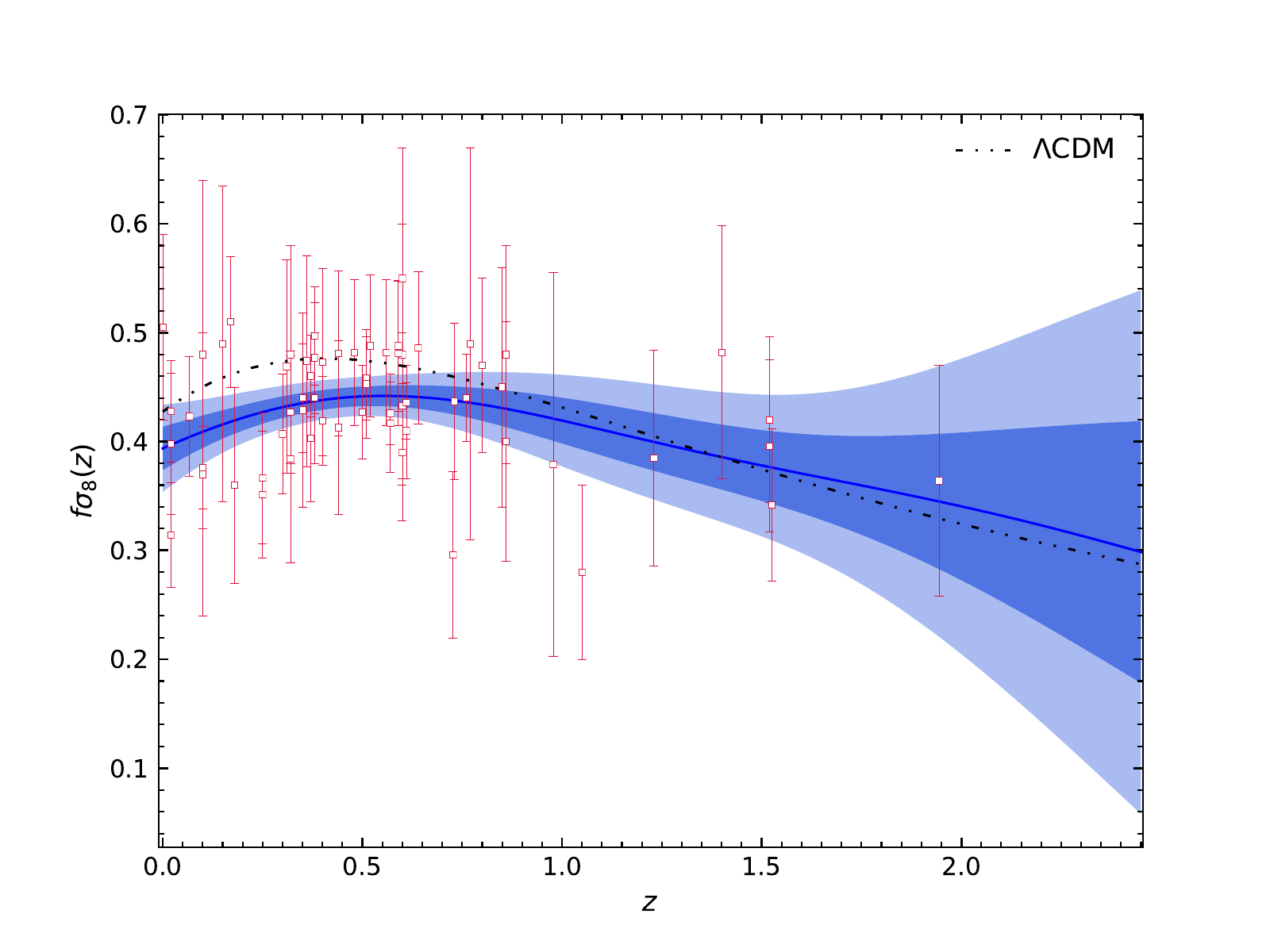}
    \caption{\label{fig:fs8_rec_sq_ca}{GP reconstructions of $f\sigma_8^{}(z)$ with  the squared--exponential (left) and the Cauchy (right) kernel functions.}}
\end{center}
\end{figure*}

\subsubsection{Growth rate data}\label{sec:growth_data}
We now consider the growth of cosmic structure measurements. It is well--known that maps of galaxies where distances are measured from spectroscopic redshifts are characterised by anisotropic deviations from the true galaxy distribution. Such differences arise due to the galaxy's recessional velocities which include contributions from both the Hubble flow, and peculiar velocities from the motions of galaxies in comoving space. Although these RSD \cite{Kaiser:1987qv} can be considered as a nuisance when trying to reconstruct the true spatial distribution of galaxies, RSD unequivocally encode information on the evolution of cosmic structure formation. Since the evolution and formation of cosmic structure is implicitly governed by the underlying theory of gravitation, RSD offers another very promising avenue to test modified theories of gravity (see, for instance, Refs. \cite{Linder:2005in,Jain:2007yk,Song:2008vm,Weinberg:2012es,Baker:2014zva,Kazantzidis:2018rnb}). \medskip

Although the growth rate $f(z)$ can be obtained from RSD cosmological probes and used to constrain cosmological models (see, for instance, Refs. \cite{Gupta:2011kw,Gonzalez:2016lur}), RSD measurements are very often reported in terms of the more reliable (since it evades the issue related to the bias parameter \cite{Kazantzidis:2018rnb}) density--weighted growth rate $f\sigma_8^{}(z)\equiv f(z)\sigma_8^{}(z)$. We should also remark that RSD measurements are also sensitive to $\sigma_{8,0}^{}$ \cite{Lahav:2001sg}. The latter could be properly computed via the low--redshift RSD measurements, as it (nearly) evades the extrapolation from a fiducial cosmological model, an assumption which could not be avoided in the case of CMB data sets. \medskip

Furthermore, galaxy surveys often have to consider degeneracies between RSDs and the so--called Alcock--Paczynski
effect \cite{Alcock:1979mp}, which arises from the need to assume a cosmological model to transform redshifts into distances. At low redshifts this effect is not significant, meaning that our considered measurements are fairly independent of the choice of the fiducial cosmological model. We have adopted the rough approximation \cite{Macaulay:2013swa,Kazantzidis:2018rnb,Li:2019nux} of the Alcock--Paczynski effect by re--scaling the growth rate measurements and
uncertainties by the ratio of $H(z)D_A^{}(z)$, where $D_A^{}(z)$ is the angular diameter distance, although it should be noted that our final results were not affected by this re-scaling. \medskip

\begin{table}[t]
\centering
\begin{tabular}{|c|c|}
\hline
Kernel & $f\sigma_{8,0}$\\
\hline
Squared--exponential & $0.40\pm0.02$ \\
\hline
Cauchy & $0.39\pm0.02$ \\
\hline
\end{tabular}
\caption{\label{tab:fs8_compare} A comparison of reconstructed $f\sigma_{8,0}$ values when using the squared--exponential and Cauchy kernel functions.}
\end{table}

Our $f\sigma_8^{}(z)$ data set consists of all RSD measurements along with the respective covariance matrix as reported in Ref. \cite{Kazantzidis:2018rnb}. This data set incorporates a number of large--scale structure surveys which were ongoing from 2006 to 2018 \cite{Beutler:2012px,delaTorre:2013rpa,Wang:2017wia,Chuang:2012qt,Blake:2013nif,Blake:2012pj,Sanchez:2013tga,Anderson:2013zyy,Howlett:2014opa,Feix:2015dla,Okumura:2015lvp,Beutler:2016arn,Gil-Marin:2016wya,Huterer:2016uyq,Pezzotta:2016gbo,Feix:2016qhh,Alam:2016hwk,Howlett:2017asq,Mohammad:2017lzz,Shi:2017qpr,Gil-Marin:2018cgo,Hou:2018yny,Zhao:2018gvb}. We illustrate the GP reconstructions of $f\sigma_8^{}(z)$ with the squared--exponential and Cauchy kernel functions in Fig. \ref{fig:fs8_rec_sq_ca} (while in Fig. \ref{fig:fs8_rec_ma} we depict the reconstruction with the Mat\'{e}rn kernel function). As indicated in Table \ref{tab:fs8_compare}, the current GP reconstructed values of the density--weighted growth rates from the squared--exponential and Cauchy kernel functions are nearly identical. The $f\sigma_8^{}(z)$ reconstructions are nearly indistinguishable at low redshifts, although at higher redshifts, coinciding with the redshift range where there are less data points, the Cauchy (and Mat\'{e}rn) kernel function(s) tend to be characterised by a larger uncertainty with respect to the squared--exponential kernel function. It is also clear that the concordance model of cosmology agrees very well with the reconstructed $f\sigma_8^{}(z)$, although for $0.05\lesssim z\lesssim0.6$ the $\Lambda$CDM model with \textit{Planck}'s baseline parameters \cite{Aghanim:2018eyx} predicts a higher growth rate than the RSD's GP reconstructed evolution. \medskip

We should mention that we had to modify the standard approach as described in section \ref{sec:GP} for finding the optimal GP hyperparameters, since such a technique leads to an unrealistically flat GP reconstruction of $f\sigma_8^{}(z)$. We derived the optimal values of the hyperparameters by sampling the logarithm of the GP marginal likelihood on a grid of hyperparameters $\sigma_f^{}$ and $l_f^{}$, from 0.01 to the maximum redshift of $f\sigma_8(z)$ data points of $\simeq2$, with $\simeq300$ equally spaced points in log--space for each dimension \cite{Pinho:2018unz}. This now ensures that the typical scale $l_f^{}$, on the independent variable $z$ is always smaller than the redshift range of our data set.  We then select the pair of hyperparameters corresponding to the maximum of the log--marginal likelihood, which turn out to be different from those derived via the standard GP approach, consequently improving the determination of this observable.

\section{\label{sec:results}Results}

In our analyses, we perform a number of GP reconstructions using a combination of two choices, the first being the $H_0^{}$ prior which is selected from having no prior, or one of the $H_0^{\rm R}$ or $H_0^{\rm TRGB}$ values (we consider the $H_0^{\rm HW}$ prior in appendix \ref{sec:H0_HW_Matern}), while the second involves the squared--exponential or Cauchy GP kernel functions (refer to appendix \ref{sec:H0_HW_Matern} for a comparative analysis with the Mat\'{e}rn kernel function). Our GP analyses were implemented in a modified version of the public code \texttt{GaPP} (Gaussian Processes in Python)\footnote{\url{http://ascl.net/1303.027}} \cite{2012JCAP...06..036S}, which was specifically developed for the GP reconstruction of a function and its derivatives from a given data set. \medskip

\begin{figure*}[t!]
\begin{center}
    \includegraphics[width=0.49\columnwidth]{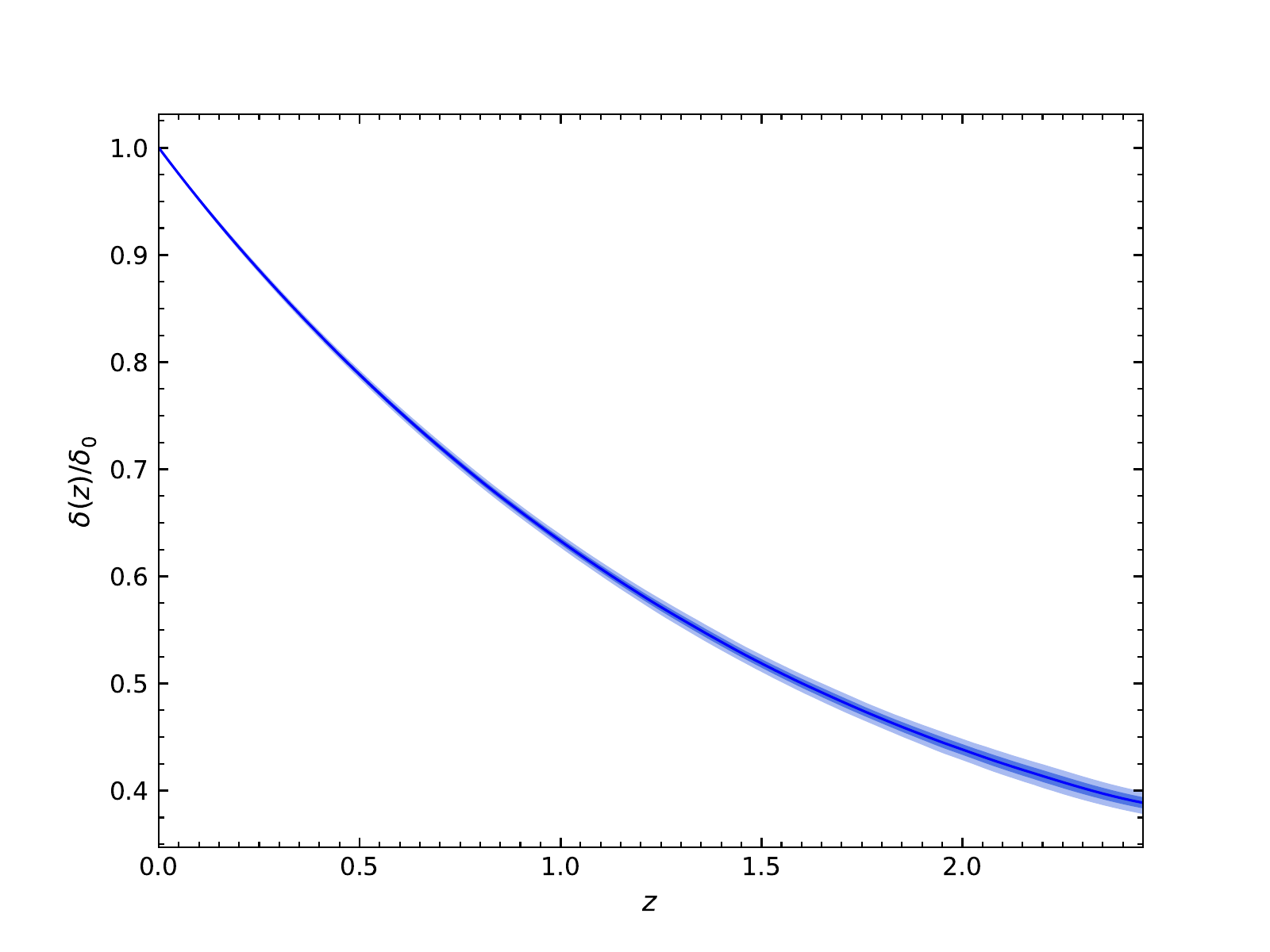}
    \includegraphics[width=0.49\columnwidth]{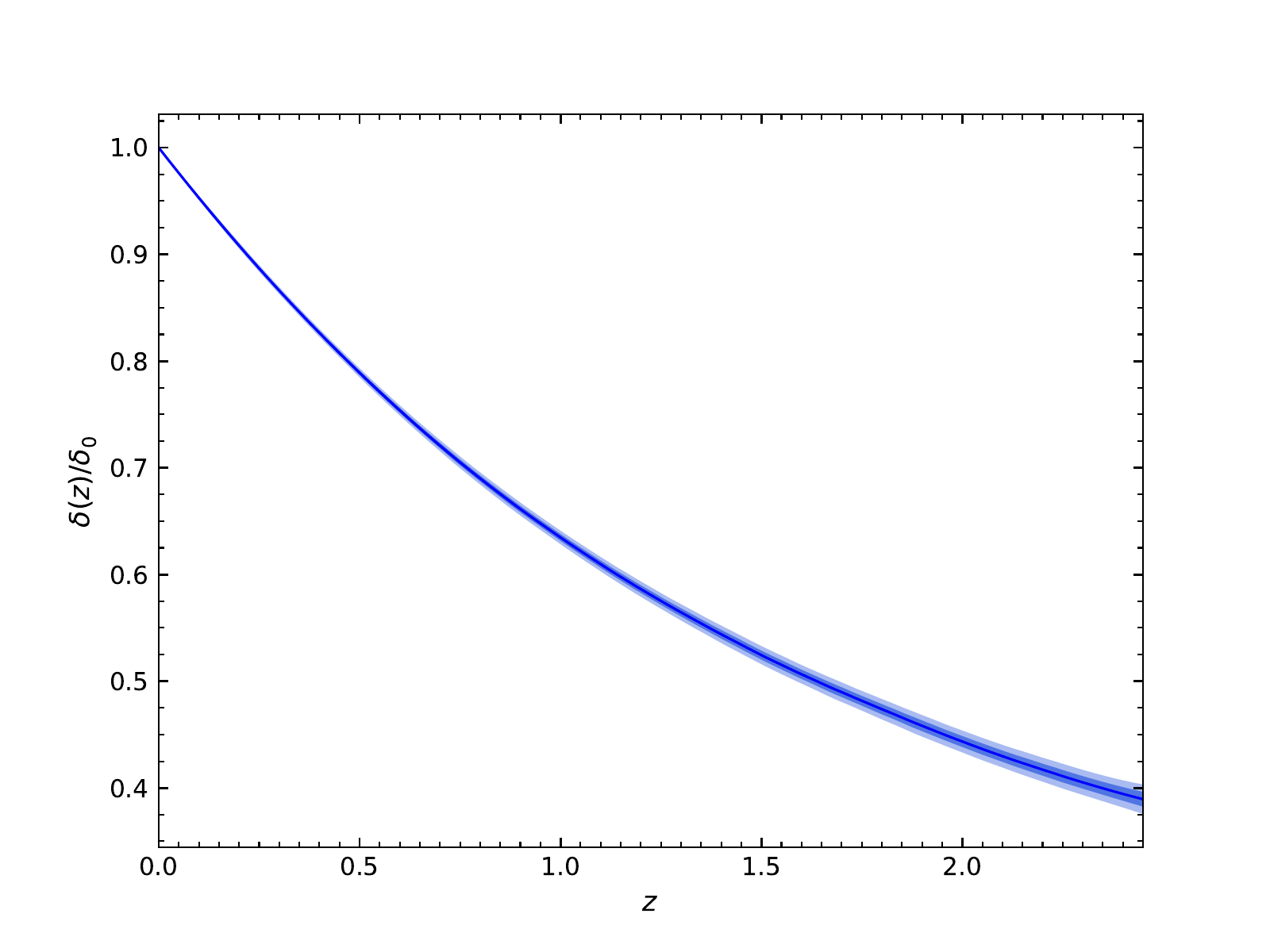}
    \includegraphics[width=0.494\columnwidth]{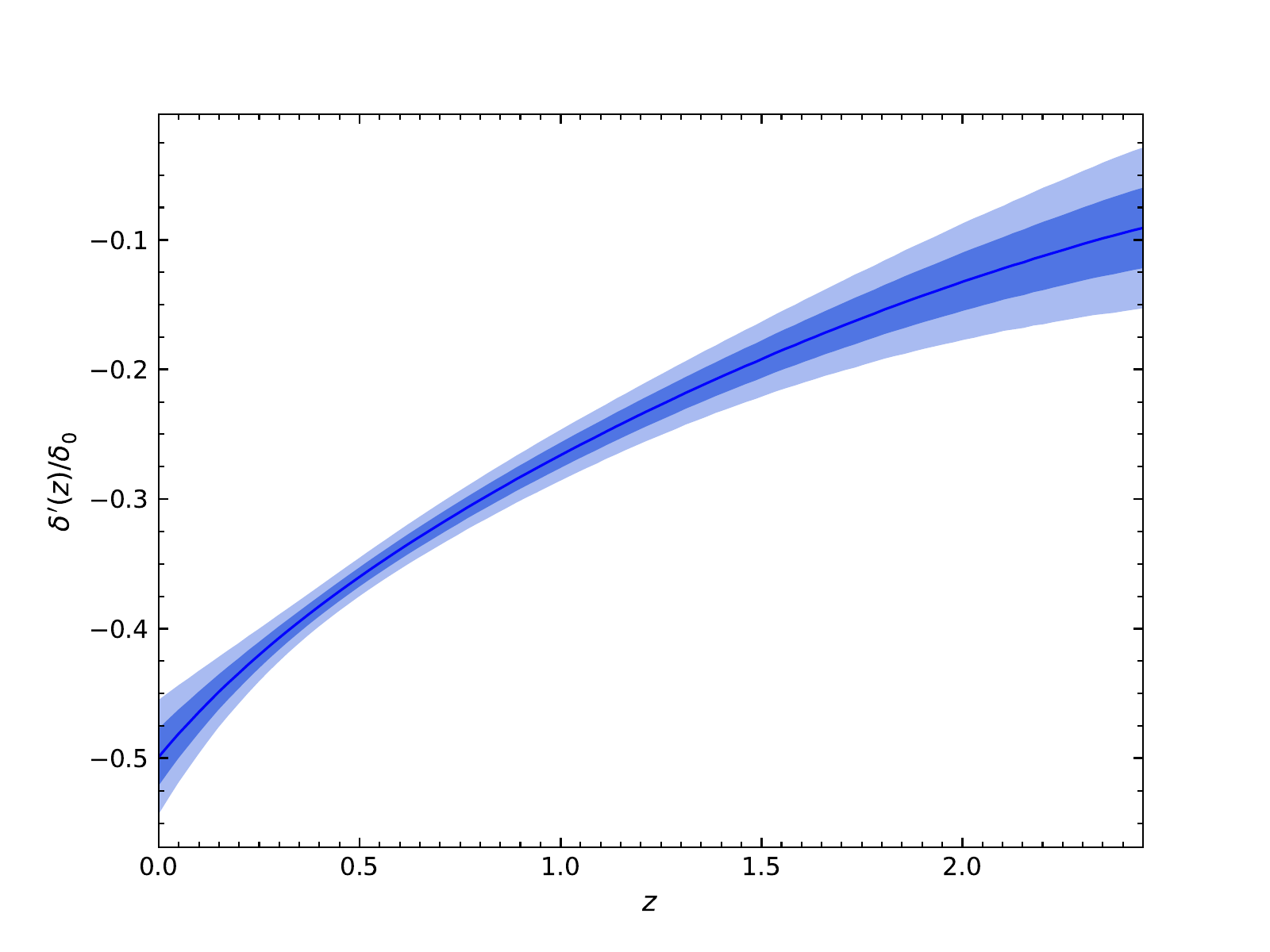}
    \includegraphics[width=0.494\columnwidth]{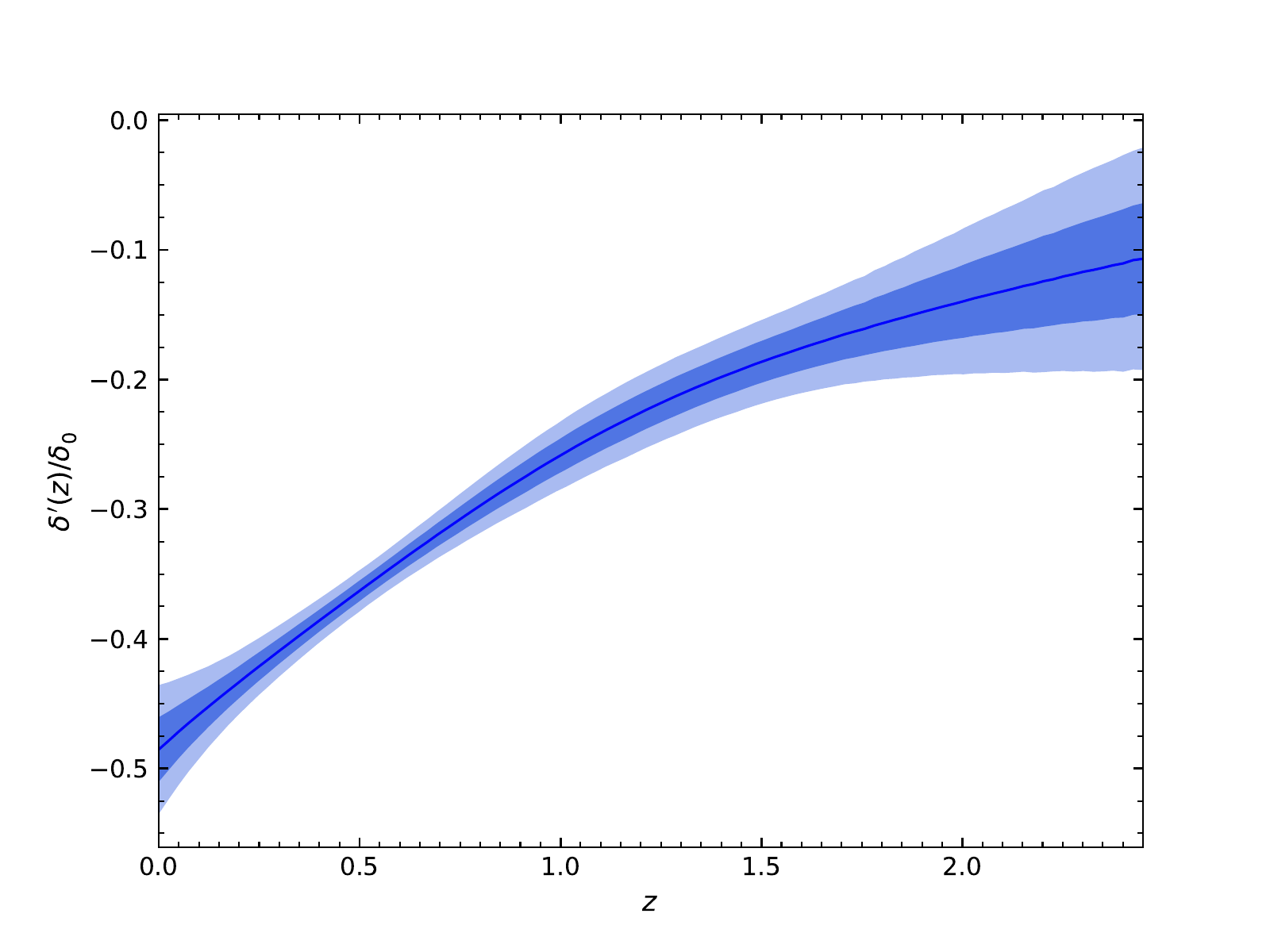}
    \includegraphics[width=0.494\columnwidth]{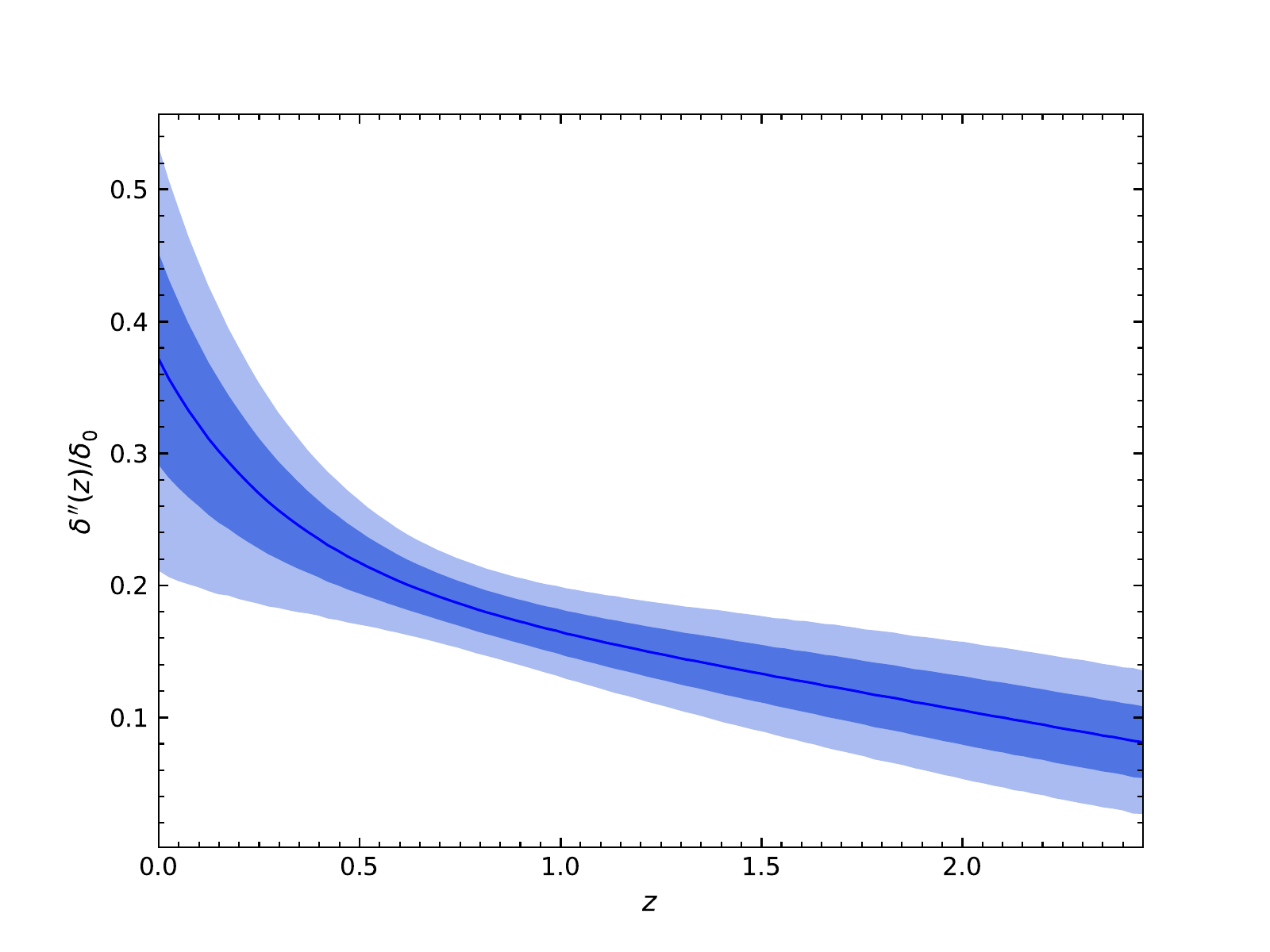}
    \includegraphics[width=0.494\columnwidth]{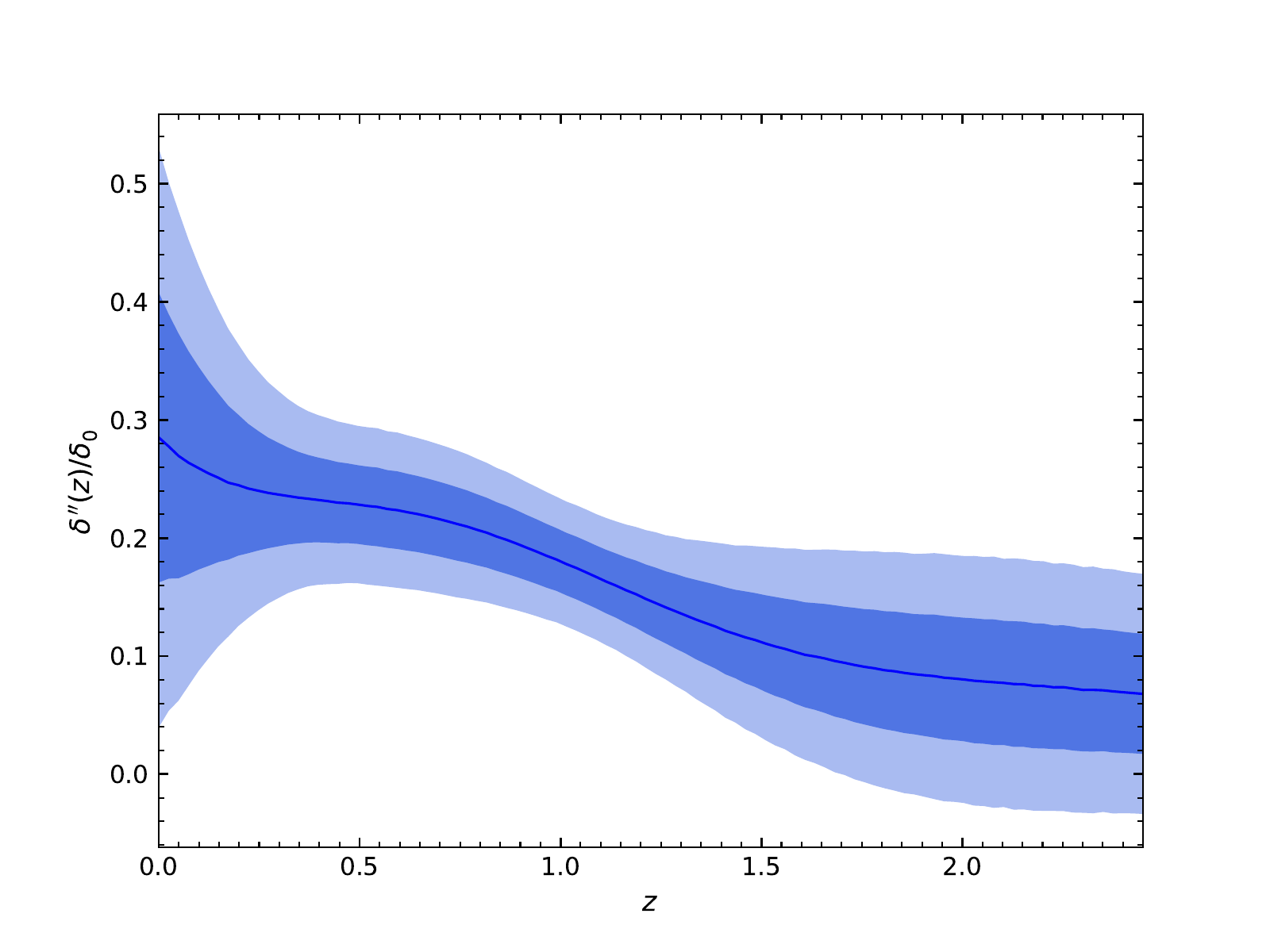}
    \caption{\label{fig:deltas_rec_sq_ca}{GP reconstructions of $\delta(z)/\delta_0^{}$, $\delta^\prime(z)/\delta_0^{}$, and $\delta^{\prime\prime}(z)/\delta_0^{}$ with  the squared--exponential (left) and the Cauchy (right) kernel functions.}}
\end{center}
\end{figure*}

In order to reconstruct the $\mathcal{F}(T)$ function from the data sets of section \ref{sec:data}, we need to reconstruct other functions describing the cosmological background evolution as well as the evolution of the matter density contrast. From the GP reconstructions of the considered $f\sigma_8^{}(z)$ data, which we illustrate in the panels of Fig. \ref{fig:fs8_rec_sq_ca}, we utilise Eqs. (\ref{eq:delta_prime})--(\ref{eq:delta_prime_prime}) to reconstruct the redshift evolution of the $\delta(z)/\delta_0$, $\delta^\prime(z)/\delta_0$ and $\delta^{\prime\prime}(z)/\delta_0$ functions. We depict the GP reconstructions of the normalised density contrast and its first and second derivatives in Fig. \ref{fig:deltas_rec_sq_ca}, where we make use of the squared--exponential and Cauchy kernel functions. For these reconstructions, we have further adopted \textit{Planck}'s baseline parameter value of $\sigma_{8,0}=0.8111 \pm 0.0060$ \cite{Aghanim:2018eyx}. From the top panels of Fig. \ref{fig:deltas_rec_sq_ca}, it is clear that there is an insignificant influence from the choice of kernel functions on the reconstruction of $\delta(z)/\delta_0$. However, this statement does not robustly hold for the reconstructions of the first and second derivative of the normalised matter density contrast. Indeed, the mean realisation and the corresponding confidence regions of $\delta^\prime(z)/\delta_0$ differ between the GP reconstruction adopting the squared--exponential kernel function with respect to the GP reconstruction using the Cauchy kernel function. Such a difference is more prominent in the reconstruction of the second derivative of $\delta(z)/\delta_0$, where we could observe that in general the squared--exponential kernel function leads to a smoother and more uniform GP reconstruction relative to the one obtained via the Cauchy kernel function. Moreover, the GP reconstructions of $\delta^\prime(z)/\delta_0$ and $\delta^{\prime\prime}(z)/\delta_0$ with the Cauchy kernel function are characterised by a more conservative confidence region than the respective GP reconstructions with the squared--exponential kernel function. \medskip

\begin{figure*}[t!]
\begin{center}
    \includegraphics[width=0.49\columnwidth]{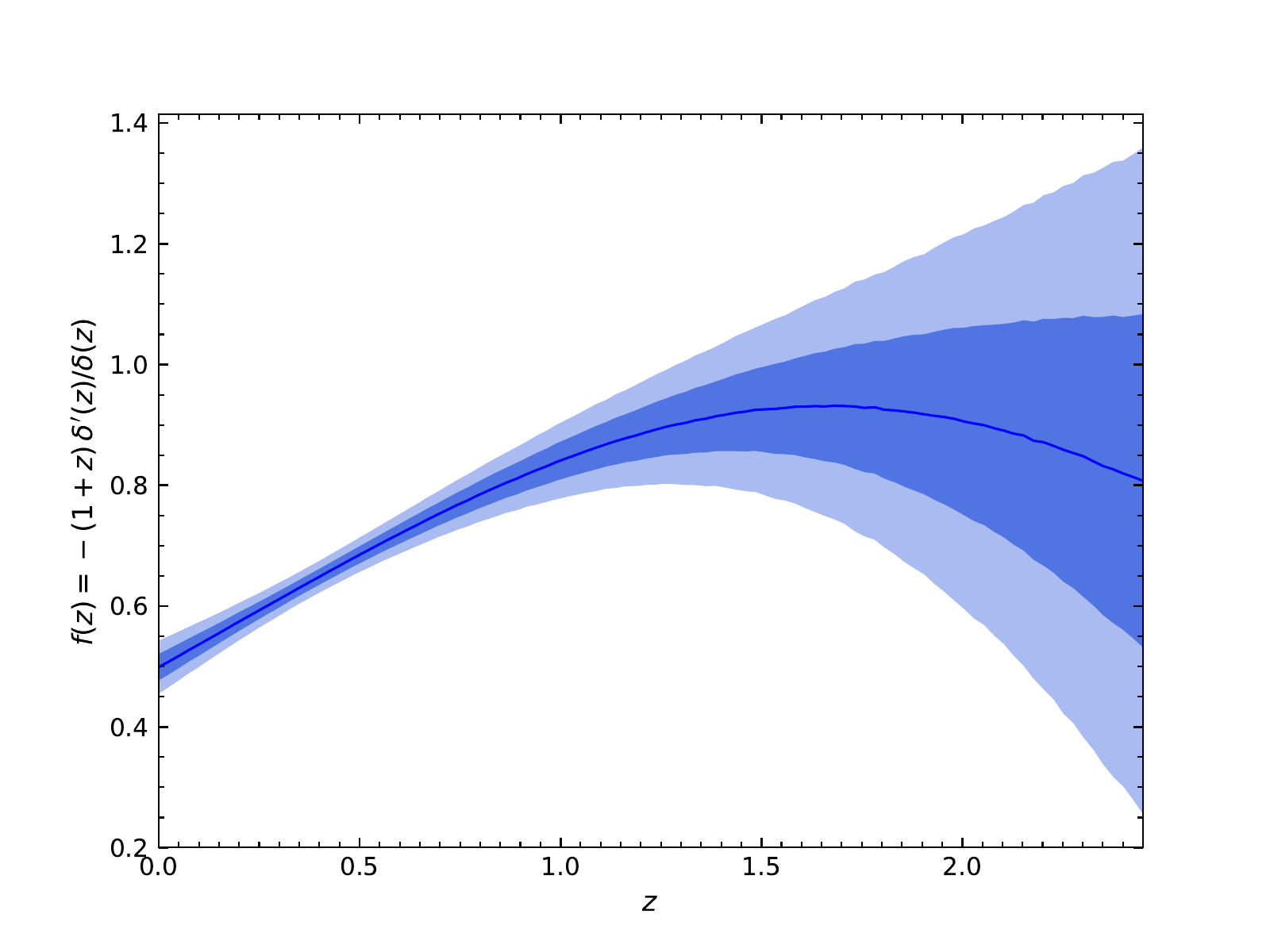}
    \includegraphics[width=0.49\columnwidth]{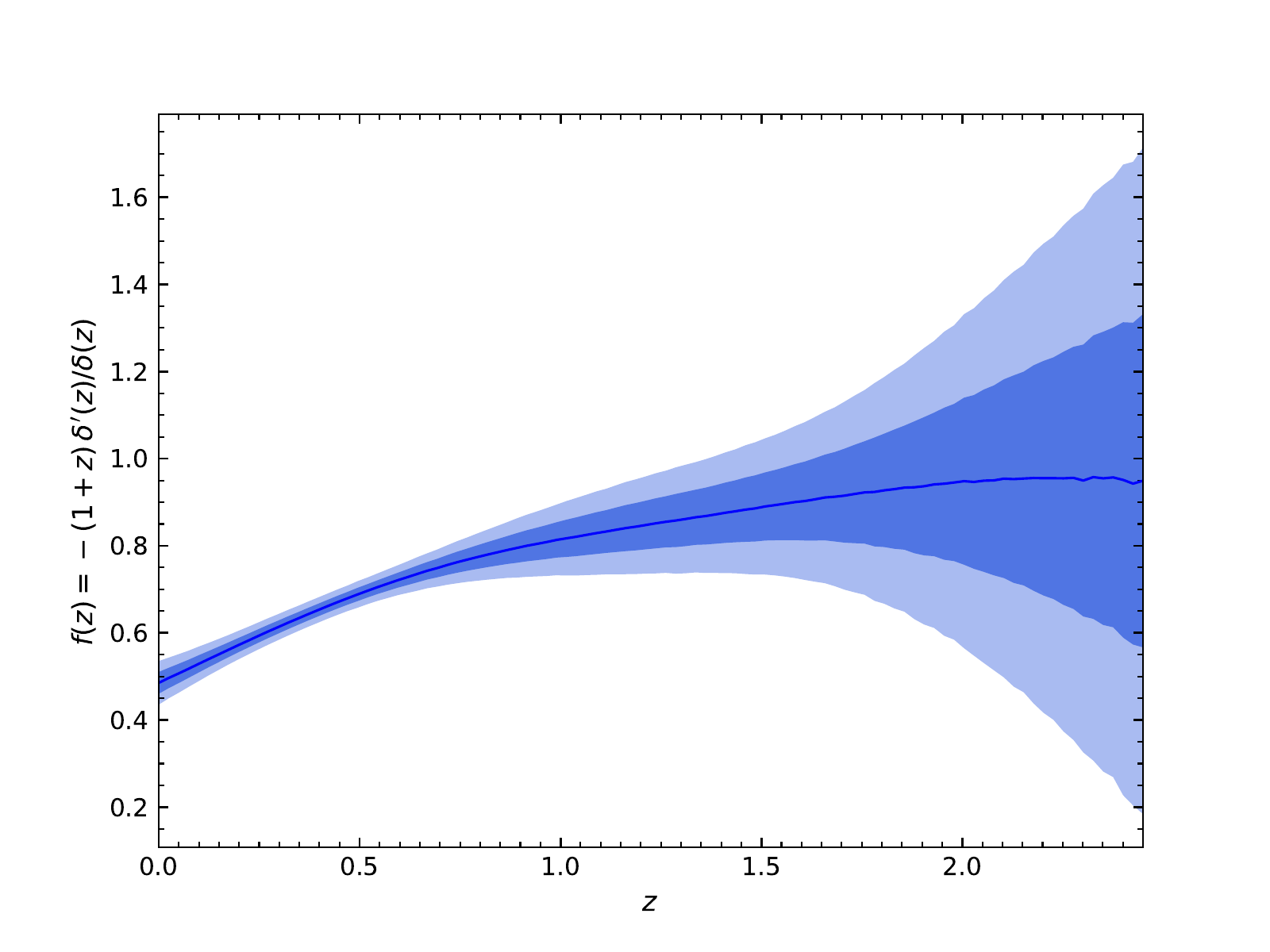}
    \includegraphics[width=0.49\columnwidth]{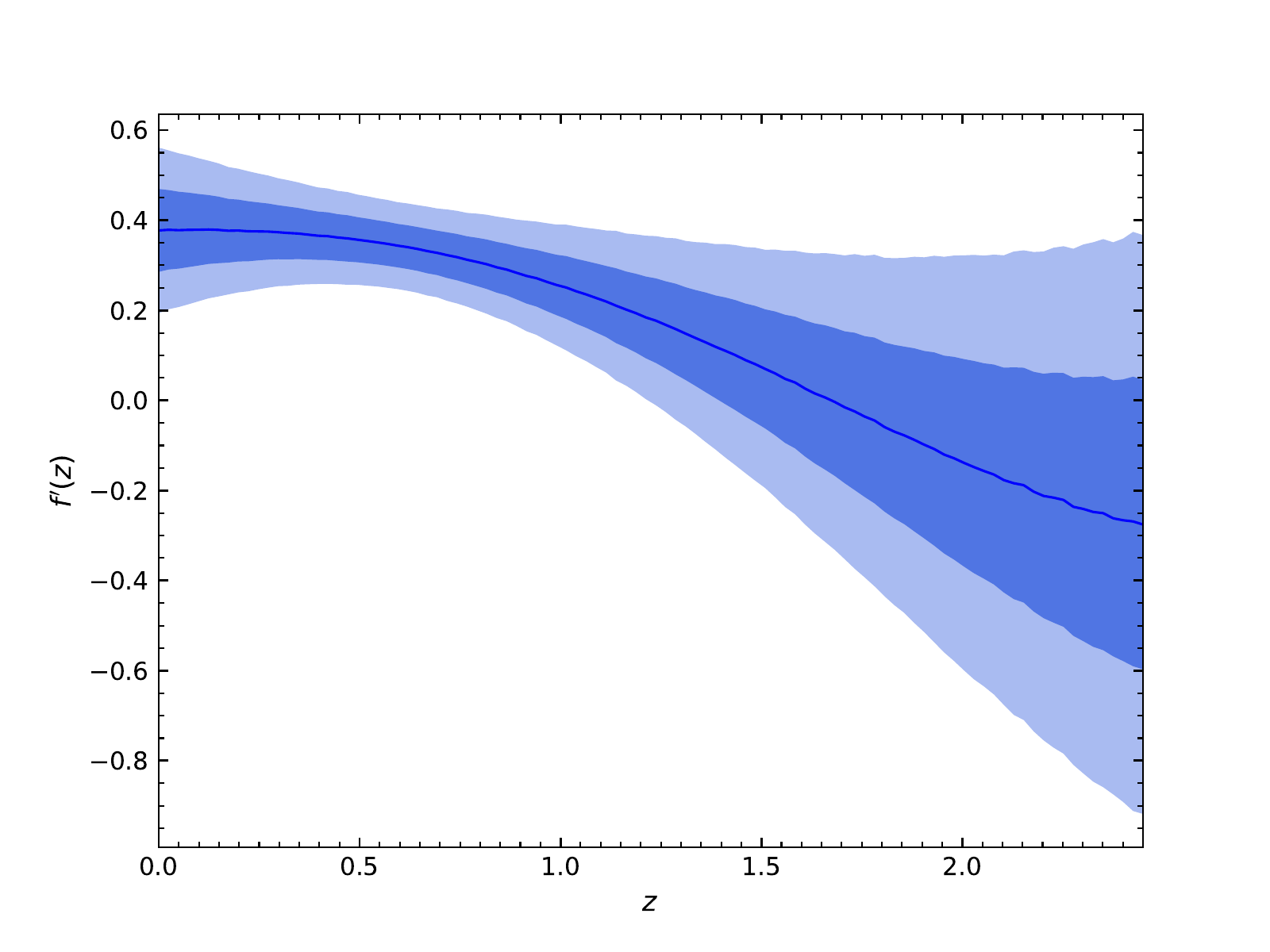}
    \includegraphics[width=0.49\columnwidth]{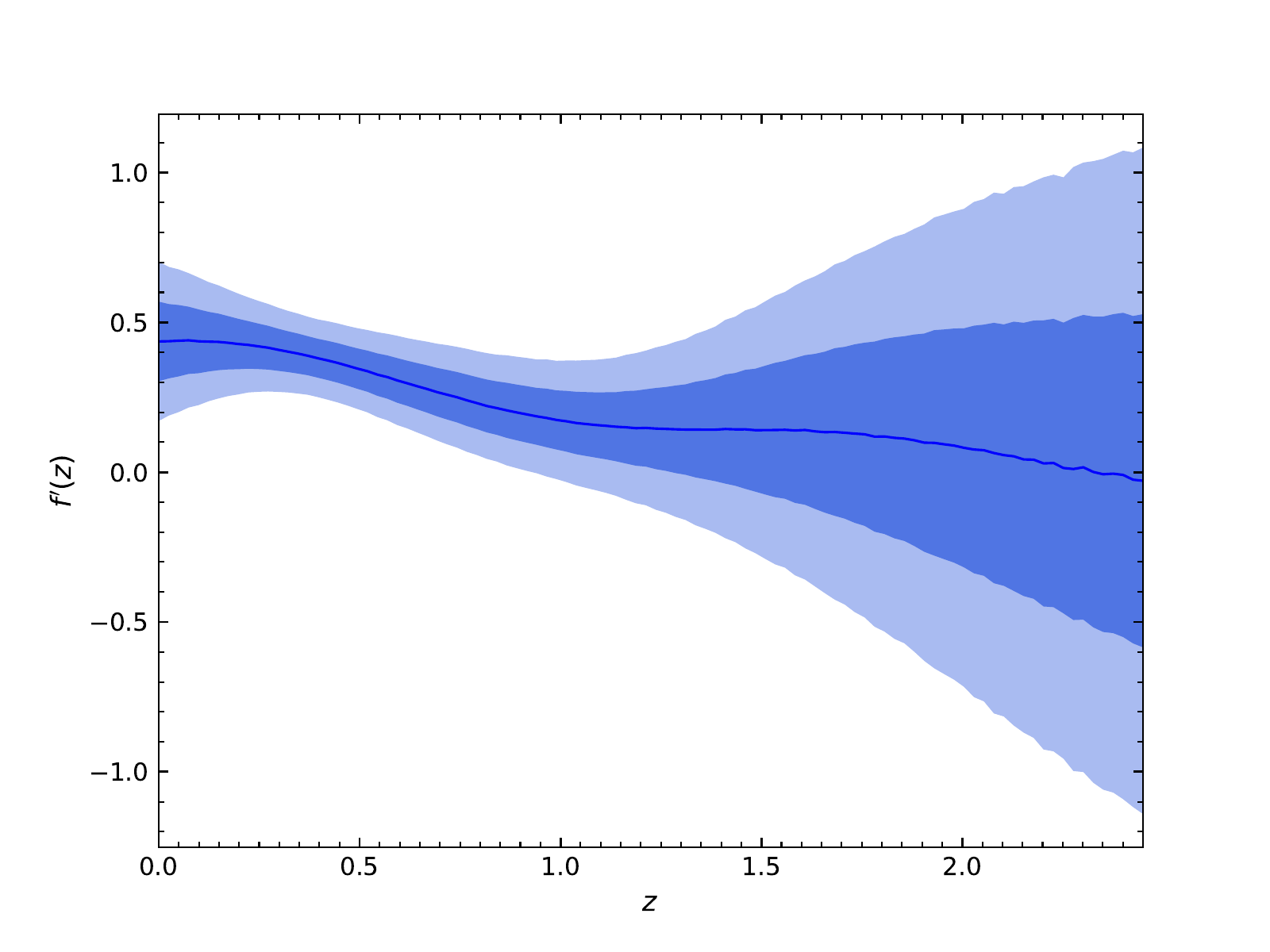}
    \caption{\label{fig:fgrowth_rec_sq_ca}{GP reconstructions of $f(z)$, and $f^\prime(z)$ with  the squared--exponential (left) and the Cauchy (right) kernel functions.}}
\end{center}
\end{figure*}

We can now map the obtained GP reconstructions of the matter density contrast and its derivatives, to the GP reconstruction of the matter growth rate function $f(z)$ with the use of Eq. (\ref{eq:fz_growth}). Furthermore, we can also infer the evolution of $f^\prime(z)$ from Eq. (\ref{eq:f_prime}), which utilises all the previously reconstructed functions of the normalised matter density contrast. The redshift evolution of the reconstructed functions of $f(z)$ and $f^\prime(z)$ are respectively depicted in the top and bottom panels of Fig. \ref{fig:fgrowth_rec_sq_ca}. Further to the above discussion, the squared--exponential kernel function led to more uniform and slightly tighter GP reconstructions with respect to those inferred via the Cauchy kernel function, although the computed GP reconstructions from the two kernel functions are consistent with one another. \medskip

\begin{figure*}[t!]
\begin{center}
    \includegraphics[width=0.49\columnwidth]{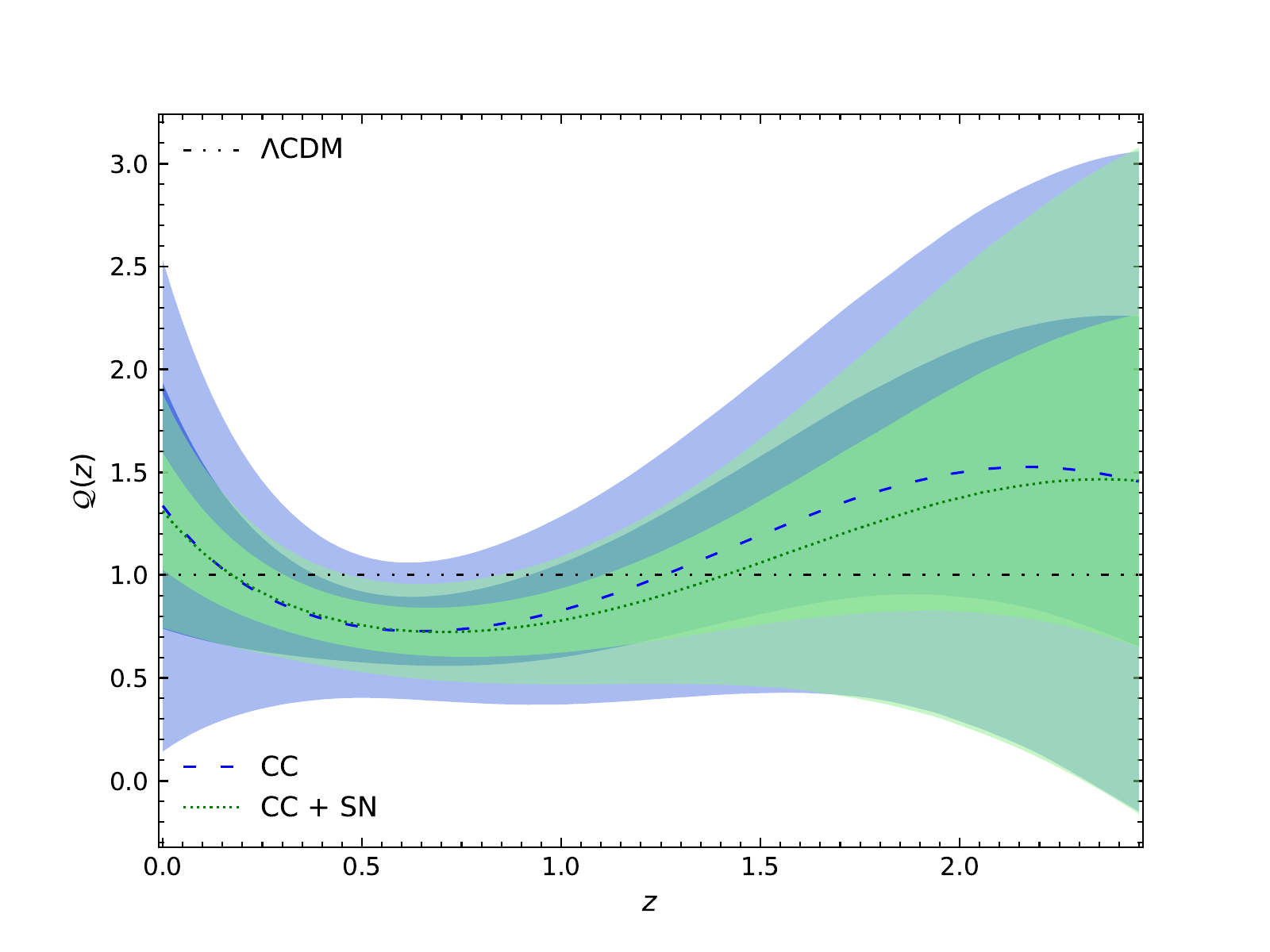}
    \includegraphics[width=0.49\columnwidth]{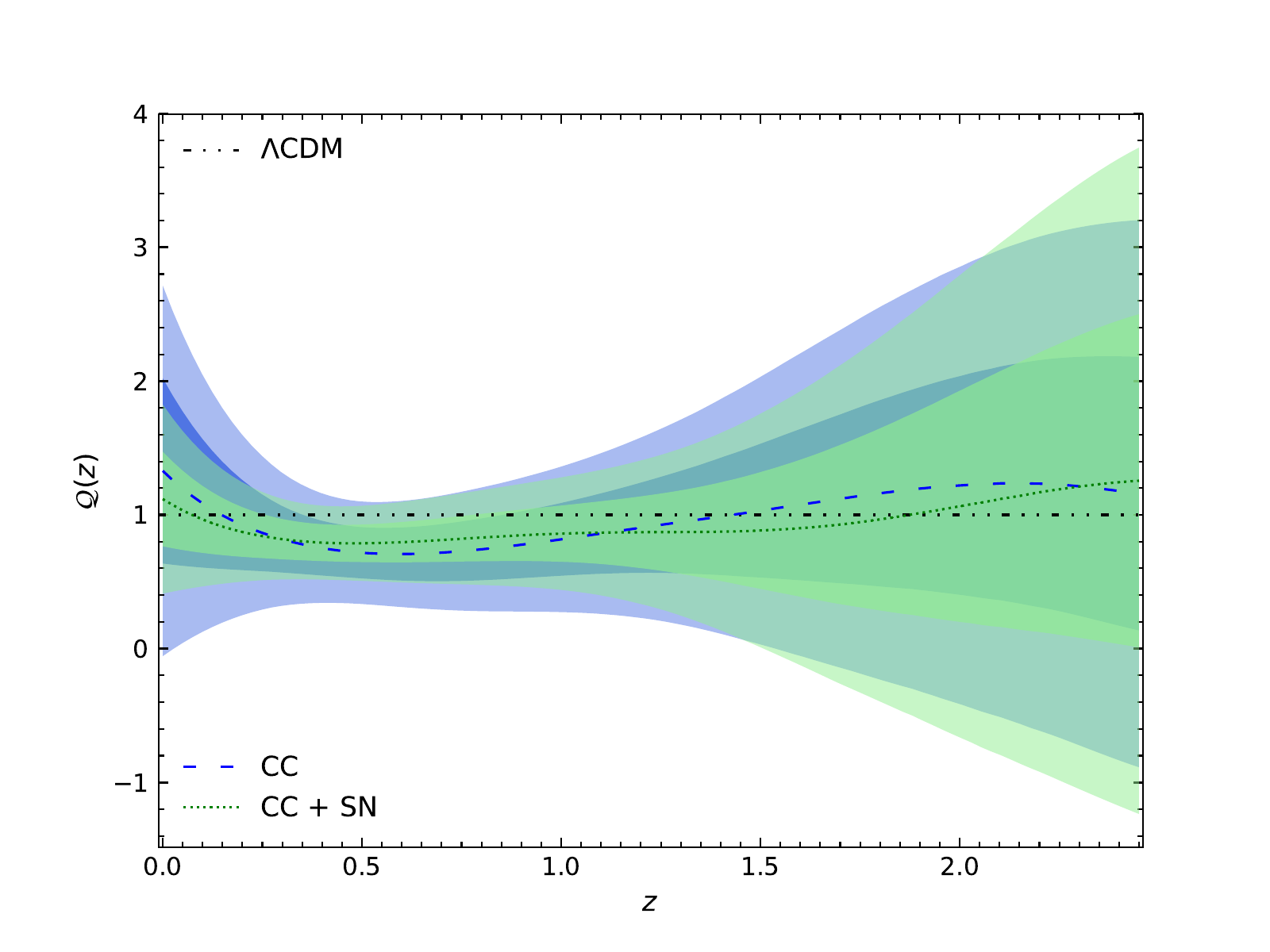}
    \includegraphics[width=0.49\columnwidth]{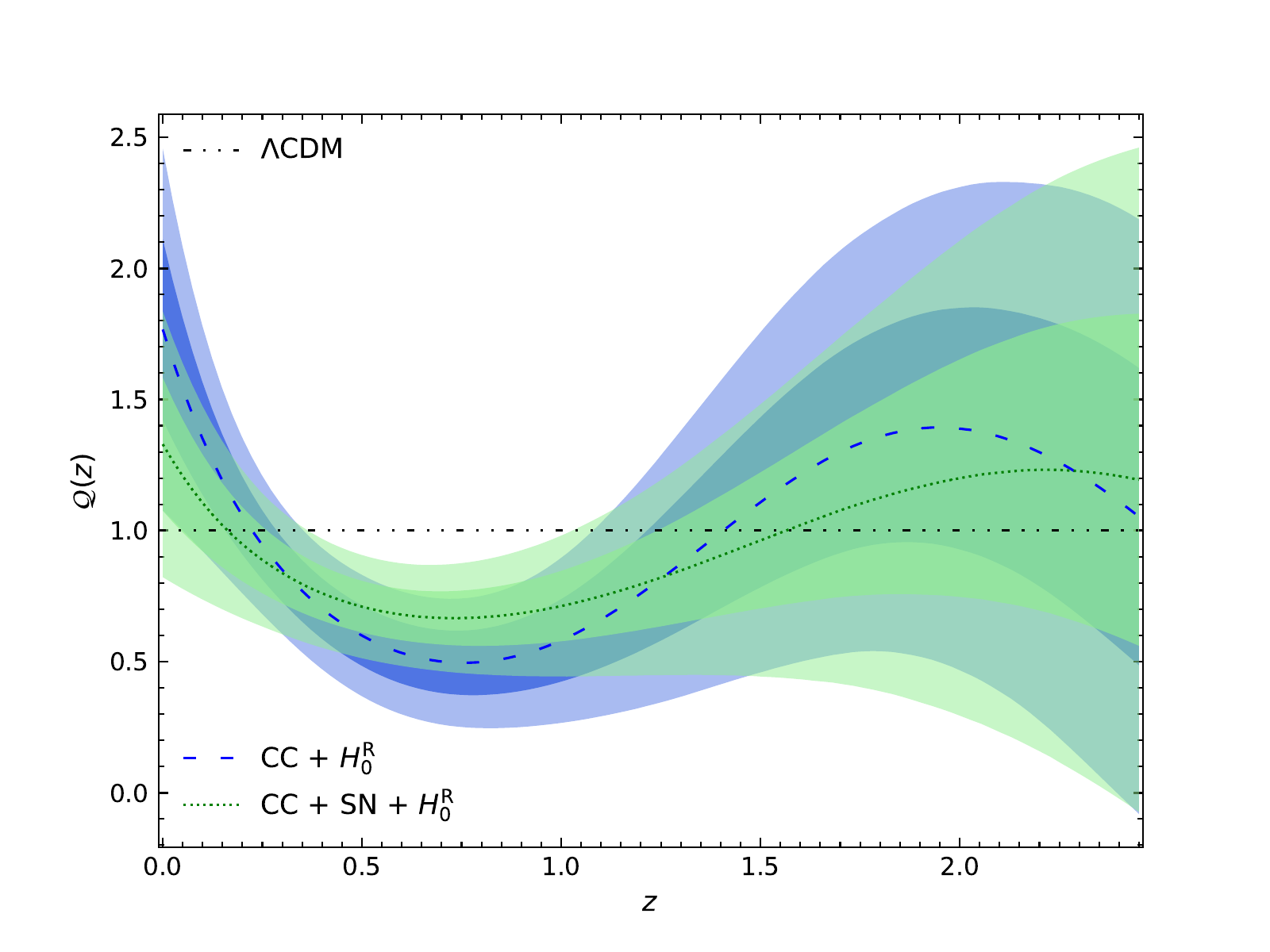}
    \includegraphics[width=0.49\columnwidth]{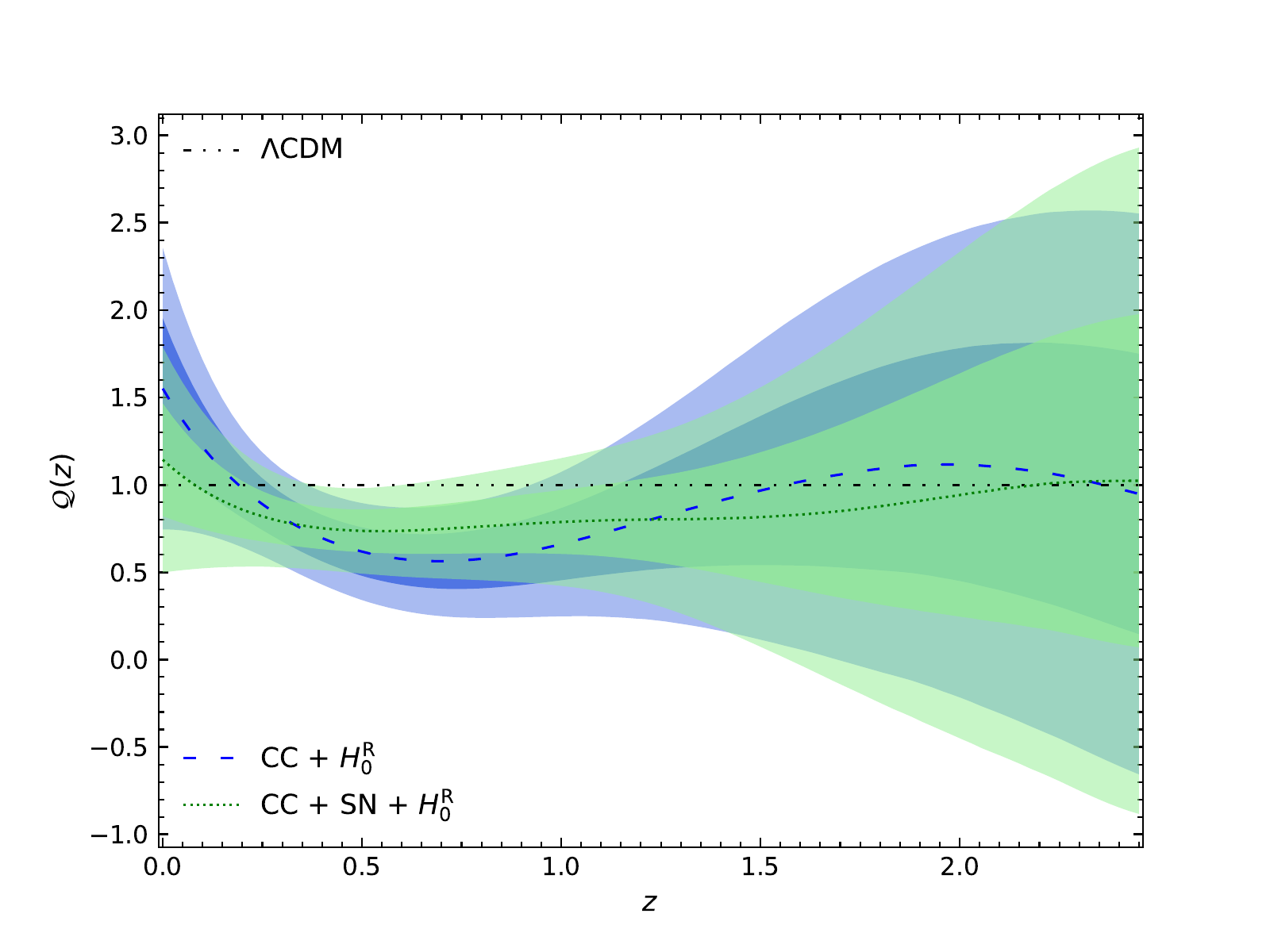}
    \includegraphics[width=0.49\columnwidth]{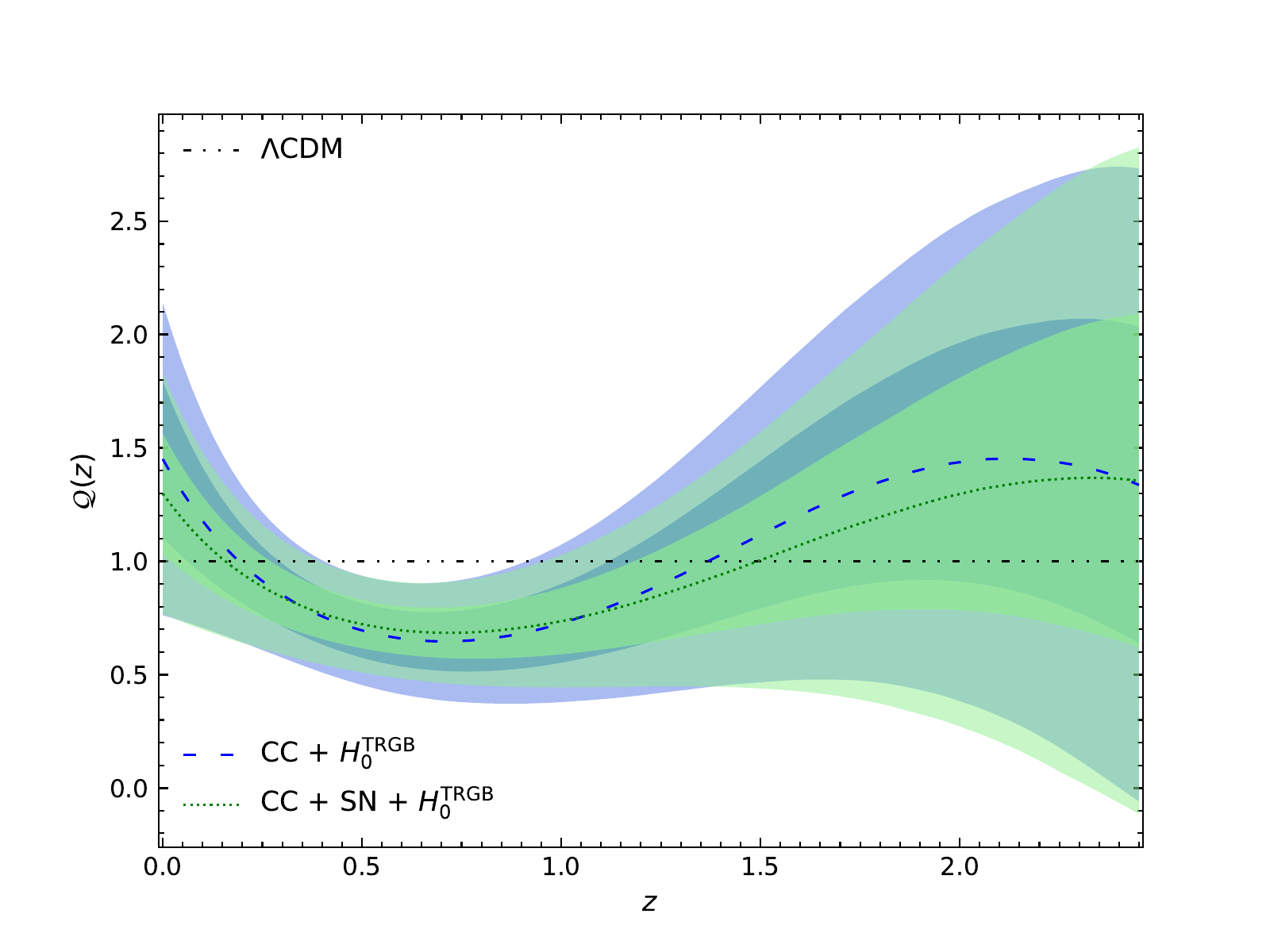}
    \includegraphics[width=0.484\columnwidth]{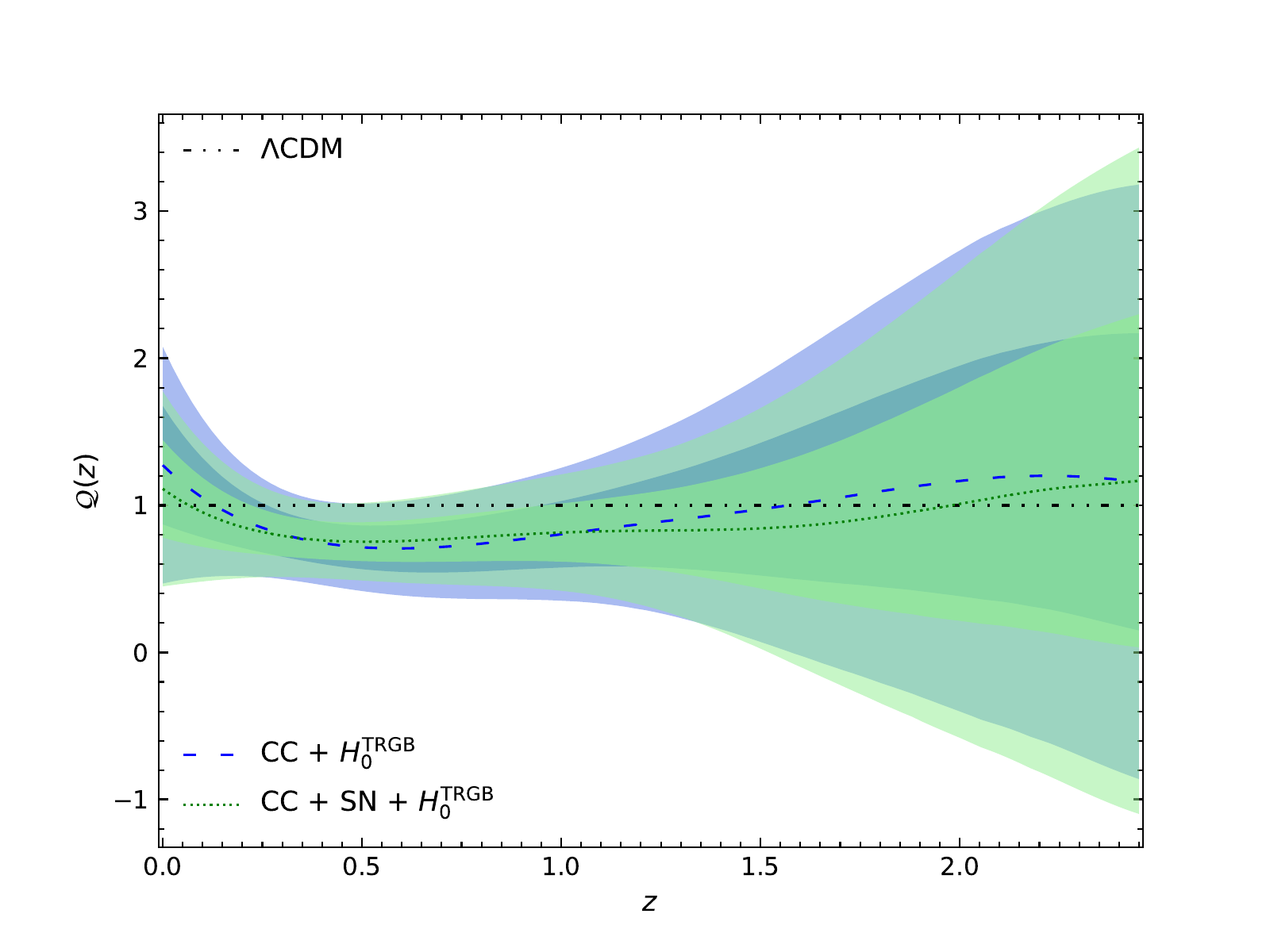}
    \caption{\label{fig:Qz_rec_sq_ca}{GP reconstructions of $\mathcal{Q}(z)$ with CC (blue) and CC + SN (green) data sets. Similar to Fig. \ref{fig:H0_rec_sq_ca}, we consider different $H_0^{}$ priors, and utilise the squared--exponential (left) and the Cauchy (right) kernel functions.}}
\end{center}
\end{figure*}

We are now in a position to reconstruct the fractional change in the effective Gravitational coupling function $\mathcal{Q}(z)$ via Eq. (\ref{eq:Qz}), where we have further assumed that $\Omega_{m,0}=0.3153 \pm 0.0073$ \cite{Aghanim:2018eyx}. For the GP reconstruction of $\mathcal{Q}(z)$ we also have to adopt a value of Hubble's constant, which we infer from the GP reconstruction of $H(z)$, and when applicable, an $H_0$ prior will also be considered. Since we focus on the reconstruction of $H(z)$ without a local $H_0$ prior, and the reconstruction of $H(z)$ in the presence of the $H_0^\mathrm{R}$ and $H_0^\mathrm{TRGB}$ priors, we will have three scenarios for each choice of kernel functions, as depicted in the left--hand and right--hand panels of Fig. \ref{fig:Qz_rec_sq_ca}.  \medskip

\begin{figure*}[t!]
\begin{center}
    \includegraphics[width=0.49\columnwidth]{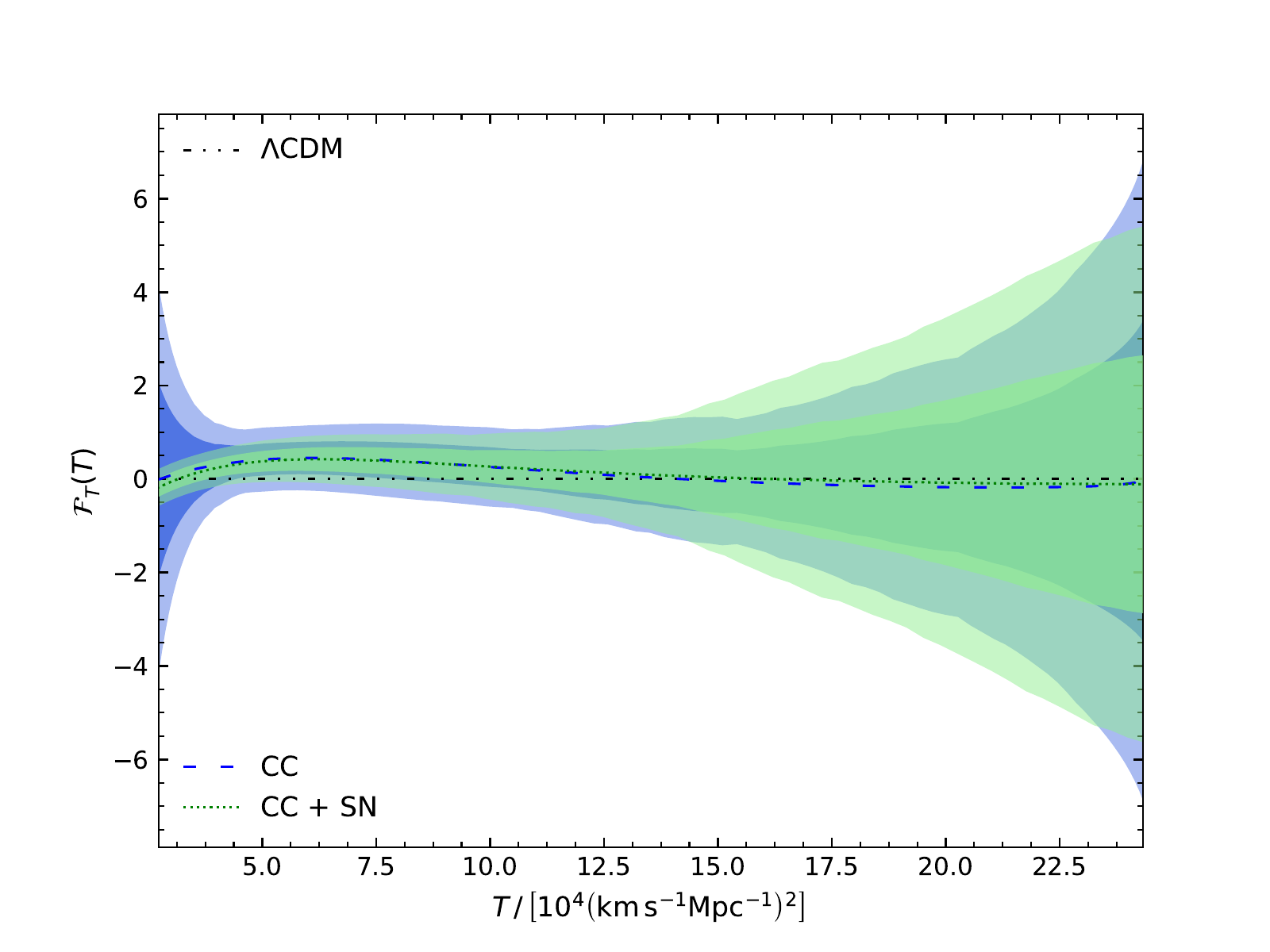}
    \includegraphics[width=0.49\columnwidth]{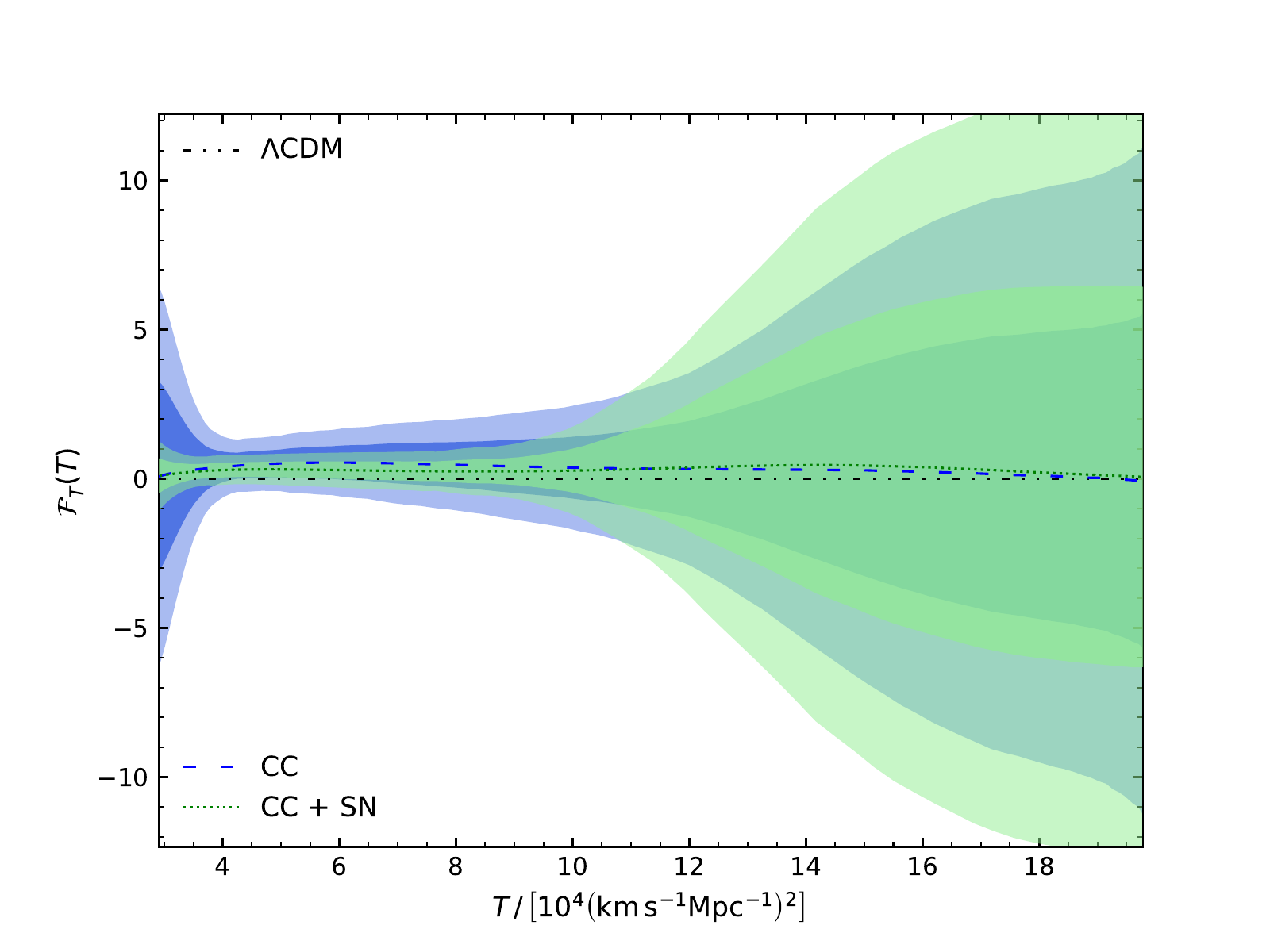}
    \includegraphics[width=0.488\columnwidth]{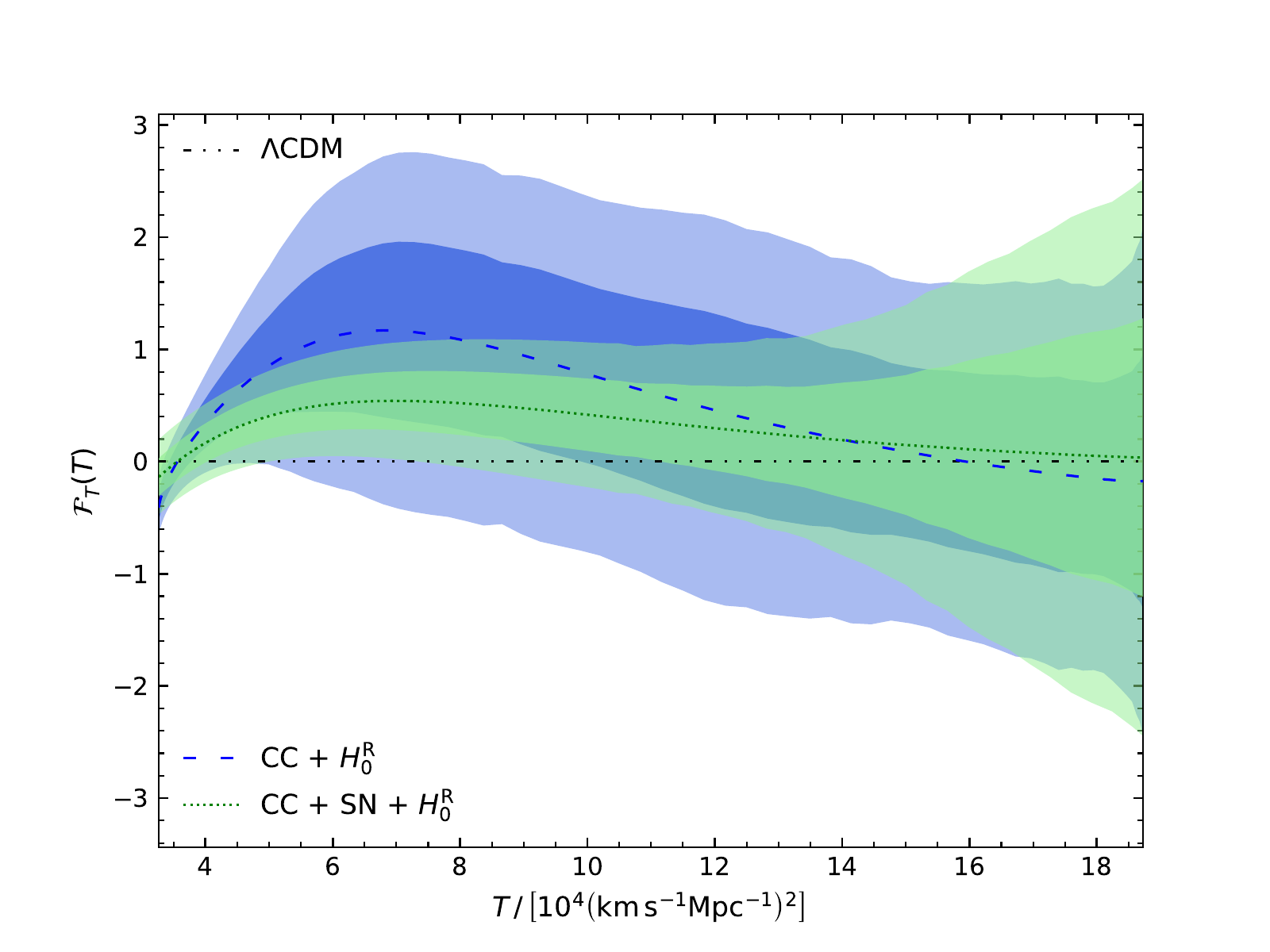}
    \includegraphics[width=0.494\columnwidth]{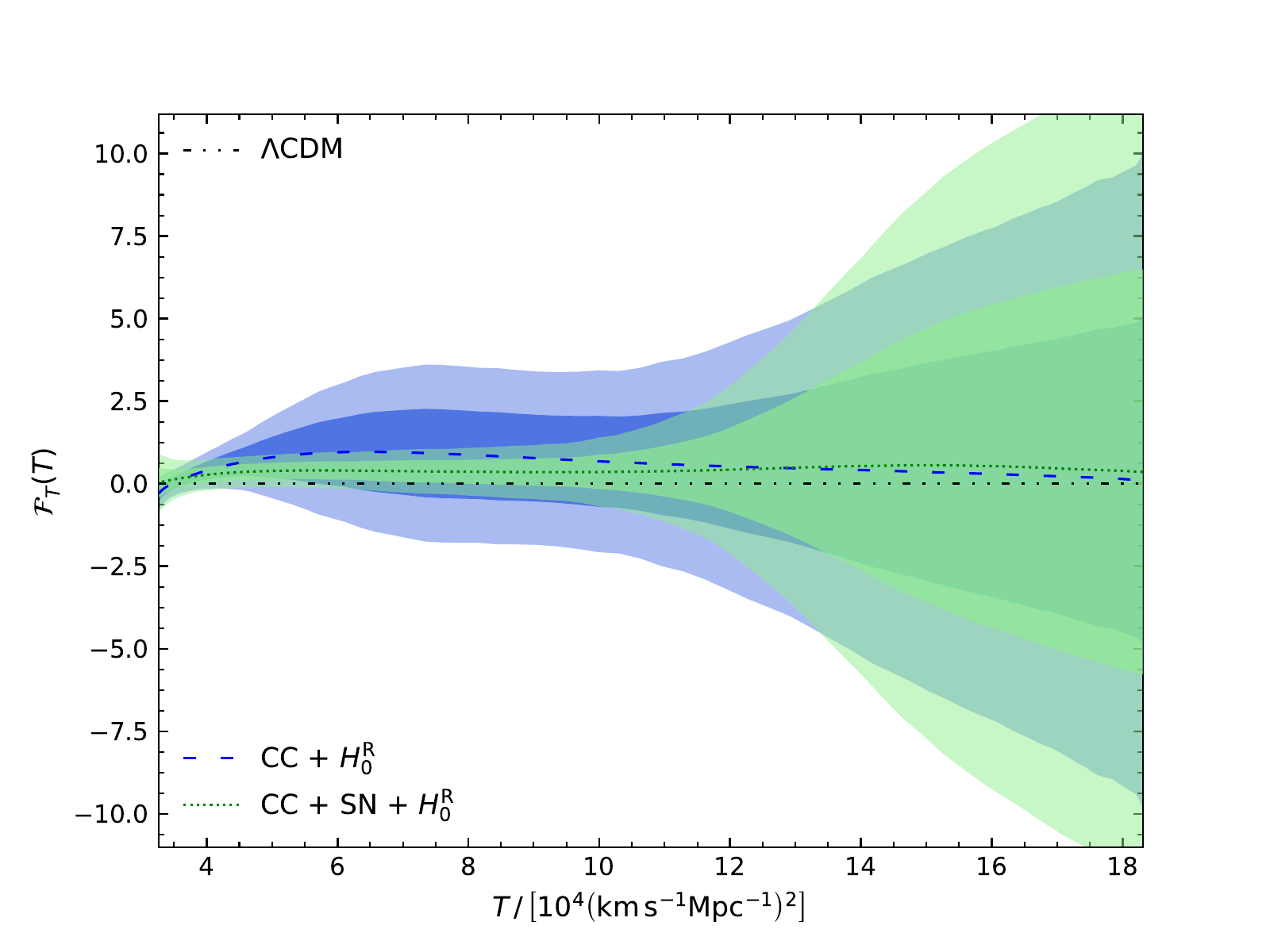}
    \includegraphics[width=0.488\columnwidth]{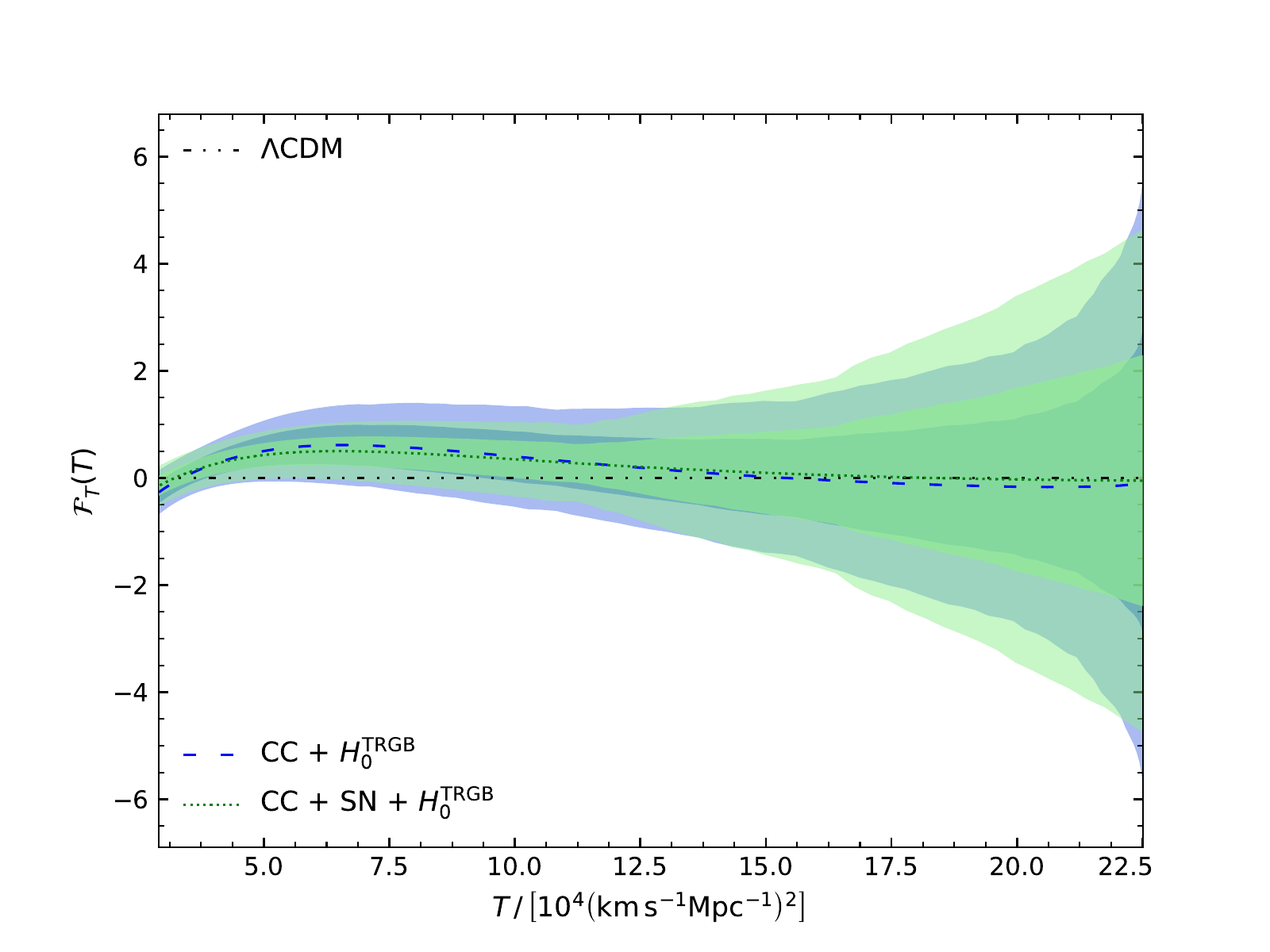}
    \includegraphics[width=0.495\columnwidth]{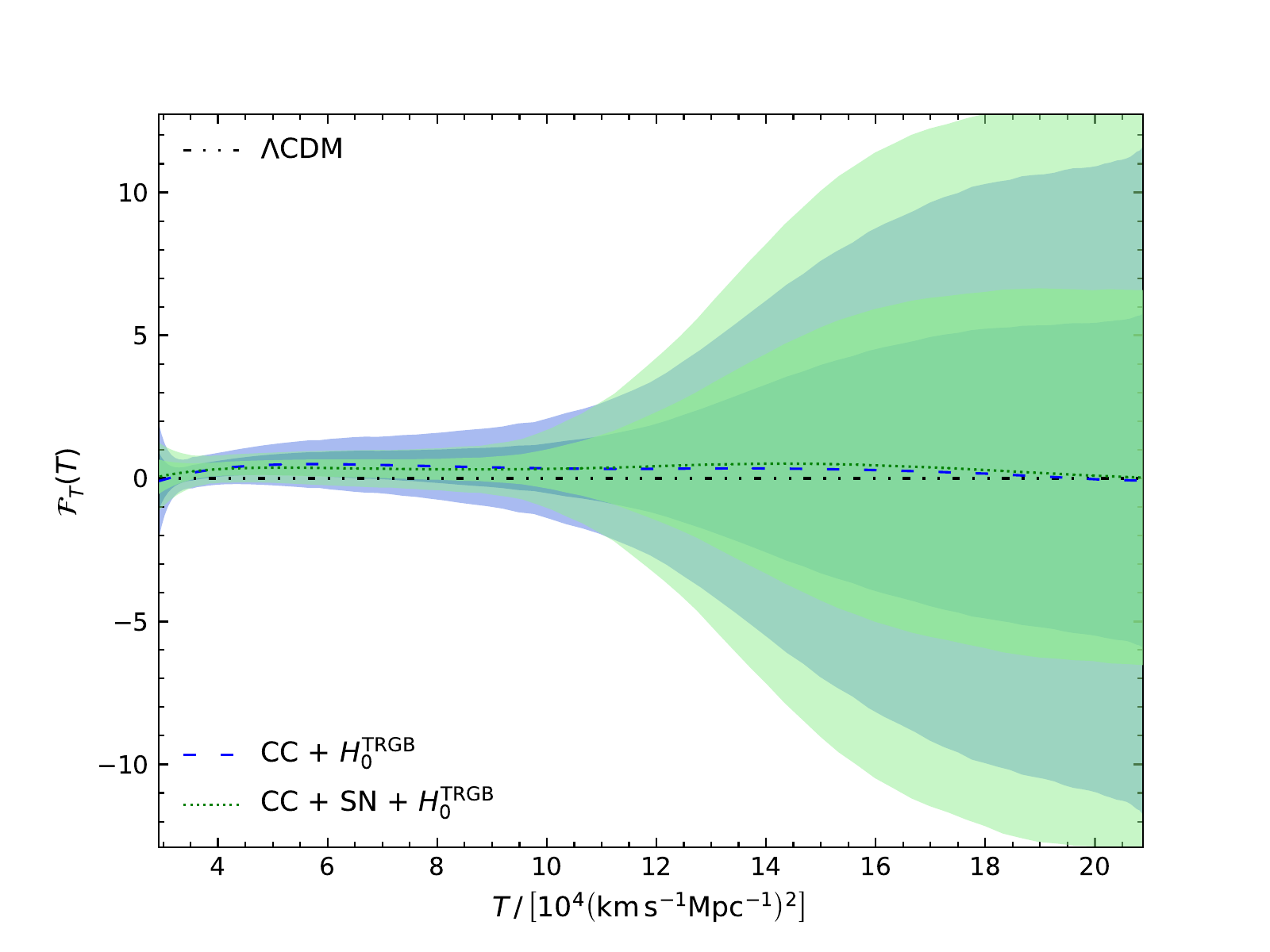}
    \caption{\label{fig:f_T_T_rec_sq_ca}{GP reconstructions of $\mathcal{F}_T^{}(T)$ with CC (blue) and CC + SN (green) data sets. Similar to Fig. \ref{fig:H0_rec_sq_ca}, we consider different $H_0^{}$ priors, and utilise the squared--exponential (left) and the Cauchy (right) kernel functions.}}
\end{center}
\end{figure*}

\begin{figure*}[t!]
\begin{center}
    \includegraphics[width=0.49\columnwidth]{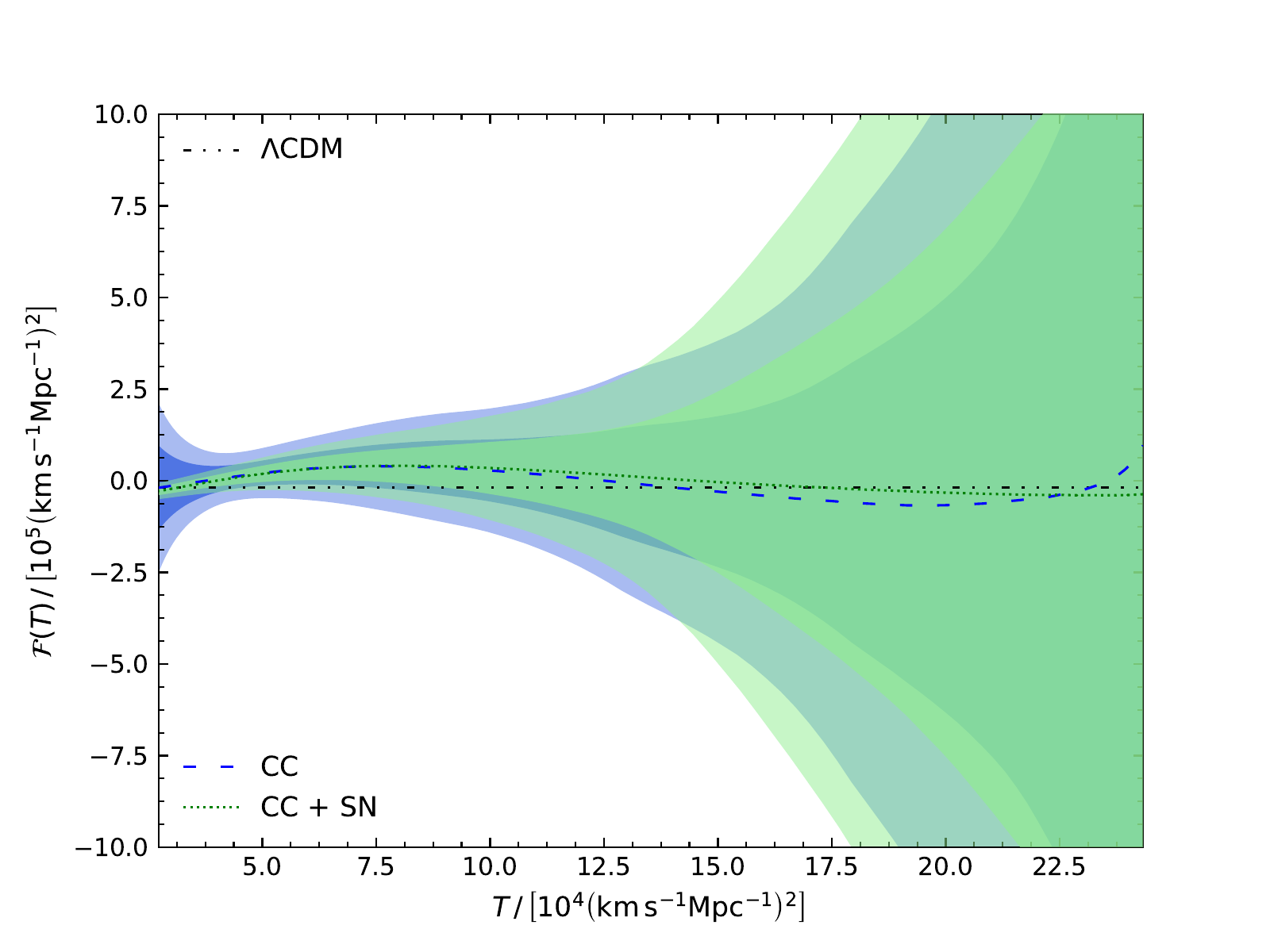}
    \includegraphics[width=0.49\columnwidth]{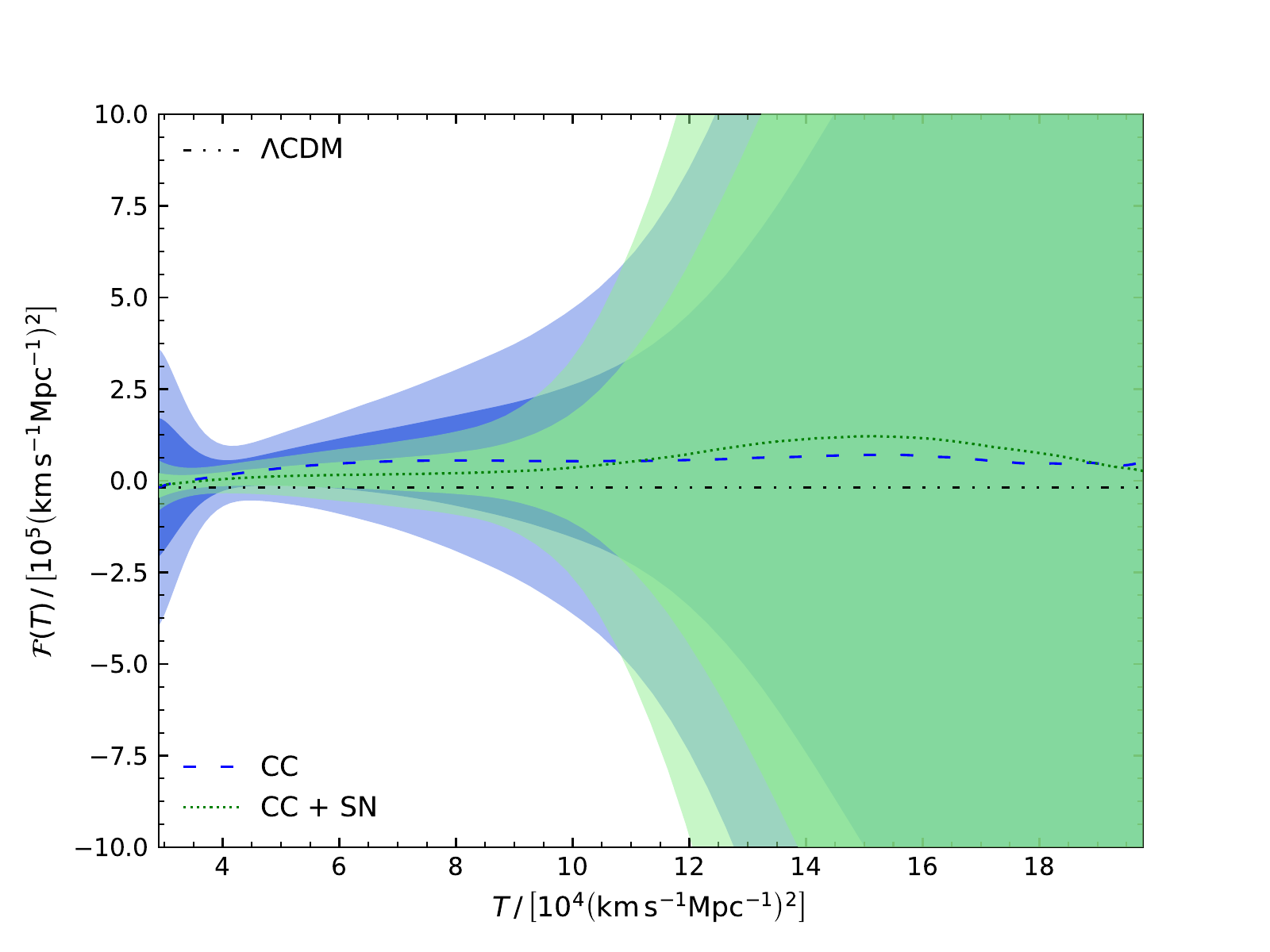}
    \includegraphics[width=0.49\columnwidth]{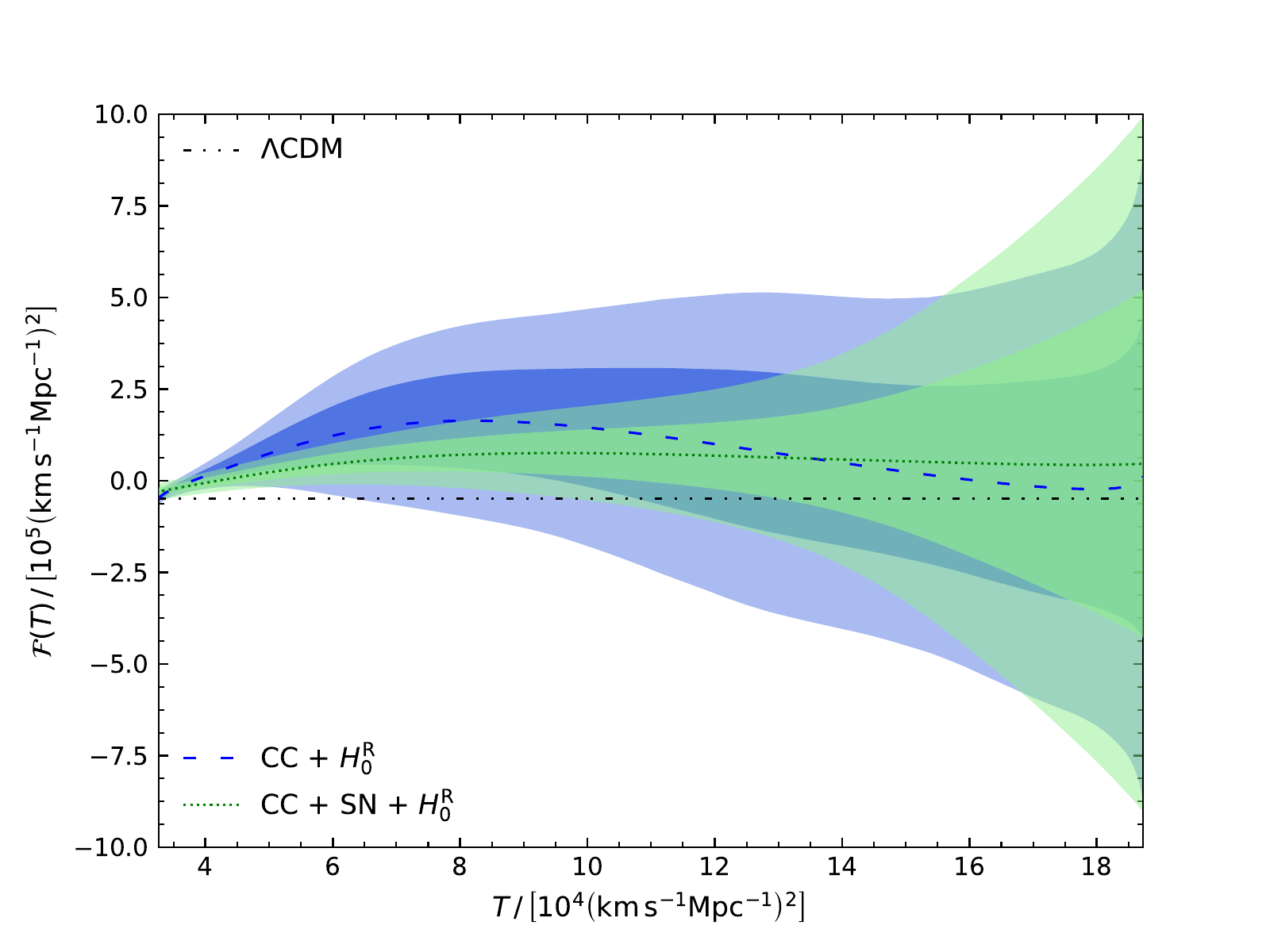}
    \includegraphics[width=0.49\columnwidth]{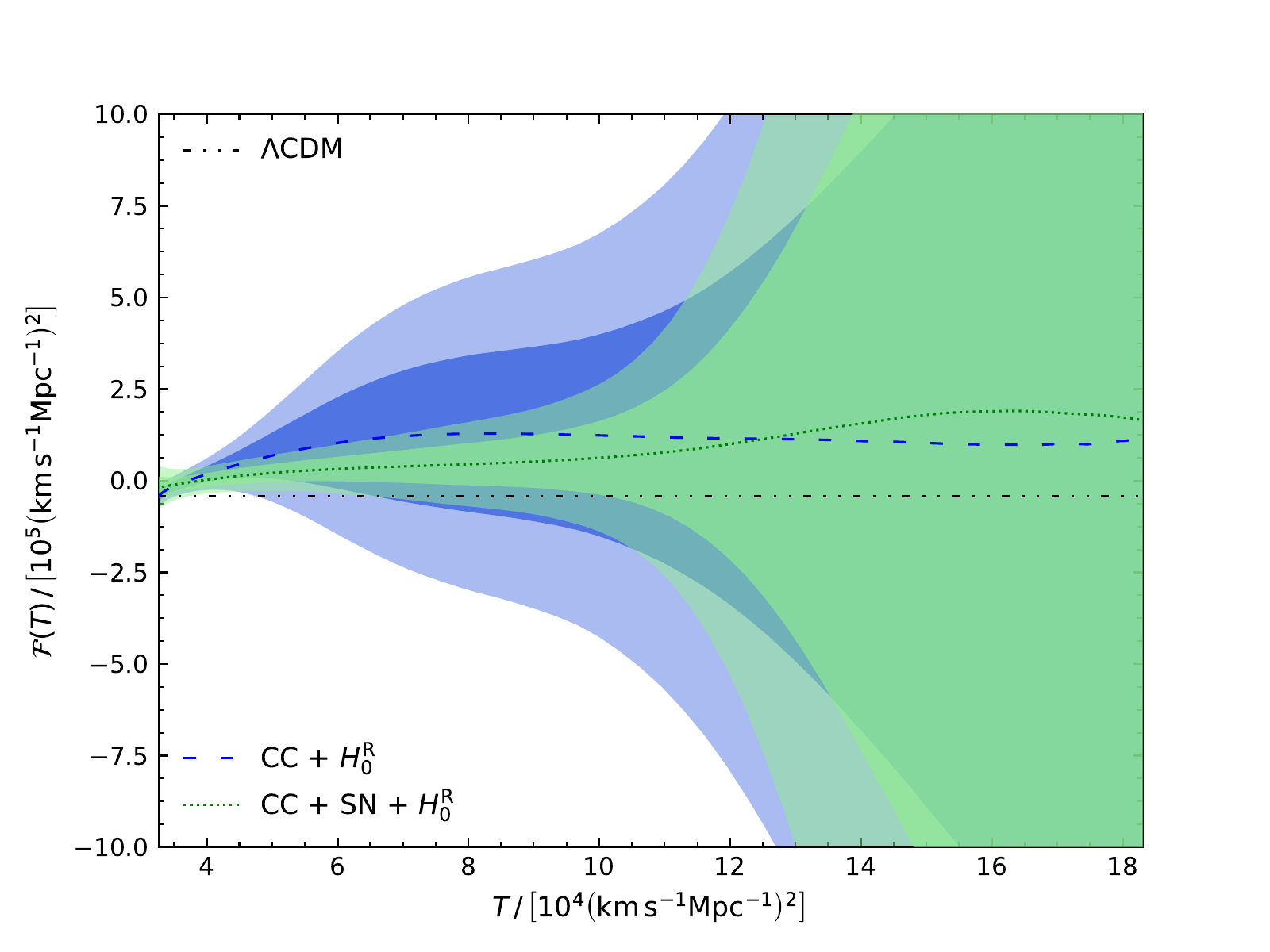}
    \includegraphics[width=0.4878\columnwidth]{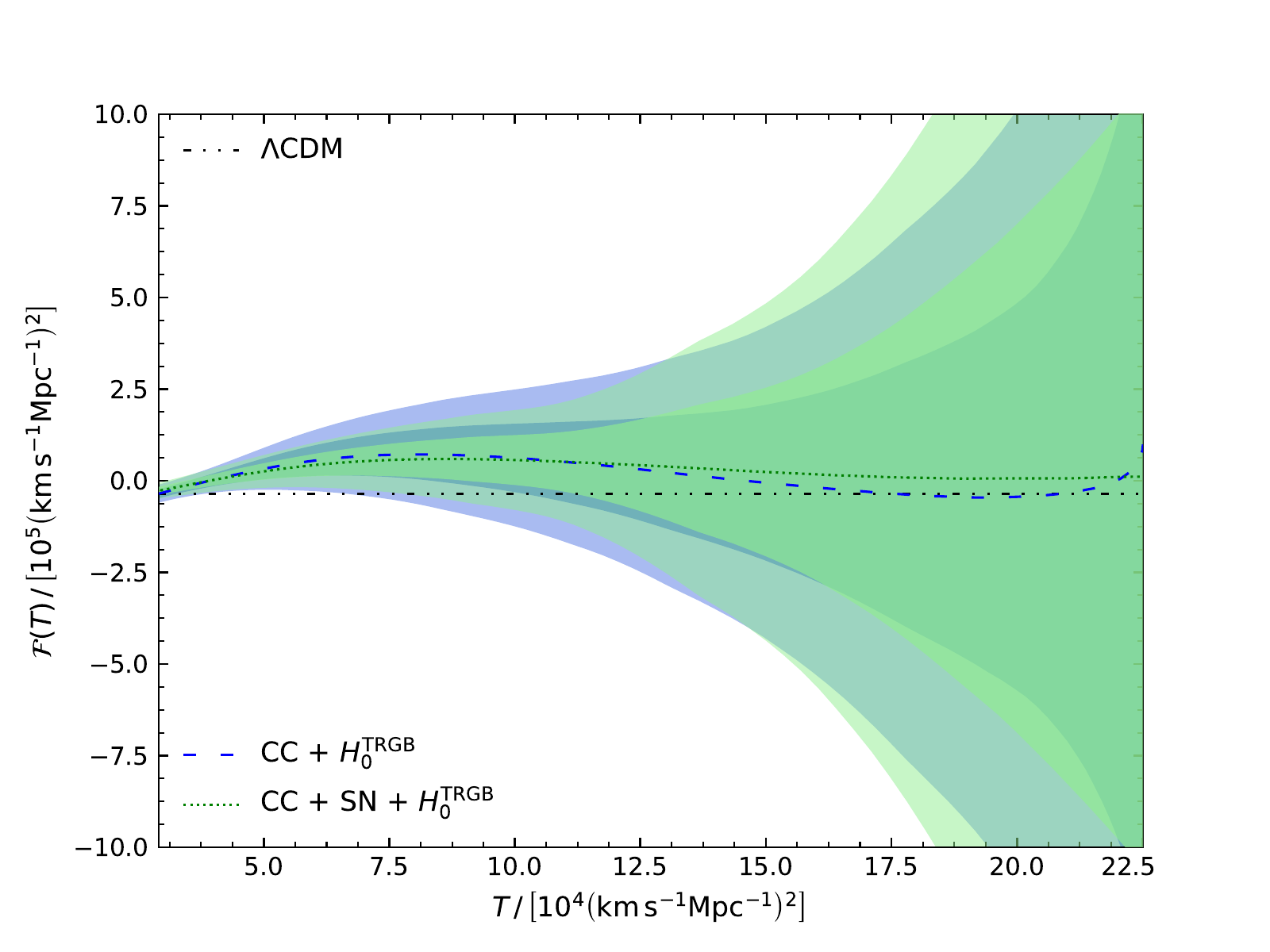}
    \includegraphics[width=0.486\columnwidth]{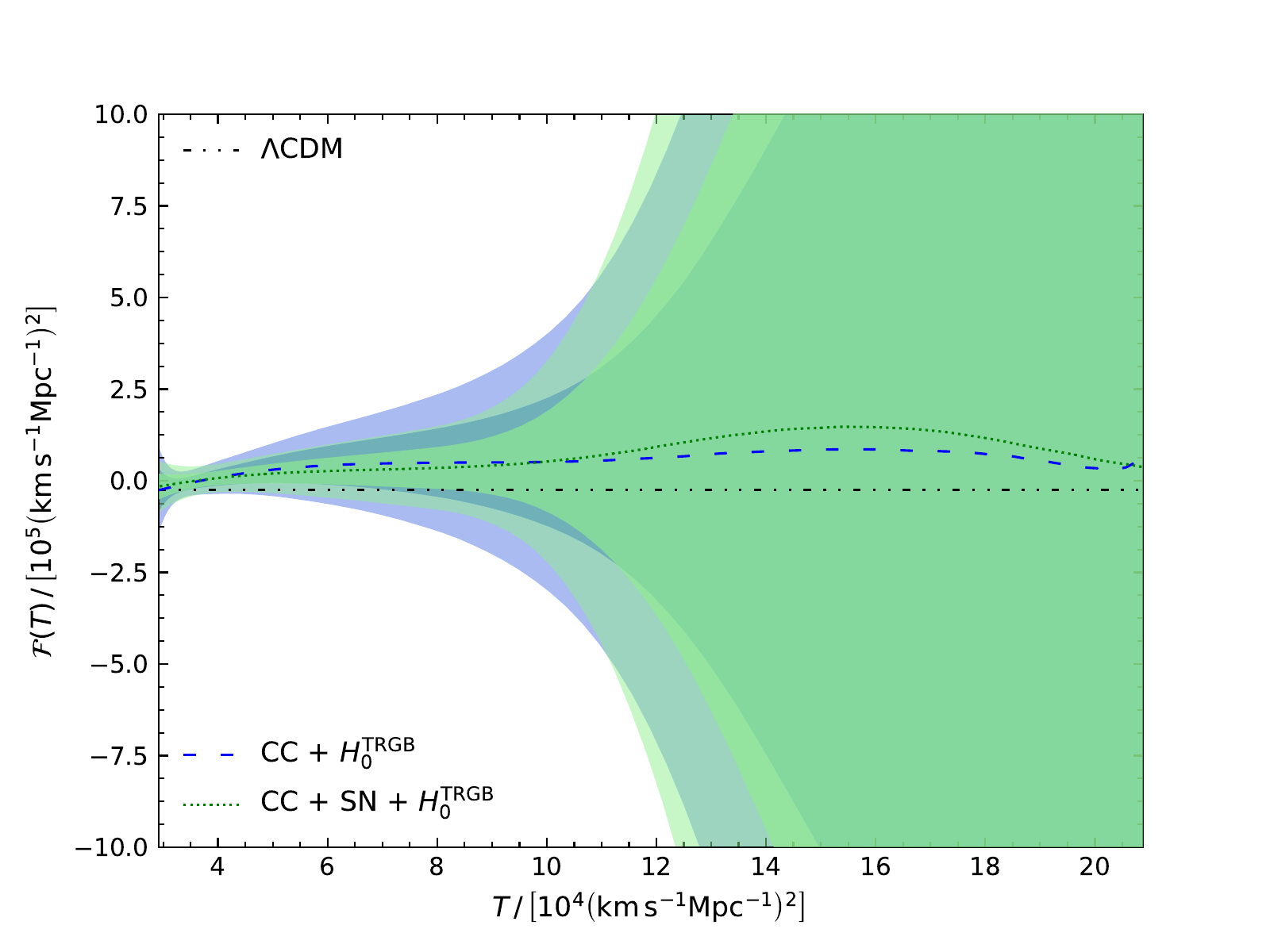}
    \caption{\label{fig:fT_rec_sq_ca}{GP reconstructions of $\mathcal{F}(T)$ with CC (blue) and CC + SN (green) data sets. Similar to Fig. \ref{fig:H0_rec_sq_ca}, we consider different $H_0^{}$ priors, and utilise the squared--exponential (left) and the Cauchy (right) kernel functions.}}
\end{center}
\end{figure*}

One could clearly notice that in the case of CC and CC + SN data sets without a local $H_0^{}$ prior, the squared--exponential and Cauchy kernel functions led to $\mathcal{Q}(z)\simeq1$, which is in full agreement with the elementary concordance model of cosmology. However, in the presence of an $H_0^{}$ prior, particularly $H_0^\mathrm{R}$, the GP reconstructed function of $\mathcal{Q}(z)$ is more dynamical and in the range of $0.4<z<0.9$ we find that $\mathcal{Q}(z)$ deviates from unity by $\gtrsim2\sigma$. As expected, such deviation is suppressed when we use the Cauchy kernel function, since all previously inferred GP reconstructions with this kernel function were always found to be more conservative than the corresponding results with the squared--exponential kernel function. Indeed, $\mathcal{Q}(z)$ is always found to be in agreement $(\lesssim2\sigma)$ with unity when we utilise the Cauchy kernel function in the presence or absence of the considered priors on the Hubble constant. \medskip 

From the above reconstructions of $\mathcal{Q}(z)$, we can easily reconstruct the redshift evolution of $\mathcal{F}_T^{}(z)$ via Eq. (\ref{eq:Qz}). We further express the GP reconstructed function of $\mathcal{F}_T^{}(z)$ as a function of $T$, by mapping the GP reconstructed $H(z)$ to the torsion scalar via Table. \ref{tab:TG_scalars}. The evolution of the GP reconstructed functions $\mathcal{F}_T^{}(T)$ are illustrated in the panels of Fig. \ref{fig:f_T_T_rec_sq_ca}, where the left--hand side and right--hand side panels correspond respectively to the reconstructions with the squared--exponential and Cauchy kernel functions. Further to the above discussion, the Cauchy kernel function was again characterised by more conservative constraints on $\mathcal{F}_T^{}(T)$ with respect to those inferred with the squared--exponential kernel function. Moreover, it is clear that the joint CC + SN data set led to tighter constraints with respect to the CC data. All reconstructions of $\mathcal{F}_T^{}(T)$ were found to be consistent with a null value; in full agreement with the $\Lambda$CDM model. We should remark that even in the case of the $H_0^\mathrm{R}$ prior, the confidence region of the GP reconstruction was always consistent $(\lesssim2\sigma)$ with a null value. \medskip

We finally reconstruct the $\mathcal{F}(T)$ function via Eq. (\ref{eq:f(T)_rec}) where we depict the GP reconstructions in Fig. \ref{fig:fT_rec_sq_ca}. The GP reconstructions of $\mathcal{F}(T)$ were found to be weakly constrained at high values of $T$ (corresponding to high--$z$ values), particularly with the Cauchy kernel function. Furthermore, the GP reconstructions of $\mathcal{F}(T)$ are consistent with the $\Lambda$CDM theoretical prediction, although at very low values of $T$ (low--$z$), there is a slight departure $(\gtrsim2\sigma)$ from the corresponding $\Lambda$CDM value when we adopt $H_0^\mathrm{R}$, and mildly with $H_0^\mathrm{TRGB}$. Such a low--redshift deviation from the $\Lambda$CDM model could be attributed to the disagreement between the high $H_0^\mathrm{R}$ value and the preferred (GP reconstructed) value of $H_0^{}$ with the CC and CC + SN data sets, leaving a small redshift window for a plausible and competitive $\mathcal{F}(T)$ model. \medskip

\section{\label{sec:conc}Conclusions}
GP is a stochastic process that offers an interesting approach to modeling the behaviour of data. In the cosmological context, GP are an effective setting in which to investigate cosmological parameters such as the Hubble and $f\sigma_8^{}$ parameters. These parameter values have led to intense debate in recent years with numerous approaches being explored \cite{divalentino2021realm}. In this work, we continue to build on recent works in which cosmological data is used to constrain the arbitrary Lagrangian within a broad cosmology beyond $\Lambda$CDM, which in this instance is $\mathcal{F}(T)$ gravity. In Refs.\cite{Cai:2019bdh,Briffa:2020qli,Ren:2021tfi}, the idea of using Hubble data to reconstruct the $\mathcal{F}(T)$ Lagrangian was explored together with other background cosmological parameters such as the deceleration and equation of state parameters (which are consistent with the current work). In the present work, this approach was extended to growth rate data wherein this data was used to constrain the values of the $\mathcal{F}(T)$ Lagrangian. This is the first time (to the best of our knowledge) that growth rate data has been used in this way. Moreover, this provides a crucial consistency check on the results from the background analysis. \medskip

To achieve this, we use the Friedmann equation in Eq. (\ref{eq:Friedmann_1}) to reconstruct Hubble data, and in conjunction with the linear perturbation equation associated with scalar perturbations, we determine values of $\mathcal{F}(T)$ using Eq. (\ref{eq:f(T)_rec}). To test the model independence of the GP approach, we utilised two main kernel functions, namely the squared--exponential and Cauchy covariance functions, as well as the Mat\'{e}rn kernel in the appendix (for comparison purposes). To within 1$\sigma$ confidence level, the ensuing results are consistent with each other. In all cases, we consider the effects of CC and CC + SN datasets separately for these kernels, which are also considered in the context of $H_0^{\rm R}$ and $H_0^{\rm TRGB}$ priors for the background setting. For the growth rate data we considered RSD data, which in combination, provides reliable restrictions on the possible values of the functional form of $\mathcal{F}(T)$. However, for larger redshift values, these constraints are not very restrictive due to the lack of data points at these redshifts. Also, the combined data set errors has an effect on the reconstruction uncertainties. \medskip

These GP reconstructions are depicted in Fig. \ref{fig:fT_rec_sq_ca}, where the Lagrangian functional form is shown against the torsion scalar argument. While $\mathcal{F}(T)$ are well constrained at low redshifts, the situation for higher redshifts becomes exceedingly unconstrained. This is most expressed for the $H_0^{\rm R}$ prior settings due to the very large value of $H_0$ in this circumstance. On the other hand, the $H_0^{\rm TRGB}$ also points to an increase in the uncertainties at lower redshifts but the impact is not as strong as under the $H_0^{\rm R}$ prior setting. We should finally remark that the inferred GP reconstructions of the $\mathcal{F}(T)$ Lagrangian would further instigate the ongoing development of the $\mathcal{F}(T)$ theoretical framework, and its confrontation with cosmological data sets.

\begin{acknowledgments}
The authors would like to acknowledge networking support by the COST Action CA18108 and funding support from Cosmology@MALTA which is supported by the University of Malta. This research has been carried out using computational facilities procured through the European Regional Development Fund, Project No. ERDF-080 ``A supercomputing laboratory for the University of Malta''.
\end{acknowledgments}

\begin{figure*}[t!]
\begin{center}
    \includegraphics[width=0.49\columnwidth]{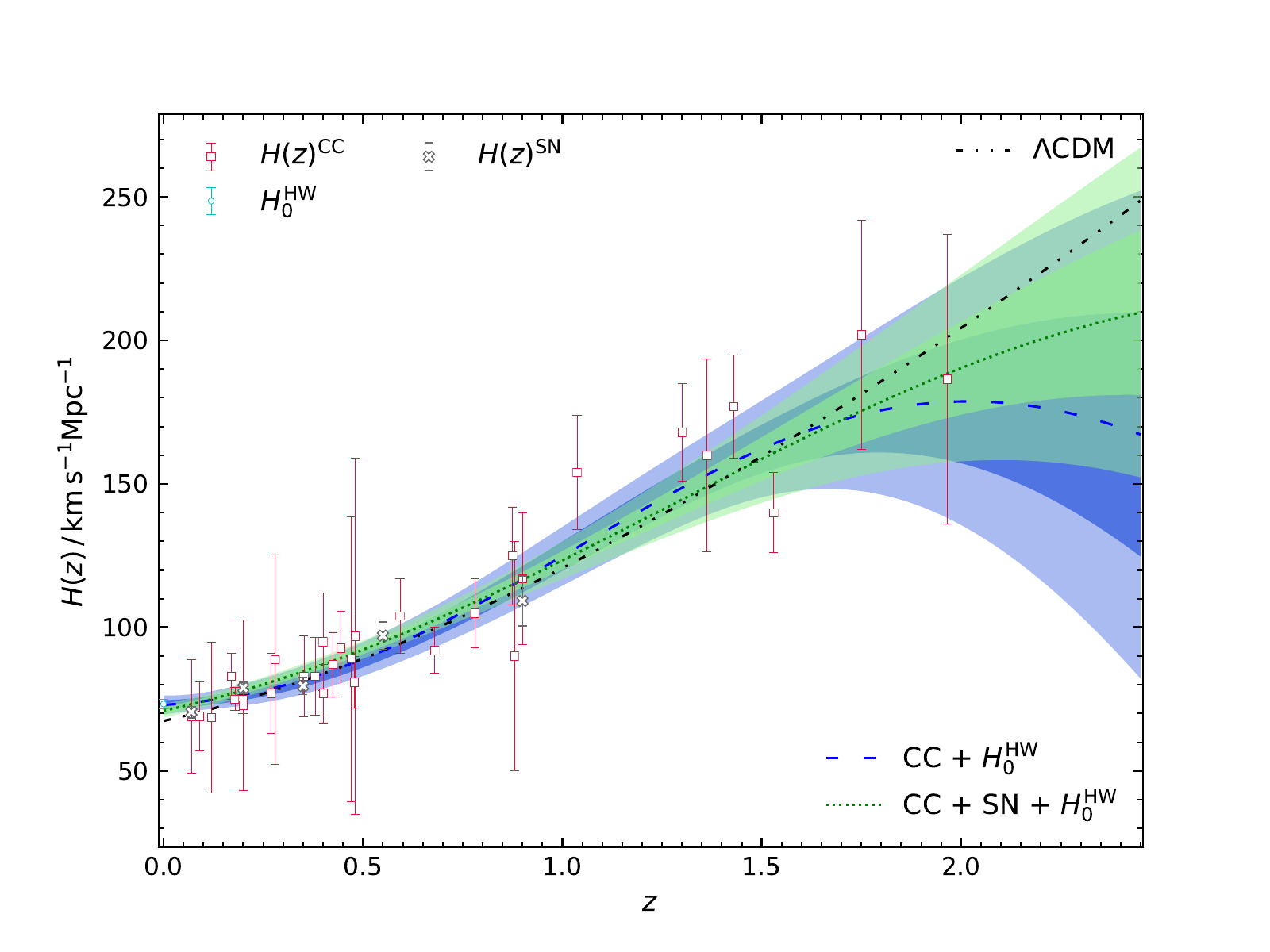}
    \includegraphics[width=0.49\columnwidth]{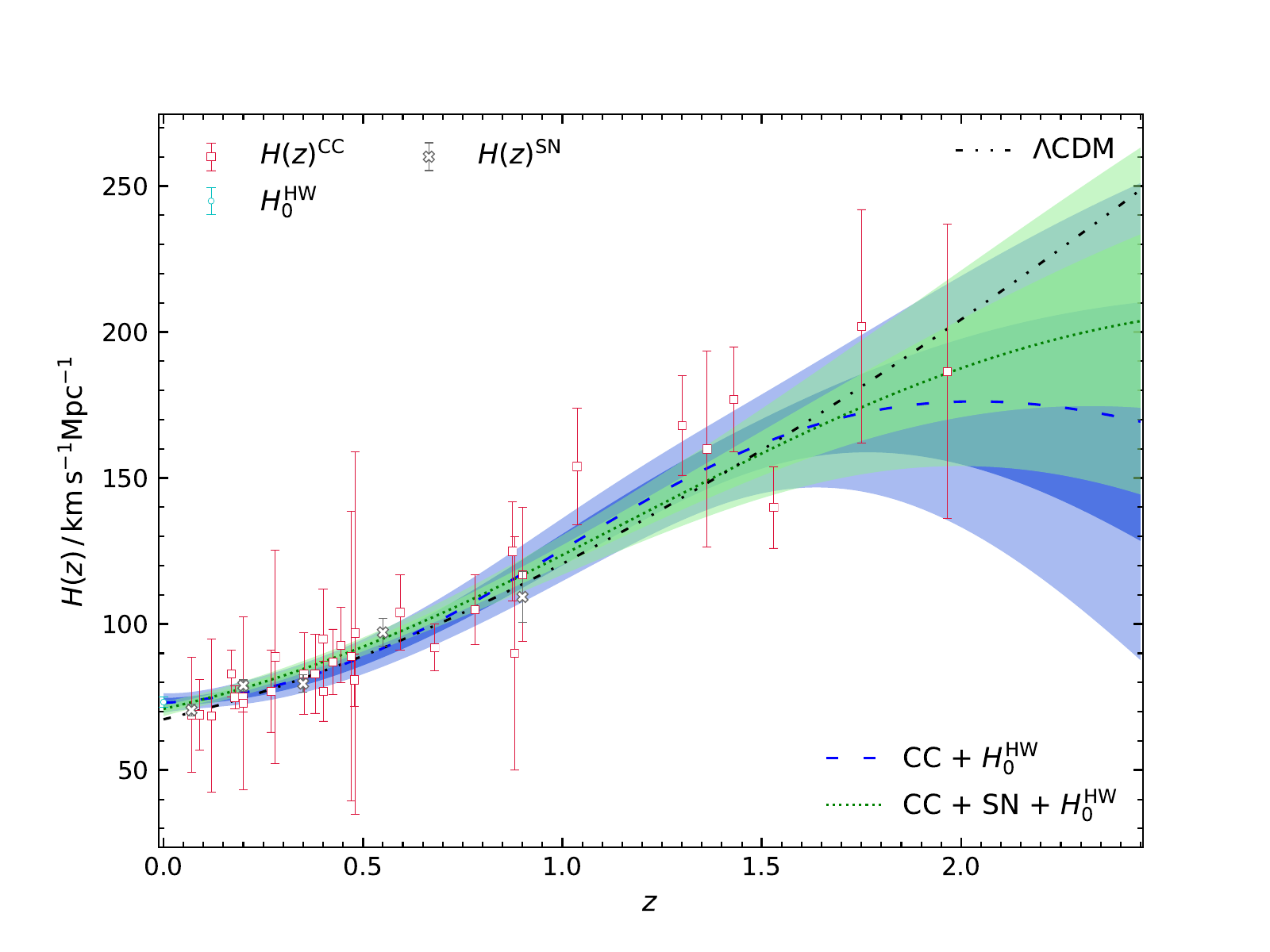}
    \includegraphics[width=0.49\columnwidth]{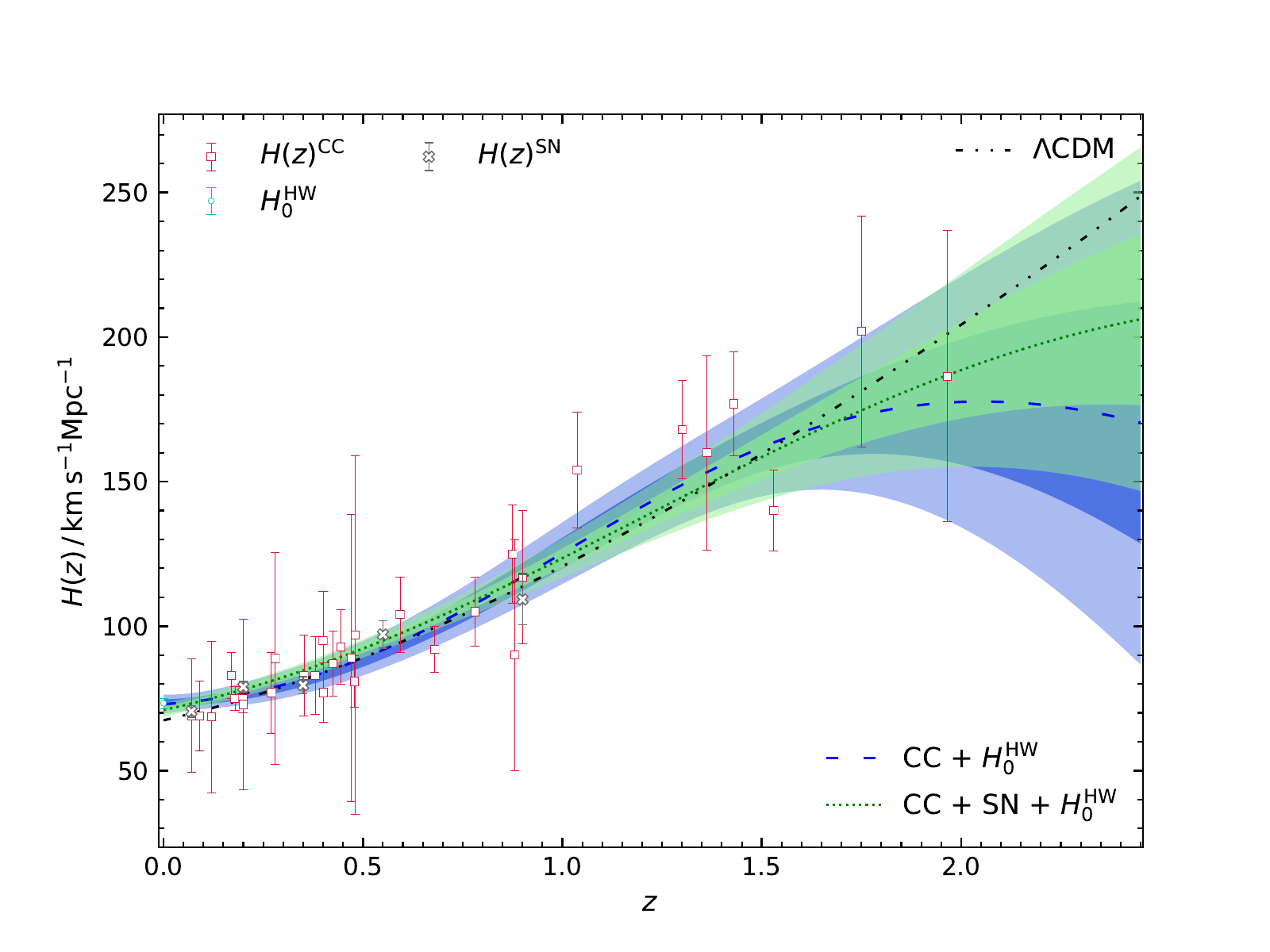}
    \caption{\label{fig:H0_HW_sq_ca_ma}{GP reconstructions of $H(z)$ with CC (blue) and CC + SN (green) data sets, and with the $H_0^{\rm HW}$ prior. We compare the GP reconstructions utilising the squared--exponential (top--left), Cauchy (top--right), and Mat\'{e}rn (bottom) kernel functions.}}
\end{center}
\end{figure*}

\begin{figure*}[h!]
\begin{center}
    \includegraphics[width=0.55\columnwidth]{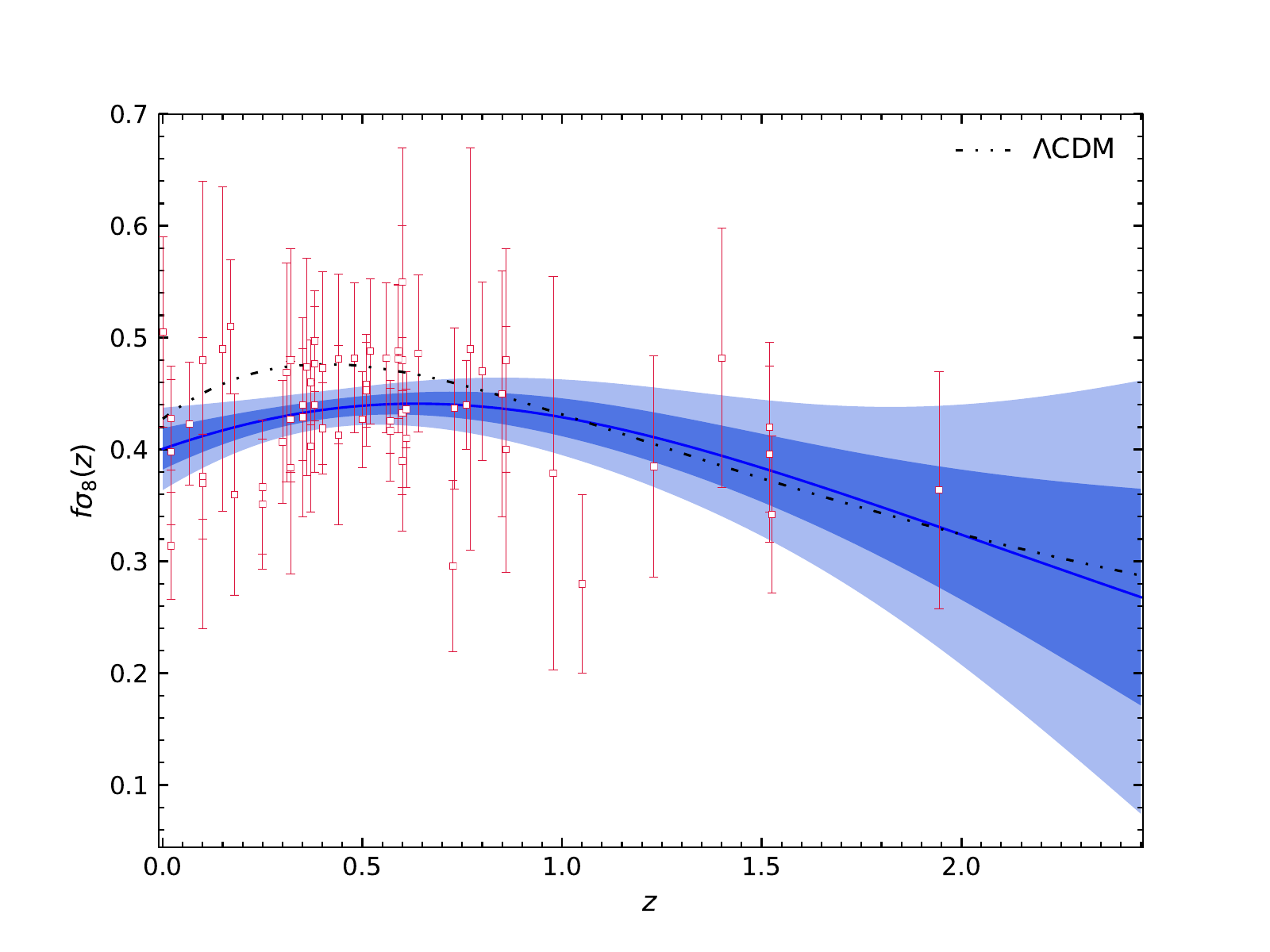}
    \caption{\label{fig:fs8_rec_ma}{GP reconstruction of $f\sigma_8^{}(z)$ with  the  Mat\'{e}rn kernel function of Eq. (\ref{eq:Matern}).}}
\end{center}
\end{figure*}

\appendix
\section{\label{sec:H0_HW_Matern}Additional comparisons}

We here illustrate the inferred GP reconstructions with the $H_0^{\rm HW}$ prior along with a comparative analysis of results obtained with the squared--exponential, Cauchy, and Mat\'{e}rn kernel functions, where the latter kernel function is given by 

\begin{equation}
\mathcal{K}\left(z,\tilde{z}\right) = \sigma_f^2 \Bigg(1 + \frac{\sqrt{3}|z-\tilde{z}|}{l_f}\Bigg)\exp\left[-\frac{\sqrt{3}|z-\tilde{z}|}{l_f}\right]\,.\label{eq:Matern}
\end{equation}

We illustrate the similarities between the $H(z)$ GP reconstructions when utilising different kernel functions in Fig. \ref{fig:H0_HW_sq_ca_ma}, while in Fig. \ref{fig:fs8_rec_ma} we depict the GP reconstruction of $f\sigma_8^{}(z)$ with the Mat\'{e}rn kernel function. From the latter figures, it should be noted that there is a very good agreement with the corresponding GP reconstructions (see, Figs. \ref{fig:H0_rec_sq_ca}--\ref{fig:fs8_rec_sq_ca}) inferred with the squared--exponential kernel function. Indeed, the current GP reconstructed value of the density--weighted growth rate with the Mat\'{e}rn kernel function is of $f\sigma_{8,0}^{}=0.40\pm0.02$, which is nearly identical to the GP reconstructed values from the squared--exponential and Cauchy kernel functions as reported in section \ref{sec:growth_data}.

\begin{figure*}[t!]
\begin{center}
    \includegraphics[width=0.42\columnwidth]{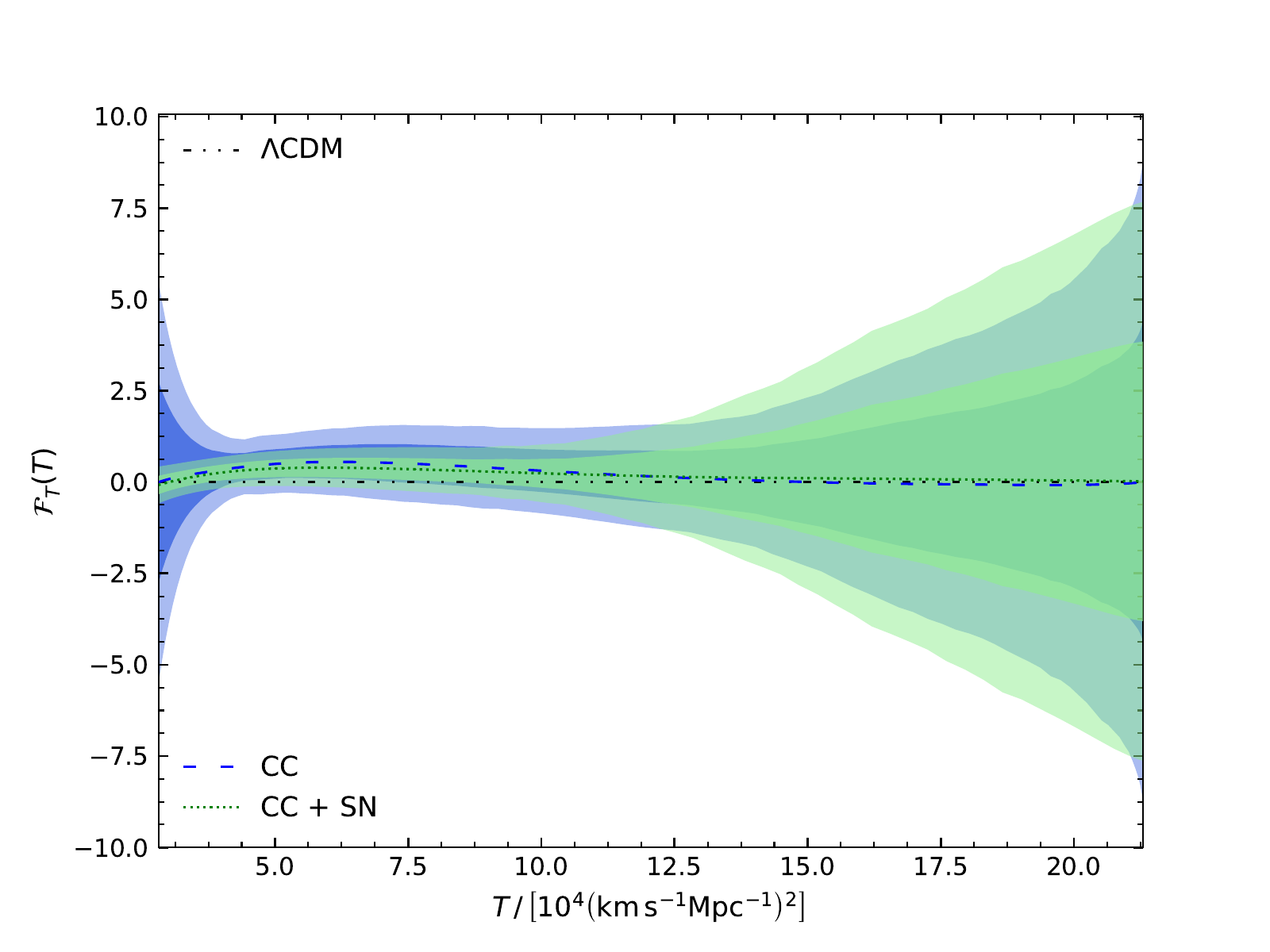}
    \includegraphics[width=0.42\columnwidth]{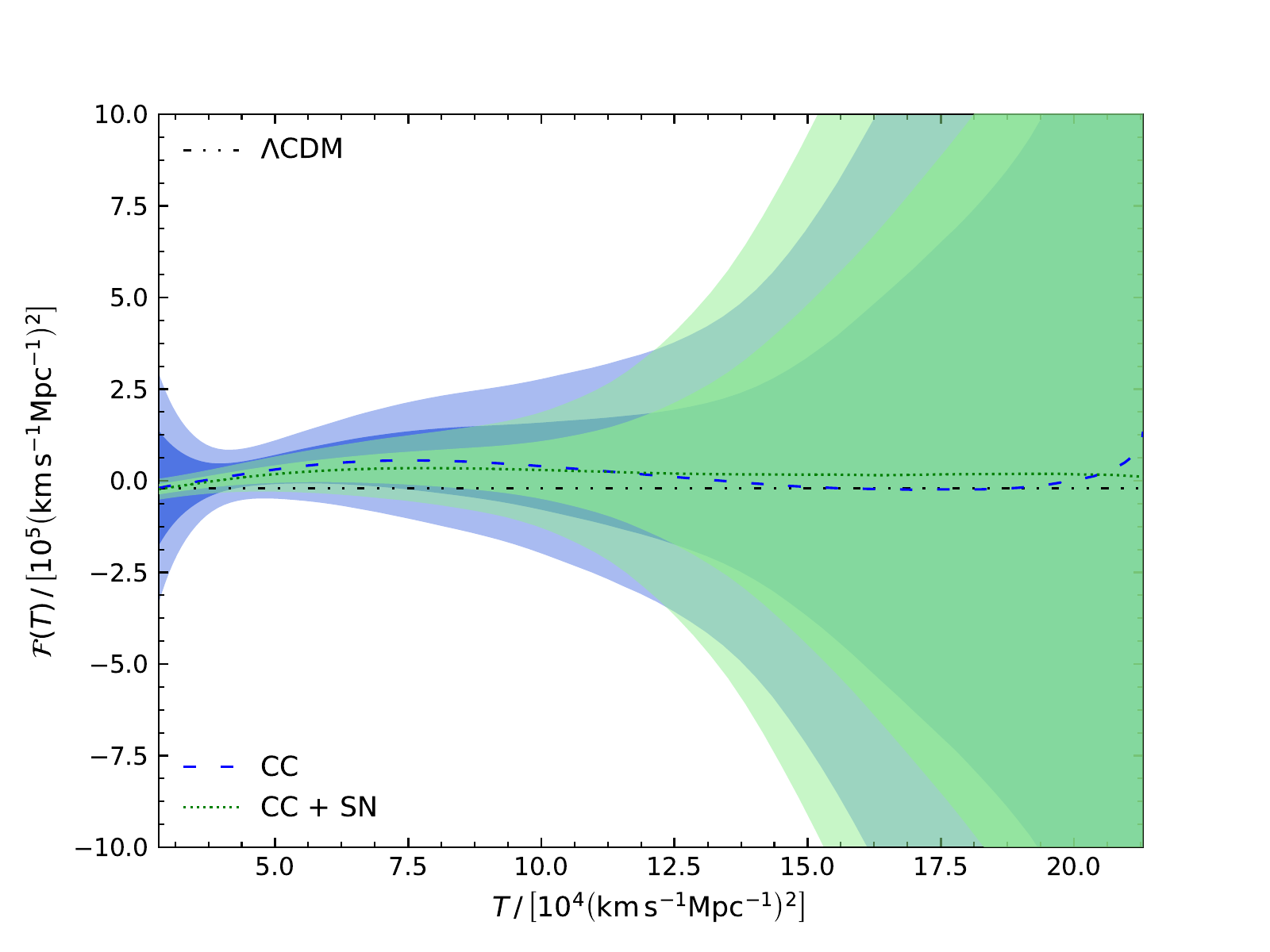}
    \includegraphics[width=0.42\columnwidth]{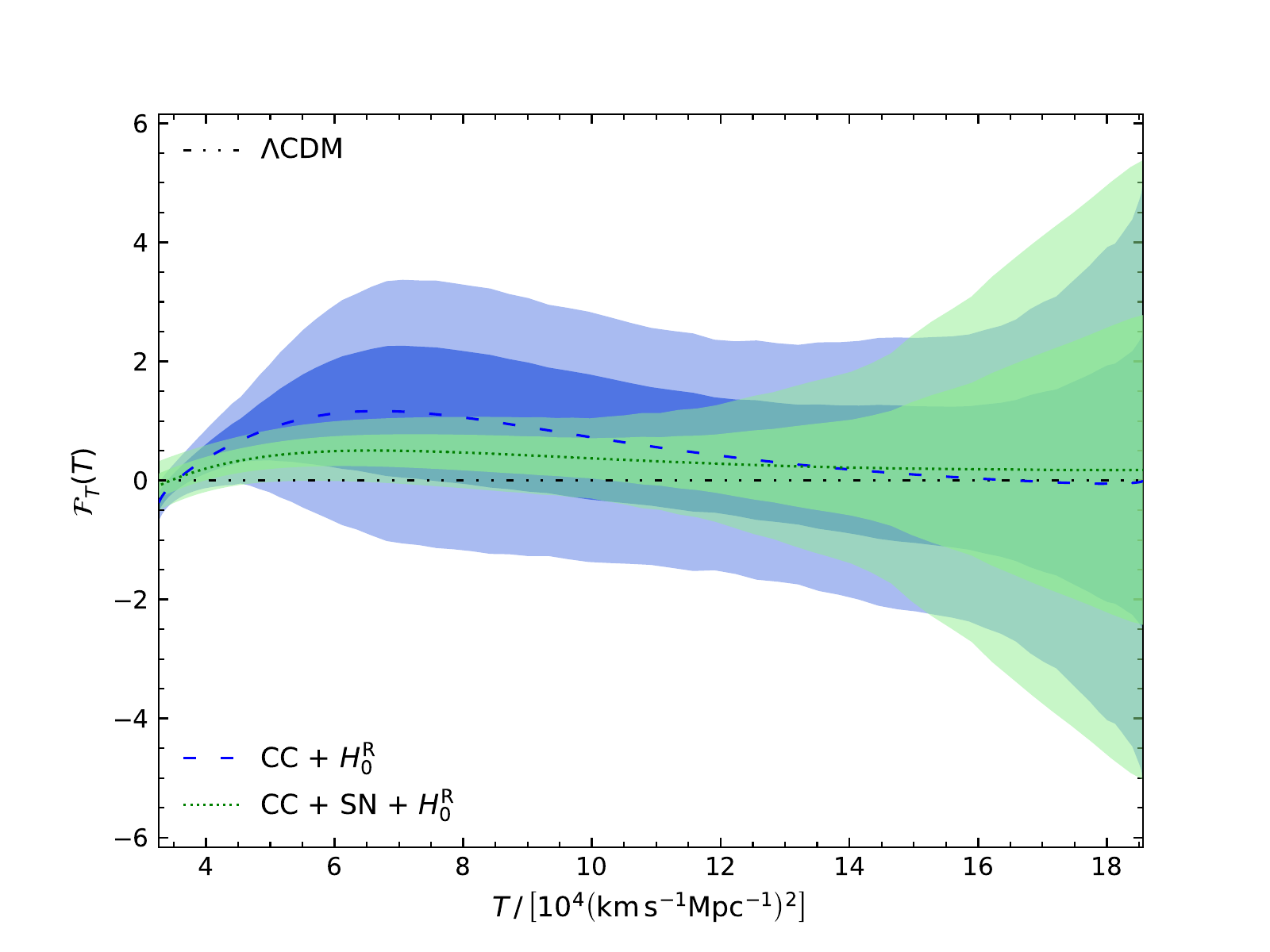}
    \includegraphics[width=0.42\columnwidth]{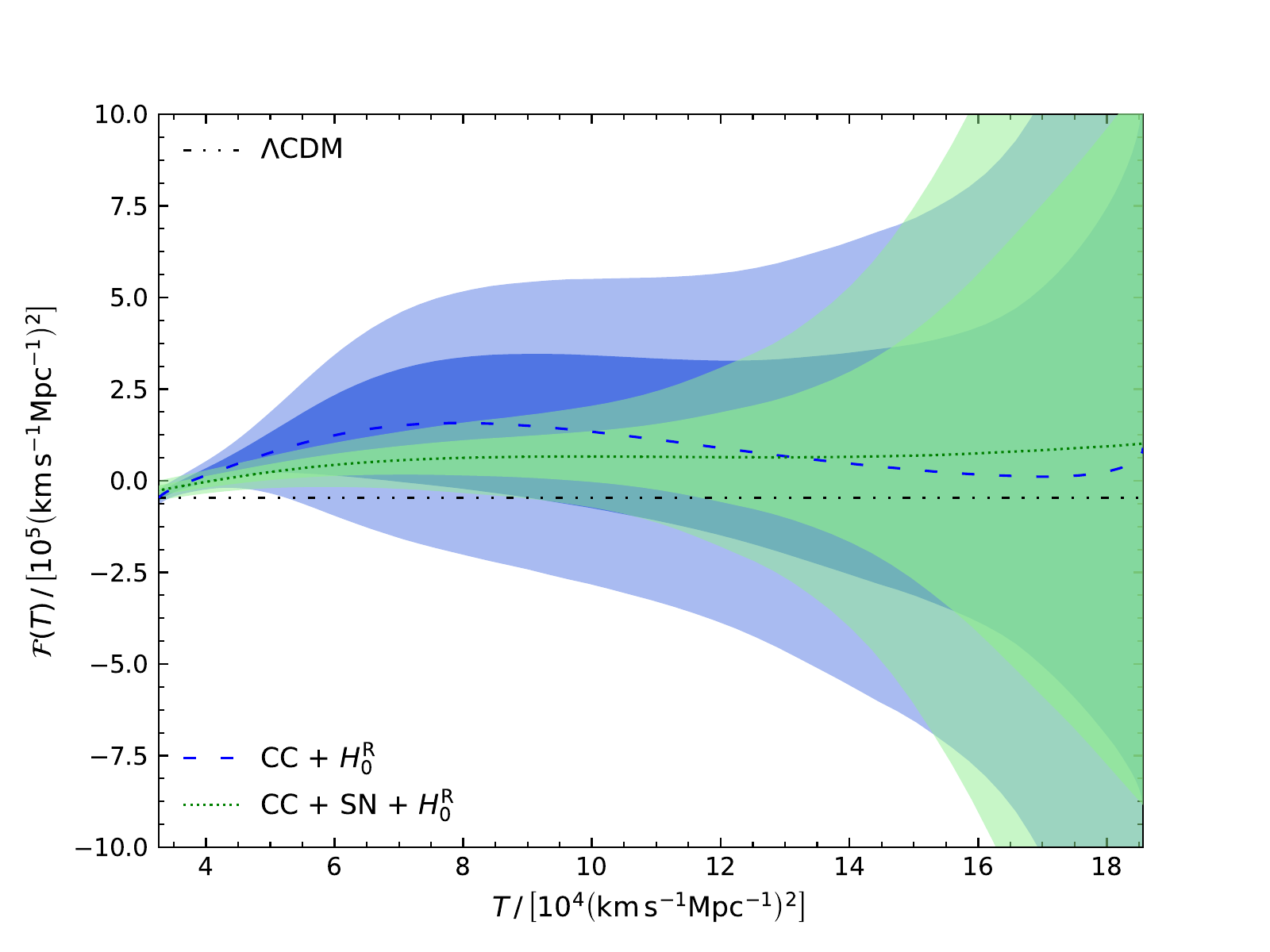}
    \includegraphics[width=0.42\columnwidth]{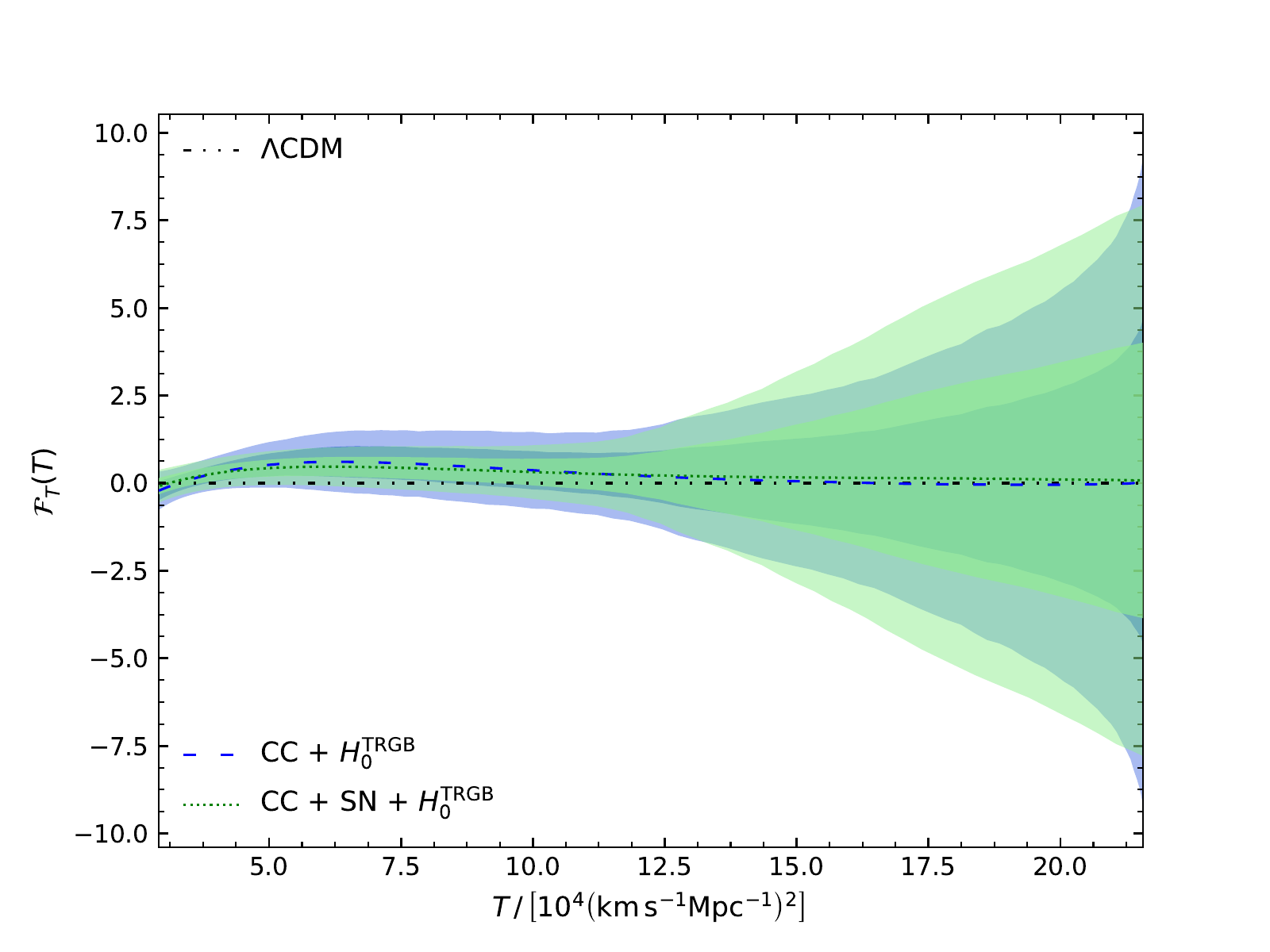}
    \includegraphics[width=0.42\columnwidth]{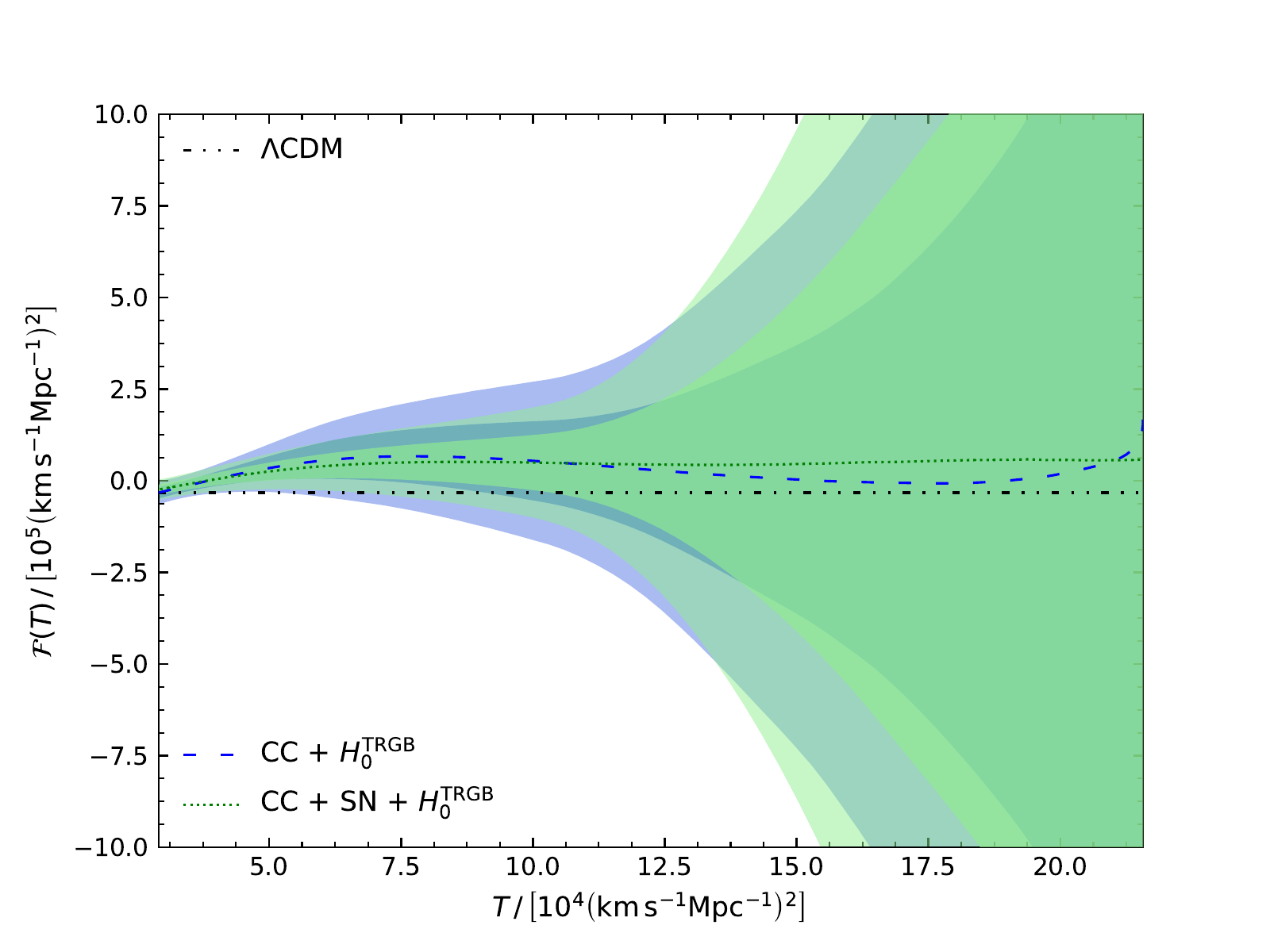}
    \includegraphics[width=0.42\columnwidth]{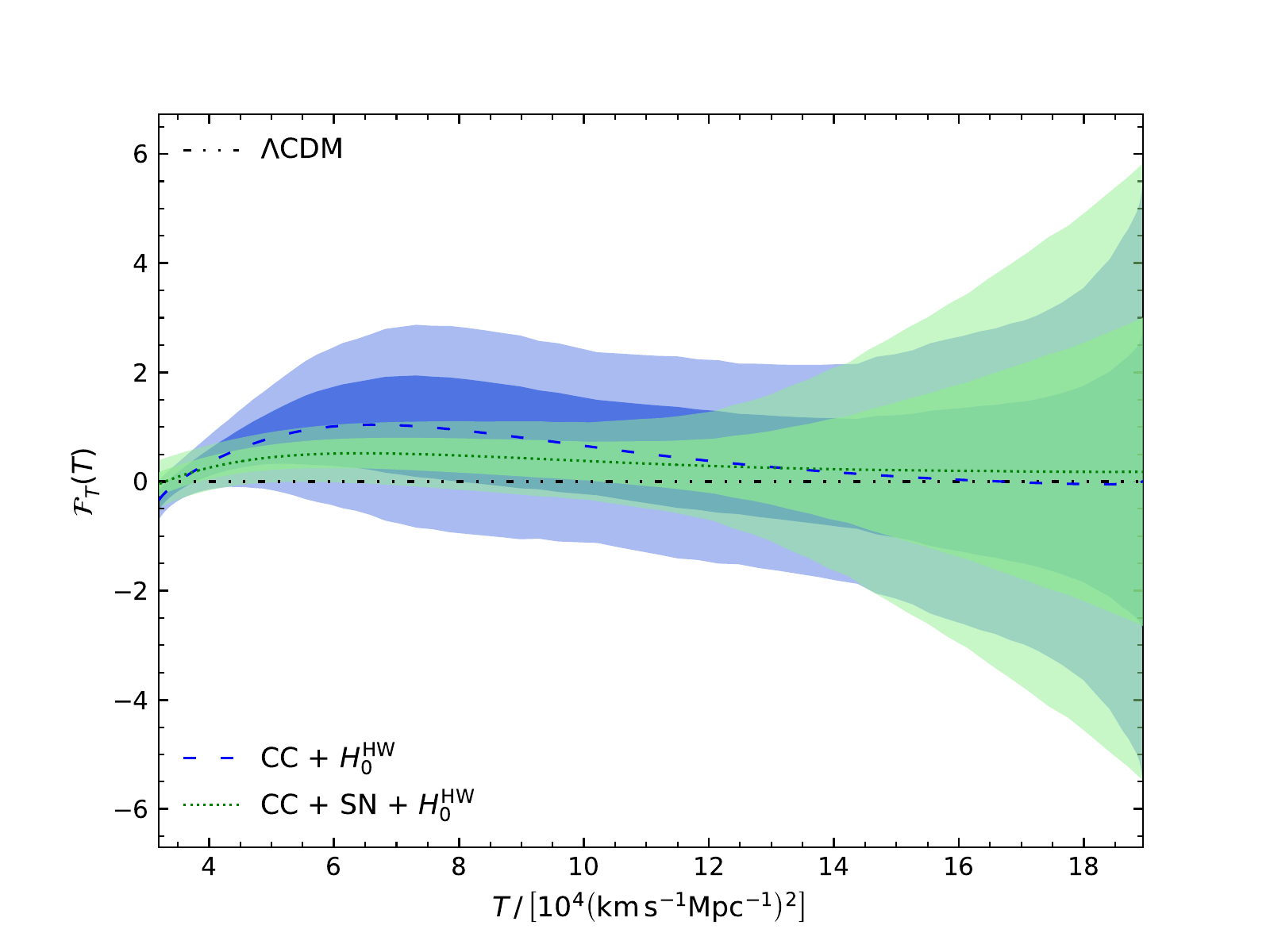}
    \includegraphics[width=0.42\columnwidth]{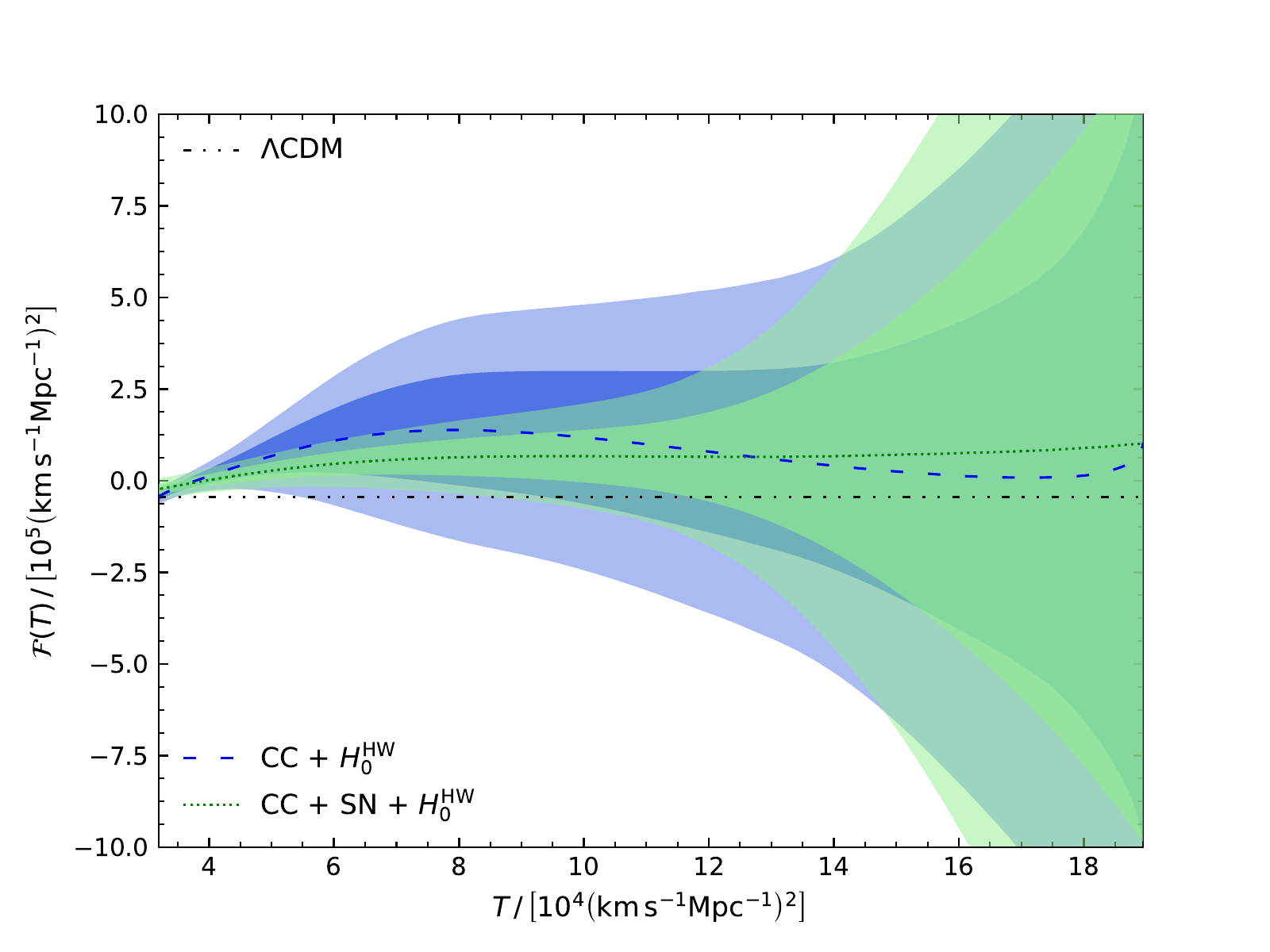}
    \caption{\label{fig:fT_rec_ma}{GP reconstructions of $\mathcal{F}_T^{}(T)$ (left) and $\mathcal{F}(T)$ (right) with CC (blue) and CC + SN (green) data sets. In all panels, we make use of the Mat\'{e}rn kernel function, and in each panel we consider one of the mentioned $H_0^{}$ priors.}}
\end{center}
\end{figure*}

Since $H_0^{\rm HW} = 73.3^{+1.7}_{-1.8} \,{\rm km\, s}^{-1} {\rm Mpc}^{-1}$ \cite{Wong:2019kwg}, lies between $H_0^{\rm R} = 74.22 \pm 1.82 \,{\rm km\, s}^{-1} {\rm Mpc}^{-1}$ \cite{Riess:2019cxk} and $H_0^{\rm TRGB} = 69.8 \pm 1.9 \,{\rm km\, s}^{-1} {\rm Mpc}^{-1}$ \cite{Freedman:2019jwv}, it is expected that the $H_0^{\rm HW}$ GP reconstructions will be in agreement with the $H_0^{\rm R}$ and $H_0^{\rm TRGB}$ GP reconstructions. Indeed, this is clearly depicted in the panels of Fig. \ref{fig:fT_rec_ma}. Moreover, since $H_0^{\rm HW}$ is closer to the mean value of $H_0^{\rm R}$ than to $H_0^{\rm TRGB}$, the resulting $H_0^{\rm HW}$ GP reconstructions are very similar to the $H_0^{\rm R}$ GP reconstructions. Consequently, we do not discuss the $H_0^{\rm HW}$ GP reconstructions in the main text, but we illustrate such similarity in this appendix for completeness purposes.

\bibliographystyle{JHEP}
\bibliography{references}

\end{document}